\documentclass[11pt,canadian]{article}
\usepackage[T1]{fontenc}
\usepackage[latin9]{inputenc}
\usepackage[letterpaper]{geometry}
\usepackage{color}
\usepackage{babel}
\usepackage{array}
\usepackage{longtable}
\usepackage{booktabs}
\usepackage{units}
\usepackage{multirow}
\usepackage{amsmath}
\usepackage{amsthm}
\usepackage{amssymb}
\usepackage{stackrel}
\usepackage{graphicx}
\usepackage{rotfloat}
\usepackage{setspace}
\usepackage[authoryear]{natbib}
\usepackage{xargs}[2008/03/08]
\usepackage[unicode=true,pdfusetitle,
 bookmarks=true,bookmarksnumbered=false,bookmarksopen=false,
 breaklinks=false,pdfborder={0 0 0},pdfborderstyle={},backref=false,colorlinks=true]
 {hyperref}
\hypersetup{
 linkcolor=pigment, citecolor=pigment, urlcolor=pigment}

\makeatletter

\newcommand{\noun}[1]{\textsc{#1}}
\providecommand{\tabularnewline}{\\}

\definecolor{pigment}{rgb}{0.2, 0.2, 0.6}
\usepackage{arydshln}
\usepackage{breakcites}
\usepackage{breakurl}
\usepackage{amsfonts}
\usepackage{booktabs}
\usepackage{floatpag}
\usepackage{authblk}

\PassOptionsToPackage{position=top}{subfig}

\@ifundefined{showcaptionsetup}{}{%
 \PassOptionsToPackage{caption=false}{subfig}}
\usepackage{subfig}
\makeatother

\begin{document}
\global\long\def\argparentheses#1{\mathopen{\left(#1\right)}\mathclose{}}%
\global\long\def\ap#1{\argparentheses{#1}\mathclose{}}%

\global\long\def\var#1{\mathbb{V}\argparentheses{#1}}%
\global\long\def\pr#1{\mathbb{P}\argparentheses{#1}}%
\global\long\def\ev#1{\mathbb{E}\argparentheses{#1}}%

\global\long\def\smft#1#2{\sum_{#1}^{#2}}%
\global\long\def\smo#1{\sum_{#1}}%
\global\long\def\prft#1#2{\prod_{#1}^{#2}}%
\global\long\def\pro#1{\prod_{#1}}%
\global\long\def\uno#1{\underset{#1}{\bigcup}}%

\global\long\def\order#1{\mathcal{O}\argparentheses{#1}}%
\global\long\def\R{\mathbb{R}}%
\global\long\def\Q{\mathbb{Q}}%
\global\long\def\N{\mathbb{N}}%
\global\long\def\Z{\mathbb{Z}}%
\global\long\def\F{\mathcal{F}}%

\global\long\def\mathtext#1{\mathrm{#1}}%
\global\long\def\mt#1{\mathtext{#1}}%

\global\long\def\maxo#1{\underset{#1}{\max\,}}%
\global\long\def\argmaxo#1{\underset{#1}{\mathtext{argmax}\,}}%
\global\long\def\minargmaxo#1{\underset{#1}{\mathtext{minargmax}\,}}%
\global\long\def\argsupo#1{\underset{#1}{\mathtext{argsup}\,}}%
\global\long\def\supo#1{\underset{#1}{\sup\,}}%
\global\long\def\info#1{\underset{#1}{\inf\,}}%
\global\long\def\mino#1{\underset{#1}{\min\,}}%
\global\long\def\argmino#1{\underset{#1}{\mathtext{argmin}\,}}%
\global\long\def\limo#1#2{\underset{#1\rightarrow#2}{\lim}}%
\global\long\def\supo#1{\underset{#1}{\sup}}%
\global\long\def\info#1{\underset{#1}{\inf}}%

\global\long\def\b#1{\boldsymbol{#1}}%
\global\long\def\ol#1{\overline{#1}}%
\global\long\def\ul#1{\underline{#1}}%

\global\long\def\argparentheses#1{\mathopen{\left(#1\right)}\mathclose{}}%
\global\long\def\mathtext#1{\mathrm{#1}}%
\global\long\def\mt#1{\mathtext{#1}}%

\newcommandx\der[3][usedefault, addprefix=\global, 1=, 2=, 3=]{\frac{\mt d^{#2}#3}{\mt d#1^{#2}}}%
\newcommandx\pder[3][usedefault, addprefix=\global, 1=, 2=]{\frac{\partial^{#2}#3}{\partial#1^{#2}}}%
\global\long\def\intft#1#2#3#4{\int\limits _{#1}^{#2}#3\mt d#4}%
\global\long\def\into#1#2#3{\underset{#1}{\int}#2\mt d#3}%
\global\long\def\intf#1#2{\int#1\mt d#2}%

\global\long\def\th{\theta}%
\global\long\def\la#1{~#1~}%
\global\long\def\laq{\la =}%
\global\long\def\normal#1#2{\mathcal{N}\argparentheses{#1,\,#2}}%
\global\long\def\uniform#1#2{\mathcal{U}\argparentheses{#1,\,#2}}%
\global\long\def\I#1#2{\mbox{I}_{#1}\argparentheses{#2}}%
\global\long\def\chisq#1{\chi_{#1}^{2}}%
\global\long\def\dar{\,\Longrightarrow\,}%
\global\long\def\dal{\,\Longleftarrow\,}%
\global\long\def\dad{\,\Longleftrightarrow\,}%
\global\long\def\norm#1{\left\Vert #1\right\Vert }%
\global\long\def\code#1{\mathtt{#1}}%
\global\long\def\descr#1#2{\underset{#2}{\underbrace{#1}}}%
\global\long\def\NB{\mathcal{NB}}%
\global\long\def\e#1{{\scriptstyle \cdot10^{#1}}}%
\global\long\def\P{\mathbb{P}}%
\global\long\def\pb#1{\Bigg(#1\Bigg)}%

\global\long\def\l#1{l\argparentheses{#1}\mathclose{}}%

\global\long\def\pb#1{\Bigg(#1\Bigg)}%
\global\long\def\cpb#1{\Big[#1\Big]}%

\global\long\def\vv#1{\boldsymbol{#1}}%

\global\long\def\True{\mathtt{True}}%
\global\long\def\False{\mathtt{False}}%

\global\long\def\mat#1{\mt{#1}}%
\global\long\def\mm#1{\underline{#1}}%

\global\long\def\pl{\ell_{\mt{PL}}}%

\global\long\def\wt#1{\widetilde{#1}}%
\global\long\def\test#1#2#3{Z_{#1+#2}(#3)}%

\global\long\def\gammadist#1#2{\mt{Gamma}\ap{#1,\,#2}}%
\global\long\def\poissondist#1{\mt{Poisson}\ap{#1}}%
\global\long\def\binomdist#1#2{\mt{Binomial}\ap{#1,\,#2}}%
\global\long\def\bernoullidist#1{\mt{Bernoulli}\argparentheses{#1}}%

\global\long\def\besJ#1#2{\mt{BesselJ}\ap{#1,\,#2}}%
\global\long\def\besY#1#2{\mt{BesselY}\ap{#1,\,#2}}%
\global\long\def\meiG#1#2{\mt{MeijerG}\ap{#1,\,#2}}%

\global\long\def\mmod{\mt{\ mod\ }}%

\global\long\def\hypGI#1#2#3{\phantom{}_{1}F_{1}\ap{#1,#2;#3}}%
\global\long\def\hypG#1#2#3#4{\phantom{}_{2}F_{1}\ap{#1,#2;#3;#4}}%
\global\long\def\ndim{n_{\mt{dim}}}%
\global\long\def\ndata{n_{\mt{data}}}%
\global\long\def\nsample{n_{\mt{sample}}}%
\global\long\def\nplots{n_{\mt{plots}}}%
\newcommandx\nseedi[1][usedefault, addprefix=\global, 1=]{N_{\mt{seed},#1}}%
\newcommandx\pmaxi[1][usedefault, addprefix=\global, 1=]{p_{\mt{max},#1}}%
\newcommandx\alphai[1][usedefault, addprefix=\global, 1=]{\alpha_{#1}}%
\newcommandx\iseedi[1][usedefault, addprefix=\global, 1=]{I_{\mt{seed},#1}}%
\global\long\def\stdGrowth{\sigma_{\mt{growth}}}%
\global\long\def\meanGrowth{\mu_{\mt{growth}}}%
\global\long\def\stdSampling{\sigma_{\mt{sampling}}}%
\global\long\def\un#1#2{\unit[#1]{#2}}%

\global\long\def\npp{\Delta B}%
\global\long\def\gpp{P}%
\global\long\def\cue{C}%
\global\long\def\entropy{S}%

\author[1,2,*]{Samuel M. Fischer}
\author[3]{Xugao Wang}
\author[1,2,4]{Andreas Huth}

\affil[1]{Helmholtz Centre for Environmental Research -- UFZ, Dept. of Ecological Modelling, Permoserstr. 15, 04318 Leipzig, Germany.}
\affil[2]{Osnabrück University, Institute of Environmental Systems Research, Barbarastr. 12, 49076 Osnabrück, Germany.}
\affil[3]{Chinese Academy of Sciences, Institute of Applied Ecology, PO Box 417, Shenyang 110016, China.}
\affil[4]{German Centre for Integrative Biodiversity Research (iDiv) Halle-Jena-Leipzig, Puschstr. 4, 04103 Leipzig, Germany.}
\affil[*]{samuel.fischer@ufz.de}

\renewcommand\Affilfont{\itshape\scriptsize\raggedright}

\date{}
\title{Distinguishing mature and immature trees allows to estimate forest
carbon uptake from stand structure}

\maketitle

\begin{abstract}
Relating forest productivity to local variations in forest structure
has been a long-standing challenge. Previous studies often focused
on the connection between forest structure and stand-level photosynthesis
(GPP). However, biomass production (NPP) and net ecosystem exchange
(NEE) are also subject to respiration and other carbon losses, which
vary with local conditions and life history traits. Here, we use a
simulation approach to study how these losses impact forest productivity
and reveal themselves in forest structure. We fit the process-based
forest model Formind to a $\un{25}{~ha}$ inventory of an old-growth
temperate forest in China and classify trees as ``mature'' (full-grown)
or ``immature'' based on their intrinsic carbon use efficiency.
Our results reveal a strong negative connection between the stand-level
carbon use efficiency and the prevalence of mature trees: GPP increases
with the total basal area, whereas NPP and NEE are driven by the basal
area of immature trees. Accordingly, the basal area entropy -- a
structural proxy for the prevalence of immature trees -- correlated
well with NPP and NEE and had a higher predictive power than other
structural characteristics such as Shannon diversity and height standard
deviation. Our results were robust across spatial scales ($\un{0.04\text{-}1}{ha}$)
and yield promising hypotheses for field studies and new theoretical
work.
\end{abstract}
\begin{description}
\item [{Keywords:}] carbon balance, carbon use efficiency, forest structure,
modelling, primary production
\end{description}

\section{Introduction}

Understanding the drivers of forest productivity is key for assessing
forests' ability to provide ecosystem services (e.g. carbon sequestration
or commercial wood production) and to gauge their resilience against
disturbances and global change \citep{costanza_value_1998,anav_spatiotemporal_2015,jha_assessment_2019,sheil_interpreting_2020}.
Forests' net primary production (NPP) may be affected via two pathways:
carbon supply, i.e., gross primary production (GPP), and carbon losses
due to respiratory costs and other limiting factors \citep{wiley_reevaluation_2012}.
Forest structure (e.g. density, species composition, age and size
distribution; \citealp{mcelhinny_forest_2005}) can be both a factor
and result of processes acting on either of these pathways \citep{waide_relationship_1999,forrester_review_2016,sheil_interpreting_2020}.
For example, denser forests may exhibit a larger total leaf area
and hence higher stand productivity. Conversely, high productivity
of individual trees may lead to denser forests.  Hence, identifying
the connection between forest structure and productivity is key for
a comprehensive understanding of forest productivity.

Several studies have established links between forest structure and
carbon supply \citep{waide_relationship_1999,forrester_review_2016}.
For example, GPP is expected to benefit from higher diversity via
improved exploitation of ecological niches and reduced competition,
and vertically stratified forests may allow for more efficient light
use due to denser leaf packaging \citep{forrester_review_2016,bohn_importance_2017}.
Nonetheless, it has proven difficult to identify clear relationships
between forest structure and NPP \citep{chisholm_scale-dependent_2013},
as several factors interact \citep{forrester_review_2016} and NPP
is not only subject to supply-related but also loss-related factors.
A unified framework for forest productivity therefore also needs to
address the corresponding role of losses. This is the subject of
this study.

A tree's ability to utilize acquired carbon to form biomass may be
expressed through its carbon use efficiency ($\text{CUE}=\text{NPP}/\text{GPP}$).
In the absence of shading by larger plants, the CUE is expected to
decline with tree size, as larger trees have a higher demand for respiration
and non-structural carbon \citep{collalti_plant_2020,binkleyAcornReviewPersistent2023}.
Such respiratory losses and other, external, factors may induce site-dependent
tree size maxima, at which biomass accumulation is significantly reduced.
The resulting decline of NPP with forest age is well documented on
the stand level \citep{gower_aboveground_1996,tangSteeperDeclinesForest2014,collaltiForestProductionEfficiency2020},
 but the extent at which loss-induced limitations drive variations
of NPP on the local scale is less understood \citep{chisholm_scale-dependent_2013,rodig_importance_2018}.
This, however, would be necessary for a mechanistic understanding
of the impact of loss-related factors in comparison to supply-related
factors.

To evaluate the impact of loss-induced limitations on forest productivity,
we suggest a simple classification framework: we divide trees into
full-grown (below: ``mature'') and growing (``immature'') trees
based on their intrinsic optimal CUE, i.e., the CUE the trees could
attain if their GPP was not limited by competition. Mature trees
have a low CUE and wood production but may compete with other trees,
reducing the overall forest productivity.  Forest productivity, in
turn, may be considered on different procedural levels: GPP, representing
forests' photosynthetic capacity; NPP, denoting their total wood production
after respiratory losses; and the net ecosystem exchange (NEE), measuring
the total forest carbon sequestration in the presence of emissions
from deadwood decomposition and soil respiration. 

Studying the impact of loss-induced growth limits, we focused on three
questions: 
\begin{enumerate}
\item How do GPP, NPP, and NEE depend on the prevalence of mature and immature
trees?
\item How can these relationships be linked to forest structure and expressed
via easily measurable forest characteristics?
\item On which spatial scales can these relationships be observed?
\end{enumerate}
To answer these questions, local carbon fluxes must be identified.
Though NPP may be estimated from inventory data, field data for GPP
and NEE, e.g. from eddy covariance measurements, are typically only
available for larger scales (about $\un{10}{ha}$). Similarly, it
can be difficult to determine which trees have reached the mature
stage. These challenges can be addressed with process-based forest
models. These models reproduce the forest dynamics under controlled
reference conditions and provide full insight into carbon fluxes as
well as the state and growth limitations of each tree.

In this study, we used the individual-based forest gap model \noun{Formind}
\citep{bohn_climate_2014,fischer_lessons_2016}. The model features
submodels on regeneration, competition, growth, and mortality and
has been successfully applied to study forest dynamics and carbon
fluxes in a variety of both temperate and tropical forests \citep{fischer_lessons_2016}.
We parameterized the model to mimic the dynamics of a species-rich
old-growth temperate forest in Changbaishan, China. Located in a natural
reserve, this forest offers unique opportunities to study long-term
forest dynamics without biases introduced by human interventions. 

We addressed the research questions by computing GPP, NPP, and NEE
on different spatial scales ($\un{0.04}{ha}$ and $\un 1{ha}$) and
setting them into relation with the basal area of mature and immature
trees as well as different measures for structural diversity. For
question (2), we suggest the DBH entropy, a measure for the diversity
of tree heights, as a general proxy for the prevalence of immature
trees and therefore also forest productivity.

\section{Materials and Methods}

\begin{figure}
\begin{centering}
\includegraphics[width=0.8\textwidth]{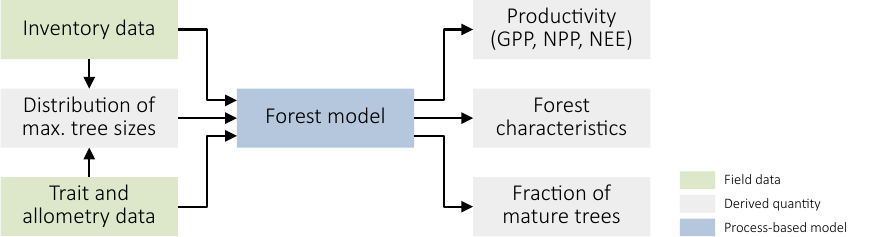}
\par\end{centering}
\caption{Summary of our approach. We use forest inventory data and data on
species' traits and allometric relationships to derive the distribution
of maximal plant sizes and parameterize a process-based forest model.
This model, in turn, yields productivity metrics (GPP, NPP, and NEE)
and different forest characteristics, including the fraction of mature
trees.\label{fig:Flow-chart}}
\end{figure}

We applied a data-driven modelling approach (Fig. \ref{fig:Flow-chart})
to analyze the relationship between forest structure and forest productivity.
We fitted the process-based forest model \noun{Formind} to forest
inventory data from Changbaishan, China, and data on species' traits
and allometric relationships. Using the model, we then linked forest
productivity to the prevalence of mature trees and other forest characteristics.
Below we describe the individual steps in detail.

\subsection{Field data }

We based our analysis on forest inventory data from an old-growth
temperate forest in the Changbaishan National Nature Reserve in northeastern
China. The surveyed area consists of $\un{25}{ha}$ of conifer/broad-leaf
mixed forest with $47$ species, a total biomass of $\unitfrac[302]{t\,ODM}{ha}$
\citep{piponiot_distribution_2022}. The inventory data contain the
position, diameter at breast height (DBH) and species of each tree
with $\mt{DBH}\geq1\mt{cm}$ for the census years $2004$, $2009$,
and $2014$. Each tree is uniquely identified with an ID number. For
trees that had multiple stems at breast height, we focused on the
main stem (maximal DBH) in our analysis and we disregarded minor stems. 

In addition to the inventory data, we used information on traits and
allometry of the species from field measurements. These data included
DBH-dependent heights, crown radii and crown base heights. Furthermore,
the dataset included the species' wood densities and shade tolerance
types (``light demanding'', ``mid-tolerant'', or ``shade tolerant'').
Not all of these data were available for all species; we provide details
in Supplementary Information (SI) \ref{sec:Data-availability}. 

\subsection{Model and parameterization }

\noun{Formind} is a process-based forest gap model featuring the main
processes regeneration, competition, tree growth, and mortality \citep{fischer_lessons_2016}.
Key idea is that trees mainly interact on a local scale \citep{shugart_gap_2018}.
Trees are mainly characterized by their DBH and species. Other properties,
such as plant height or crown size, are derived from the DBH via allometric
relationships. Below we summarize the parameterization of the model
and highlight changes to the version described before in \citet{fischer_lessons_2016}.
Details can be found in SI \ref{sec:Parameterization}.

\subsubsection*{Basic parameterization}

To reduce model complexity, we aggregated species into plant functional
types (PFTs) based on their maximal DBHs (below / above $30\mt{cm}$)
and light demand (light demanding, mid-tolerant, and shade tolerant).
When data necessary for the classification were not available, we
assigned species via a likelihood-based cluster analysis based on
shade tolerance and observed tree growth (SI \ref{subsec:Classification-of-species}).
We obtained six PFTs: small light demanding, large light demanding
1 and 2, large mid-tolerant, small shade tolerant, and large shade
tolerant species. Because \emph{Q. mongolica} had a significantly
different size structure than the other light-demanding species, we
divided the large light demanding into two PFTs, one with all other
large light demanding species and one for \emph{Q. mongolica} only.
There were no small mid-tolerant species.

We estimated mean traits and allometric relationships for the PFTs
based on the trait and allometry data. When computing the means, we
weighted species according to their shares in the inventory to best
reflect the species composition in the study area. Details can be
found in SI \ref{subsec:Allometric-relationships} and \ref{subsec:Plant-traits}.
We modelled the forest under constant climatic conditions, which we
derived based on data from the literature (evapotranspiration: \citealp{sun_spatial_2004};
temperature: \citealp{wang_study_2020}) and the WFDEI forcing dataset
(irradiance, \citealp{weedon_wfdei_2014}). See SI \ref{subsec:Climate}
for details.

We estimated the DBH-dependent base mortality for each PFT applying
a likelihood-based approach to the inventory data (SI \ref{subsec:Mortality}).
To model tree growth, we focused on the carbon use efficiency ($\text{CUE}=\text{NPP}/\text{GPP}$)
of trees under optimal growth conditions (SI \ref{subsec:Growth}).
We modelled the CUE based on the following observations and assumptions:
(1) the CUE decreases as plants grow in size, (2) the CUE under optimal
conditions suffices for the observed DBH increments, (3) the CUE of
trees in the inventory suffices to satisfy their respiratory needs,
and (4) the order of magnitude of the CUE on stand level matches field
measurements approximately (see SI \ref{subsec:CUE}). 

With the maximal tree growth estimated from the census data and the
modelled CUE under optimal conditions, we computed the GPP and respiration
of the trees. We reconciled these results with \noun{Formind}'s internal
submodel for primary production by allowing trees to flexibly allocate
biomass to stem and crown dependent on their DBH. The mean ratio of
stem and crown biomass remained fixed to values chosen so that \noun{Formind}'s
estimate of the Changbaishan forest biomass matched an estimate based
on DBH-biomass relationships from the literature (\citealp{chojnacky_updated_2014,piponiot_distribution_2022};
see SI \ref{subsec:Growth-allocation}).

We assumed that trees compete for light only, but included crown defoliation
as an additional process to account for the limited capacity of a
forest. Trees whose GPP is insufficient to satisfy their respiratory
needs loose crown biomass until all remaining parts can be maintained.
Here, we assumed that -- for a tree of given DBH -- the maintenance
respiration is proportional to the biomass. We decreased the leaf
area index (LAI) of stressed trees along with their crown completeness,
i.e., the ratio between current (reduced) and healthy crown biomass.
Trees that have lost all their crown biomass die.

\subsubsection*{Model fitting}

Some of the modelled processes depend on parameters not directly inferable
from the available data. This included the following PFT-specific
parameters: (1) the external influx of new seeds, (2) the saturation
parameters of the light response curves, (3) the magnitudes of carbon
losses other than maintenance respiration, and (4) the light required
for seedling establishment. Furthermore, we fitted a parameter controlling
the magnitude of DBH growth under optimal conditions and the sharpness
of the light threshold for seedling input. 

We fitted these $18$ parameters using a likelihood-based approach
maximizing the approximate likelihood of the inventory data, estimated
from a sample of simulation results. We determined each PFT's biomass
and stem count in $20\,\mt m\times20\,\mt m$ forest patches. The
combined information of stem count and biomass yields basic insight
into the size distribution of trees: a large stem count with small
biomass indicates a young forest with many small trees, and a small
stem count with high biomass indicates an old forest with few large
trees. The inventory covered $625$ forest patches, providing us with
a similarly-sized sample of forest states. 

To generate a forest state sample from the model, we first simulated
$\un 1{ha}$ of forest  for a burn-in period of $\un{2000}{yr}$.
Then, we sampled the forest $500$ times in $\un 5{yr}$ intervals.
We repeated this procedure $67$ times in parallel, equivalent to
simulating $\un{67}{ha}$ of forest, obtaining a sample of $837,500$
forest states. 

We estimated the likelihood of the field data via kernel density estimation
\citep[KDE;][]{wand_kernel_1995}. In KDE, the probability density
of an observation is estimated based on how many model-generated sample
points are similar to the observation. Here, similarity is measured
via kernel functions, which depend on bandwidth parameters. We used
Gaussian kernels with bandwidths chosen corresponding to the scales
of the stem counts and biomasses in the inventory data (see Table
\ref{tab:Bandwidths} in SI \ref{subsec:Fitting-procedure}). To correct
for the bias introduced when log-transforming the KDE so as to compute
the log-likelihood, we applied a bias correction function derived
via a first-order Taylor approximation (SI \ref{subsec:Fitting-procedure}). 

The resulting likelihood estimate converges to the true likelihood
as the size of the generated sample increases and the bandwidth parameters
decrease. Hence, optimizing the KDE likelihood yields consistent parameter
estimates and avoids potential biases arising if the model was fitted
via a deterministic modelling framework \citep[e.g.][]{lehmannFastCalibrationDynamic2015,rodig_spatial_2017}.
However, as the log-likelihood estimate is based on a sample of stochastic
model results, it is stochastic as well, making it difficult to optimize.
We reduced the stochasticity by decreasing the dimension of the sample
space, avoiding the ``curse of dimensionality'' \citep{wand_kernel_1995}
by considering the different PFTs as mutually independent. The parameter
estimates remain consistent despite this composite likelihood approach
\citep{varin_composite_2008}. 

We maximized the likelihood by repeatedly applying a derivative-free
optimization algorithm based on non-local quadratic approximations
\citep{cartis_improving_2019}. To avoid getting stuck in local minima,
we used the basin-hopping algorithm \citep{wales_global_1997}, which
applies multiple local optimizations with randomly perturbed initial
conditions. Throughout the fitting process, we constrained the parameters
to ecologically reasonable ranges. Details on model fitting can be
found in SI \ref{subsec:Fitting-procedure}. The fitted parameter
values are provided in SI \ref{sec:Parameterization}. 

\subsubsection*{Size limitations}

We assumed that each tree has a maximal DBH at which it stops growing.
As this maximal DBH may depend on local conditions and the tree's
species, we drew the DBH limit randomly for each tree individually
(details below). Trees that have reached their DBH limit are called
``mature'' below and are assumed to use their entire primary production
for respiration.

We constructed the distributions of the DBH limits based on the maximal
DBHs of the species in each PFT: for each species, we assumed that
the site-dependent DBH limits are uniformly distributed between the
overall maximal DBH and a value $20\%$ below this maximum. We aggregated
these species-specific distributions, weighted according to the species'
respective shares in the basal area of the inventory. That way, we
obtained the joint distribution of DBH limits for each PFT. In SI
\ref{subsec:Maximal-DBH}, we describe the approach in greater detail.

\subsection{Model validation}

\begin{figure}
\subfloat[\hspace*{\fill}]{\hspace{-0.3cm}\includegraphics[scale=0.6]{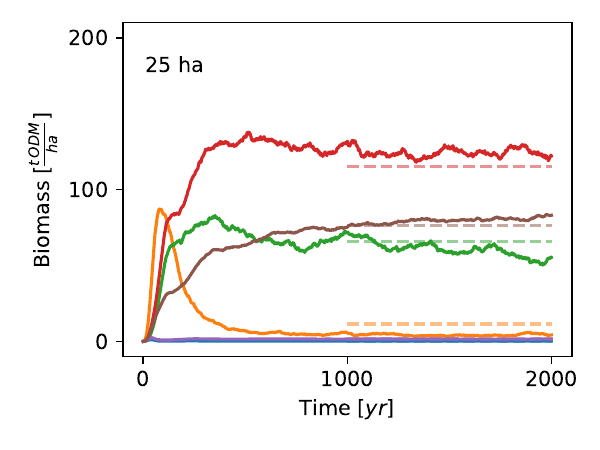}}\subfloat[\hspace*{\fill}]{\hspace{-0.2cm}\includegraphics[scale=0.6]{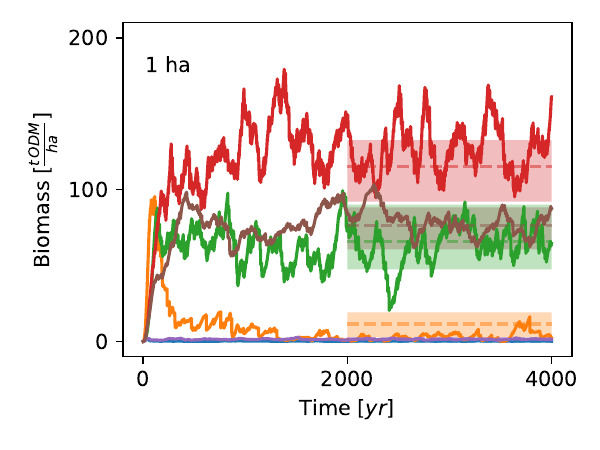}}\subfloat[\hspace*{\fill}]{\hspace{-0.2cm}\includegraphics[scale=0.6]{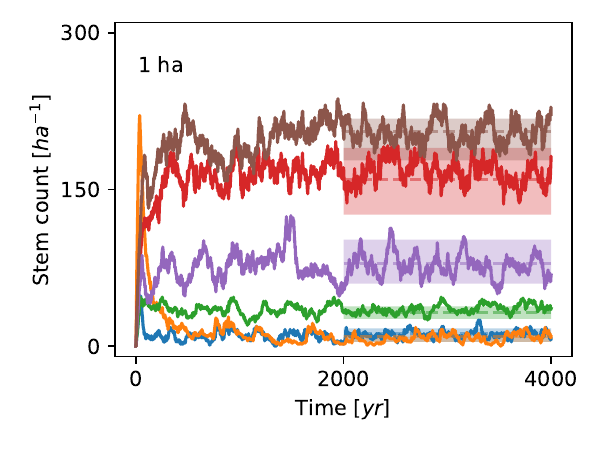}}
\noindent \begin{centering}
\subfloat{\noindent \begin{centering}
\vspace{-0.5cm}
\par\end{centering}
\includegraphics[scale=0.6]{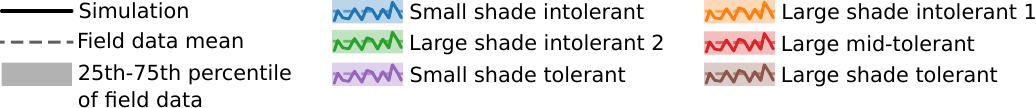}}
\par\end{centering}
\noindent \begin{centering}
\vspace{0.2cm}
\par\end{centering}
\caption{Temporal evolution of (a, b) biomass and (c) stem count of the six
PFTs on (a) the $\protect\un{25}{ha}$ scale and (b, c) the $\protect\un 1{ha}$
scale. The solid lines show the trajectory of the model simulation.
For comparison, the shaded areas depict the ranges between the 25th
and the 75th percentiles of the biomasses and stem counts from the
inventory data. The dashed lines represent the corresponding mean
values. \label{fig:succession}}
\end{figure}

We validated the fitted model by visually comparing the respective
marginal and joint distributions of the biomass and stem count values
for the considered PFTs with the corresponding distributions observed
in the field data. We created corresponding one- and two-dimensional
histograms based on both samples generated via simulations and computed
based on the forest inventory data. We observed that the simulated
trajectory and distribution of biomass and stem count matched the
values from the inventory (Fig. \ref{fig:succession}, SI \ref{sec:apx-validation}). 

To ensure the fitting algorithm did not terminate at a suboptimal
local likelihood maximum, we repeated the model fitting procedure
three times. We compared the resulting parameter estimates to assess
how well the individual parameters are estimable. The differences
between the corresponding parameter were moderate for most parameters
except the light threshold for seedling establishment (SI \ref{sec:apx-validation}). 

To validate the results on a broader scale ($\un{25}{ha}$), we furthermore
compared the modelled biomass, NPP, GPP, and LAI with values obtained
for the same forest plot in independent studies \citep{piponiot_distribution_2022}.
The simulated forest had a mean biomass of $\unitfrac[270.5]{t\,ODM}{ha}$
(estimated standard deviation for $\unit[25]{ha}$: $\unitfrac[4.38]{t\,ODM}{ha}$).
Our biomass estimates from the allometric equations by \citet{chojnacky_updated_2014}
were $\unitfrac[270.52]{t\,ODM}{ha}$ if we only considered the major
stems and $\unitfrac[284.48]{t\,ODM}{ha}$ for all stems in the inventory.
This is below the estimate by \citet{piponiot_distribution_2022}:
$\unitfrac[302]{t\,ODM}{ha}$. The simulated forest had an aboveground
wood production of $\unitfrac[2.22]{t\,ODM}{ha\cdot yr}$ (standard
deviation: $\unitfrac[0.07]{t\,ODM}{ha}$; \citealp{piponiot_distribution_2022}:
$\unitfrac[3.55]{t\,ODM}{ha\cdot yr}$) and GPP of $\unitfrac[23.39]{t\,ODM}{ha\cdot yr}$
(standard deviation: $\unitfrac[0.2]{t\,ODM}{ha}$; \citealp{wu_estimation_2009}:
$\unitfrac[29.82\text{-}33.86]{t\,ODM}{ha\cdot yr}$). The LAI of
the simulated forest was $5.18$ (standard deviation $0.05$; \citealp{liu_application_2007}:
$5.08$). See SI~\ref{sec:apx-validation} for details.

\subsection{Analysis}

To analyze the effect of mature trees on forest productivity, we simulated
$\un 1{ha}$ of the Changbaishan forest  and sampled forest characteristics
and forest productivity over time on the $\un{0.04}{ha}$ and the
$\un 1{ha}$ scale. After a burn-in period of $\un{2000}{yr}$, we
analyzed the forest $1000$ times in $\un 5{yr}$ time intervals.
We obtained a sample of $25,000$ forest states on the smaller and
$1,000$ states on the larger scale, corresponding to $\un{1000}{ha}$. 

To measure forest productivity, we computed the GPP, NPP, NEE, and
carbon use efficiency \linebreak{}
($\text{CUE}=\text{NPP}/\text{GPP}$) of the considered forest areas.
We characterized the corresponding forest states by determining the
basal area $A_{\mt{all}}$ of all trees in the forest area and the
basal area $A_{\mt{grow}}$ of only those trees that had not reached
their individual DBH limits. Based on these measures, we also determined
the basal area proportion $A_{\mt{grow}}/A_{\mt{all}}$ of immature
trees and the corresponding proportion of mature trees. Furthermore,
we computed the DBH entropy (a measure for the diversity of DBH values;
detailed explanation in section \ref{subsec:DBH-entropy}), basal-area-weighted
height standard deviation, and the Shannon diversity of PFTs on the
two considered scales. We weighted the plant heights by the basal
areas when computing the height standard deviation so as to account
for small plants having a minor impact on forest productivity.

For both considered spatial scales ($\un{0.04}{ha}$ and $\un 1{ha}$),
we plotted GPP, NPP, and NEE against the mentioned forest characteristics
and computed the respective coefficients of determination ($R^{2}$)
to quantify the strengths of the relationships. In a similar manner,
we analyzed the relationship between the basal area proportion of
mature trees and the CUE. To understand the role of the DBH entropy,
we furthermore assessed its relationship with the basal area of mature
and immature trees.

\subsection{DBH entropy as a proxy for the prevalence of mature trees\label{subsec:DBH-entropy}}

\begin{figure}
\subfloat[\hspace*{\fill}\label{fig:entropy-low}]{\includegraphics[width=0.33\textwidth]{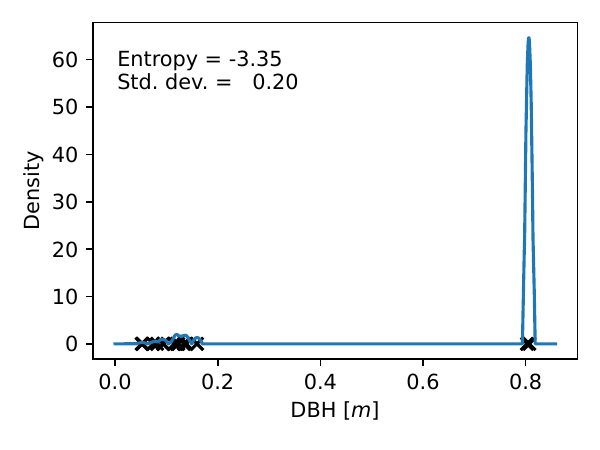}

}\hspace*{\fill}\subfloat[\hspace*{\fill}\label{fig:entropy-mid}]{\includegraphics[width=0.33\textwidth]{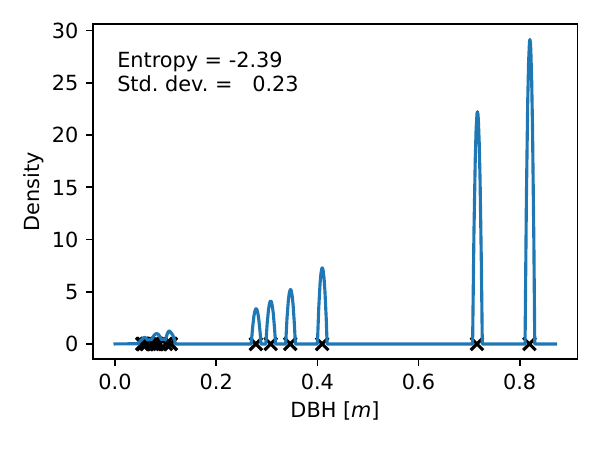}

}\hspace*{\fill}\subfloat[\hspace*{\fill}\label{fig:entropy-high}]{\includegraphics[width=0.33\textwidth]{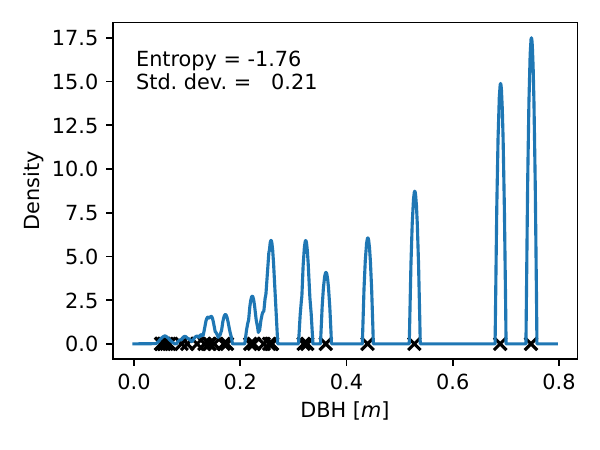}

}

\caption{Basal-area-weighted DBH distributions for $\protect\un{0.04}{ha}$
forest patches with (a) low, (b) intermediate, and (c) high entropy.
Each black cross depicts the DBH of a plant. The height of the corresponding
spike in the density function (blue line) corresponds to the plant's
share in the basal area; the contributions of trees with similar DBH
add up. The width of the spikes ($2h$; here: $2\protect\mt{cm}$)
is the scale on which different plants are considered similarly sized.
The entropy is higher the more uniformly the basal area is distributed
across plants with different DBHs. In (a), two similarly large plants
dominate the forest patch, whereas in (c), there are many medium-sized
plants with different DBHs. Note that the standard deviation of the
DBH distribution is not related to the DBH entropy.\label{fig:dbh-dists}}
\end{figure}

It is difficult to know which trees have reached their site-dependent
growth limits in field studies. Hence, a proxy for the prevalence
of mature trees is needed in practice. Such a proxy should be easy
to compute from inventory data and may account for the following working
hypotheses: (1) forest patches dominated by mature trees consist of
a small number of large individuals preventing the existence of medium-sized
trees; (2) in old-growth forests, individuals typically differ in
age and size, but mature individuals of the same species may have
similar DBH values. The proxy should also reflect that large trees
have a higher impact on forest dynamics than small trees.

As a proxy satisfying these requirements, we propose the basal-area-weighted
DBH entropy $\entropy_{\mt{DBH}}$ (below simply ``DBH entropy''),
defined as the entropy of the distribution of DBHs in a forest patch
\citep[cf.][]{staudhammer_introduction_2001,park_influence_2019}.
If we split the range of occurring DBH values into equally sized intervals
$I$ and determined the basal area share $p_{I}$ of trees in each
size class $I$ relative to the total total basal area,  the DBH
entropy could be approximated via 
\begin{equation}
\entropy_{\mt{DBH}}=-\smo{I\in\mathcal{I}}p_{I}\ln\ap{p_{I}}.\label{eq:entropy-discrete}
\end{equation}
Here, $\mathcal{I}$ is the set of DBH classes and 
\begin{equation}
p_{I}=\frac{\smo{d\in I}d^{2}}{\smo{I\in\mathcal{I}}\smo{d\in I}d^{2}}\label{eq:d-int-prob}
\end{equation}
is the basal area share of trees in size class $I$. 

The weights $p_{I}$ can be interpreted as probabilities indicating
how likely we would obtain a tree from size class $I$ if we randomly
selected trees from the forest patch with probabilities proportional
to their basal areas. The entropy is higher the more evenly the the
DBHs are distributed (Fig. \ref{fig:dbh-dists}). If the forest patch
is dominated by one or a few large trees, it is likely that we draw
one of their size classes, making the entropy small. Similarly, if
two trees have a similar DBH, the probability to pick a tree from
their size class increases, decreasing the entropy. Since we weight
the DBH distribution by the basal areas, adding small trees to the
forest patch does not change the entropy significantly. 

As the approach presented above is sensitive to the specific choice
of interval bounds, we used a more robust definition of the DBH entropy
in our analysis (SI \ref{subsec:DBH-entropy-APX}). We applied kernel
smoothing \citep{wand_kernel_1995} with an Epanechnikov kernel to
obtain a continuous estimate of the DBH distribution instead of discrete
probabilities $p_{I}$ (cf. Fig. \ref{fig:dbh-dists}), and we exchanged
the sum in equation (\ref{eq:entropy-discrete}) with an integral.
Kernel smoothing requires a bandwidth parameter (here: $1\mt{cm}$),
which is comparable to the width of the DBH intervals $I$ and defines
the scale on which two trees are regarded similar.

\section{Results}

\begin{sidewaysfigure}
\includegraphics[width=1\textheight]{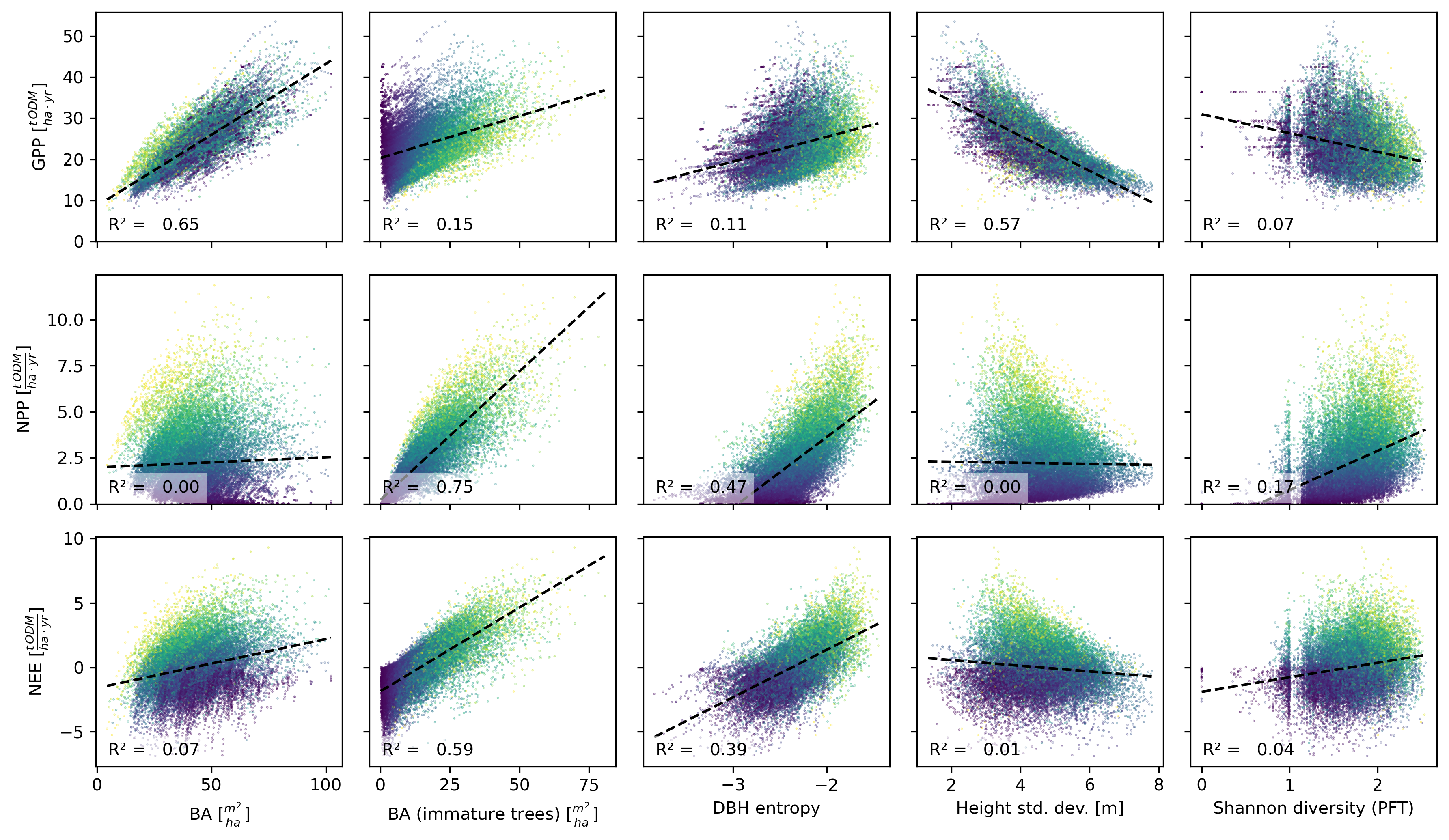}\caption{Productivity measures (GPP, NPP, and NEE) dependent on different measures
of basal area (BA) and heterogeneity. Each dot corresponds to a $\protect\un{0.04}{ha}$
forest patch (sample size: $25,000$). The colour indicates the basal
area proportion of mature trees (blue: only mature trees; yellow:
no mature trees). The GPP is mainly driven by the basal area, whereas
NPP and NEE are driven by the basal area of immature trees. The heterogeneity
measures are generally poorer predictors than the basal area measures.
Among the former, the DBH entropy has the best predictive capacity
for NPP and NEE and may serve as a valuable proxy if distinguishing
mature and immature trees is not possible.\label{fig:ProductivityBA-diversity-patch}}
\end{sidewaysfigure}

\begin{sidewaysfigure}
\includegraphics[width=1\textheight]{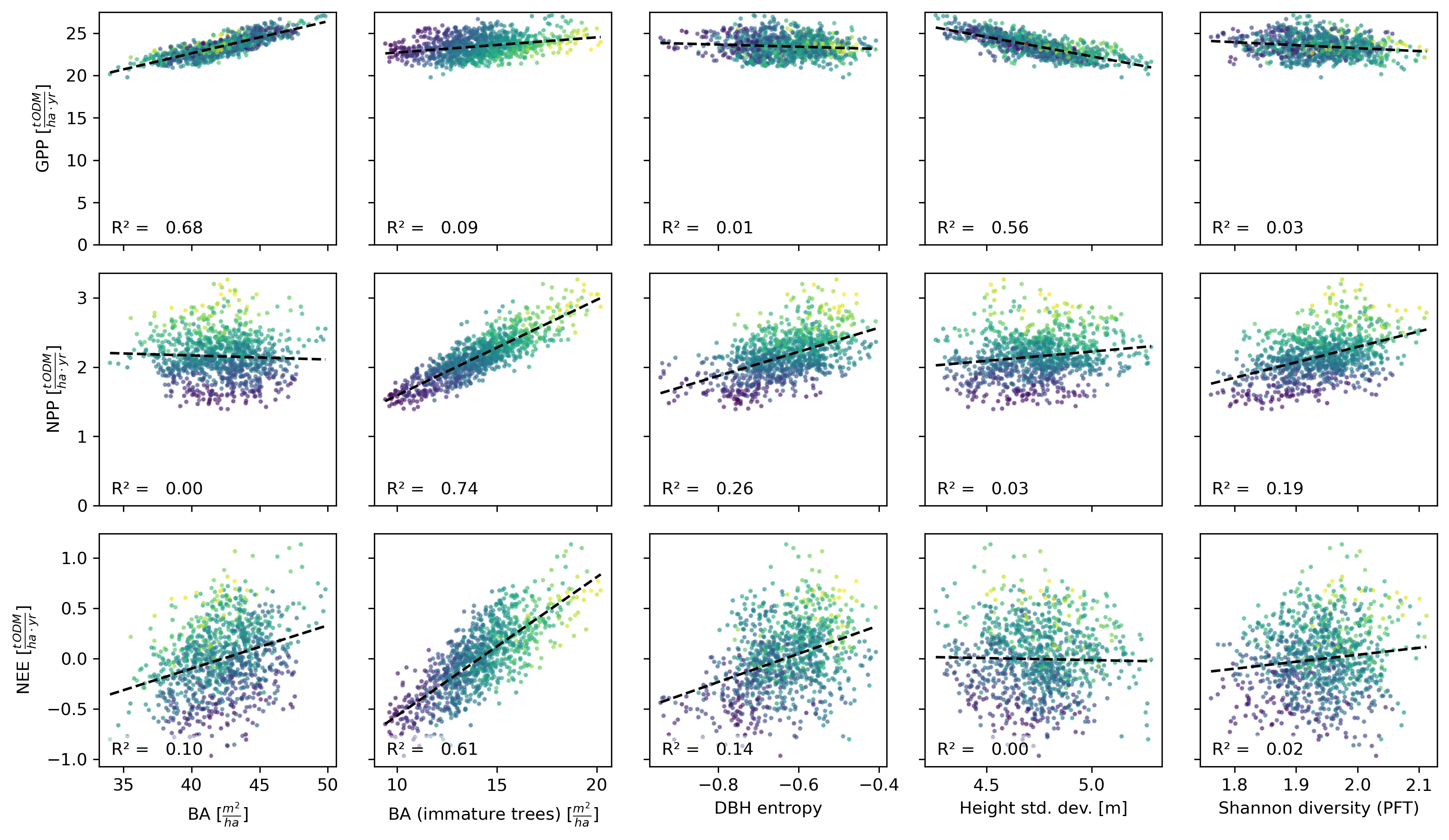}\caption{Productivity measures dependent on different measures for basal area
and heterogeneity. Each dot corresponds to a $\protect\un 1{ha}$
forest patch (sample size: $1,000$). The colour indicates the basal
area proportion of mature trees (blue: only mature trees; yellow:
no mature trees). The correlation patterns resemble those observed
on the finer scale (Fig. \ref{fig:ProductivityBA-diversity-patch}).
Only the DBH entropy looses predictive power.\label{fig:Productivity-BA-diversity-ha}}
\end{sidewaysfigure}

The basal area of the forest stand was strongly correlated with the
GPP, irrespective of the spatial scale ($R^{2}\geq0.65$). For the
NEE, these correlations were much weaker ($R^{2}\leq0.1$) and for
the NPP merely existent ($R^{2}=0$). This contrasts with the basal
area of immature trees. Here, the correlations were small for the
GPP ($R^{2}\leq0.15$) but large for the NPP ($R^{2}\geq0.74$) and
the NEE ($R^{2}\geq0.59$). We obtained a similar but slightly weaker
result for the DBH entropy. On the small scale ($\un{0.04}{ha}$),
it was weakly correlated with the GPP ($R^{2}=0.11$) but strongly
correlated with NPP ($R^{2}=0.47$) and NEE ($R^{2}=0.39$). These
correlations decreased on the larger scale ($\un 1{ha}$; $R^{2}\leq0.26$). 

The weighted tree height standard deviation was strongly negatively
correlated with the GPP ($R^{2}=0.56$) but almost uncorrelated with
NPP and NEE ($\left|R^{2}\right|\leq0.03$) on both spatial scales.
The Shannon diversity of PFTs was moderately correlated with the NPP
($R^{2}\in\left[0.17,0.19\right]$), weakly correlated with the NEE
($R^{2}\leq0.04$), and weakly negatively correlated with the GPP
($R^{2}\leq0.07$). These results are depicted in Figures \ref{fig:ProductivityBA-diversity-patch}
and \ref{fig:Productivity-BA-diversity-ha}. 

\begin{figure}
\begin{centering}
\subfloat[\hspace*{\fill}]{\includegraphics[width=0.24\textwidth]{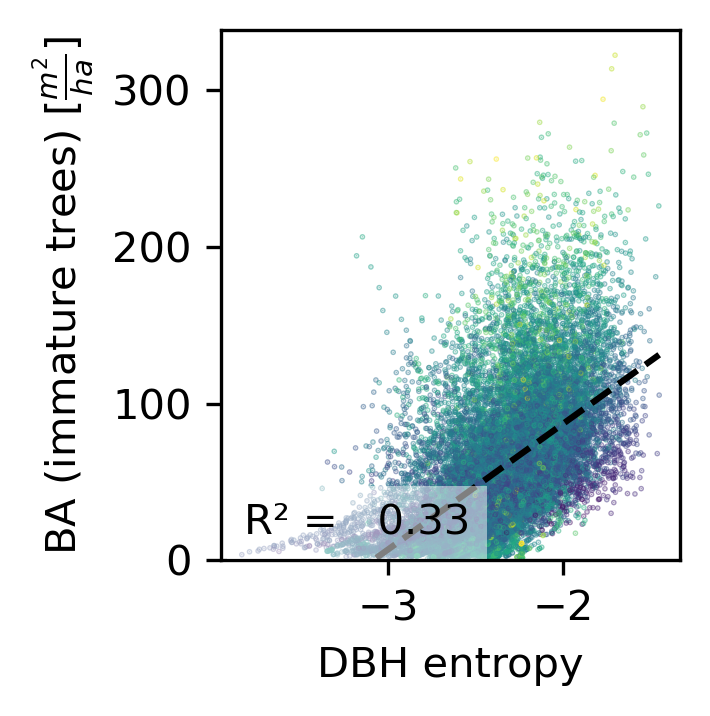}}\hspace*{\fill}\subfloat[\hspace*{\fill}]{\includegraphics[width=0.24\textwidth]{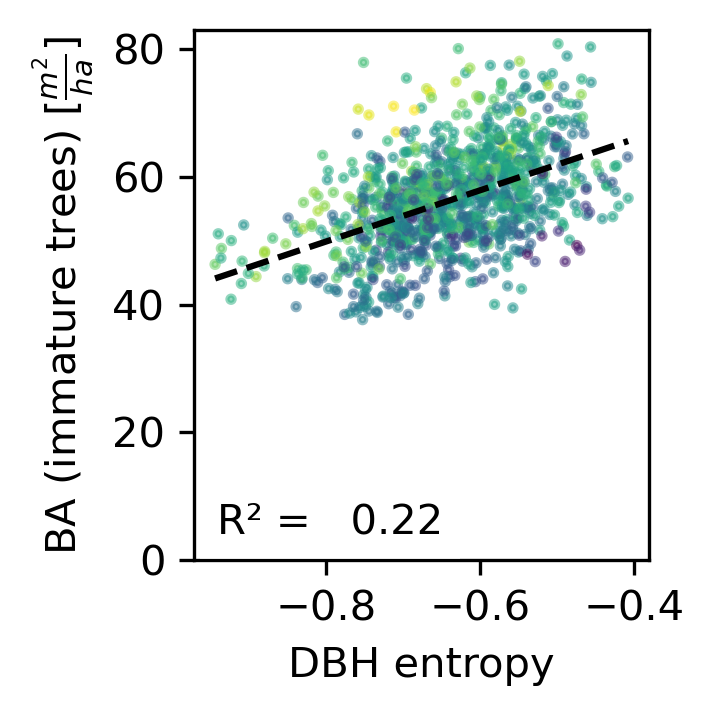}}\hspace*{\fill}\subfloat[\hspace*{\fill}]{\includegraphics[width=0.24\textwidth]{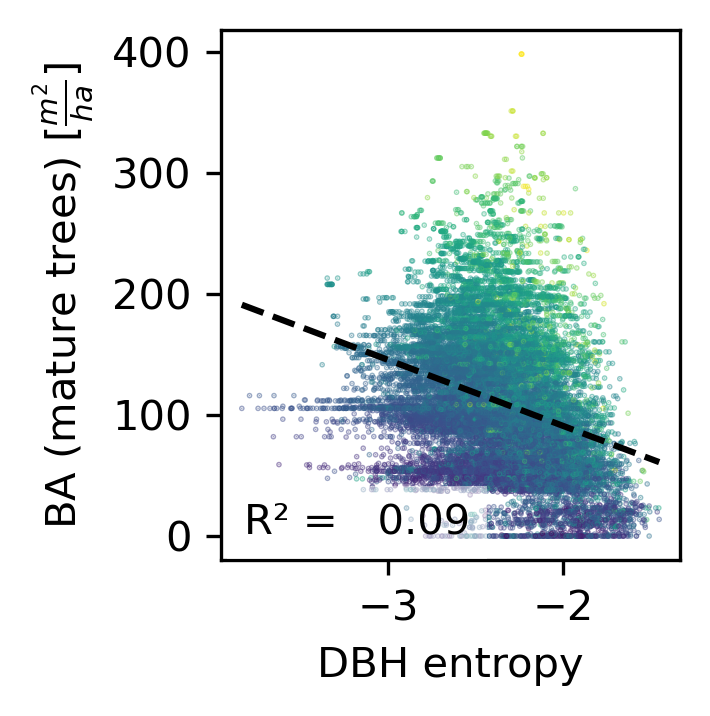}

}\hspace*{\fill}\subfloat[\hspace*{\fill}]{\includegraphics[width=0.24\textwidth]{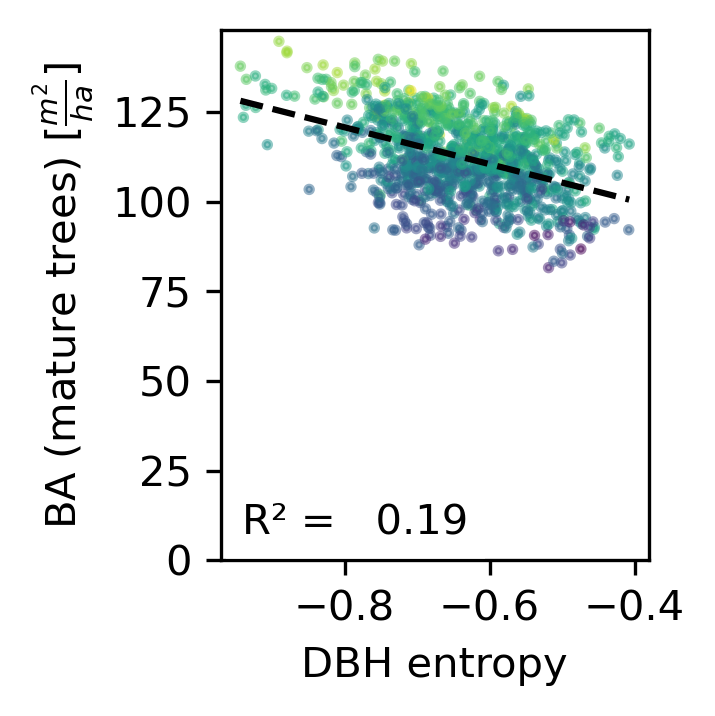}

}
\par\end{centering}
\caption{Relationship between the DBH entropy and (a, b) the basal area of
immature and (c, d) mature trees,depicted on (a, c) the $\protect\un{0.04}{ha}$
and (b, d) the $\protect\un 1{ha}$ scale. Each dot corresponds to
a forest patch of the respective scale. The colour corresponds to
the total basal area (dark: low, light: high). The DBH entropy correlates
positively with the basal area of immature trees, which drive the
NPP, and correlates negatively with the basal area of mature trees,
which do not contribute to the NPP and compete with immature trees.
The relationships are stronger on the small scale. \label{fig:Entropy-BA}}
\end{figure}

The DBH entropy was positively correlated to the basal area of immature
trees ($R^{2}=0.33$ on the small scale) and weakly negatively correlated
to the basal area of mature trees ($R^{2}=0.09$). For the latter,
the DBH entropy was a poor predictor in forest patches with large
overall basal area. On the hectare scale, the relationships became
weaker for immature trees ($R^{2}=0.23$) but stronger for mature
trees ($R^{2}=0.19$). These results are shown in Fig. \ref{fig:Entropy-BA}. 

\begin{figure}
\centering{}\subfloat[\hspace*{\fill}\label{fig:BA_growing-CUE-Patch}]{\includegraphics[width=0.24\textwidth]{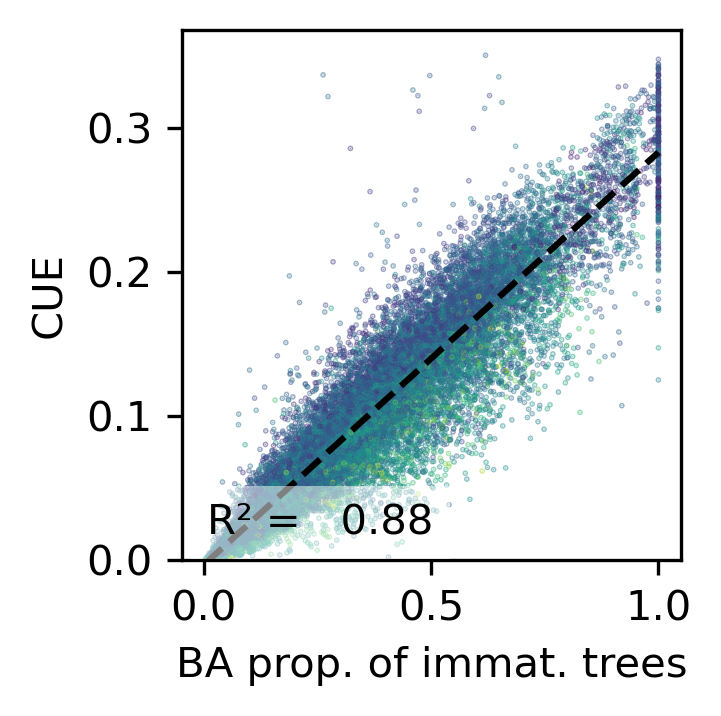}}\hspace*{\fill}\subfloat[\hspace*{\fill}\label{fig:BA_growing-CUE-Ha}]{\includegraphics[width=0.24\textwidth]{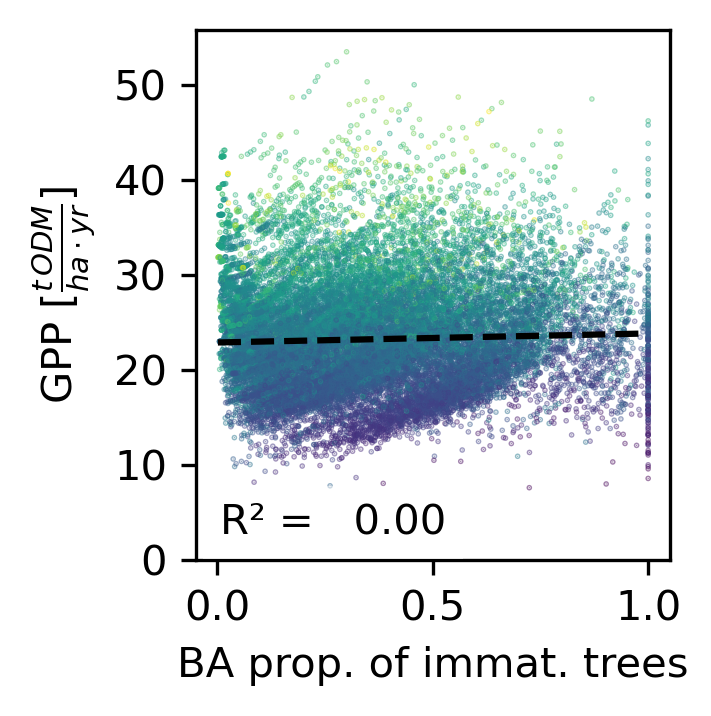}}\hspace*{\fill}\subfloat[\hspace*{\fill}\label{fig:Entropy-CUE-Ha}]{\includegraphics[width=0.24\textwidth]{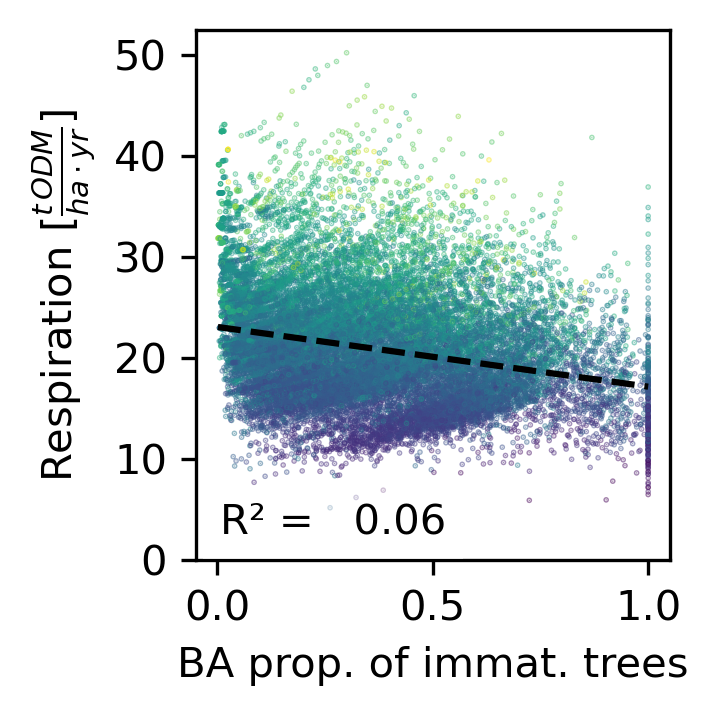}}\hspace*{\fill}\subfloat[\hspace*{\fill}\label{fig:Entropy-CUE-Patch}]{\includegraphics[width=0.24\textwidth]{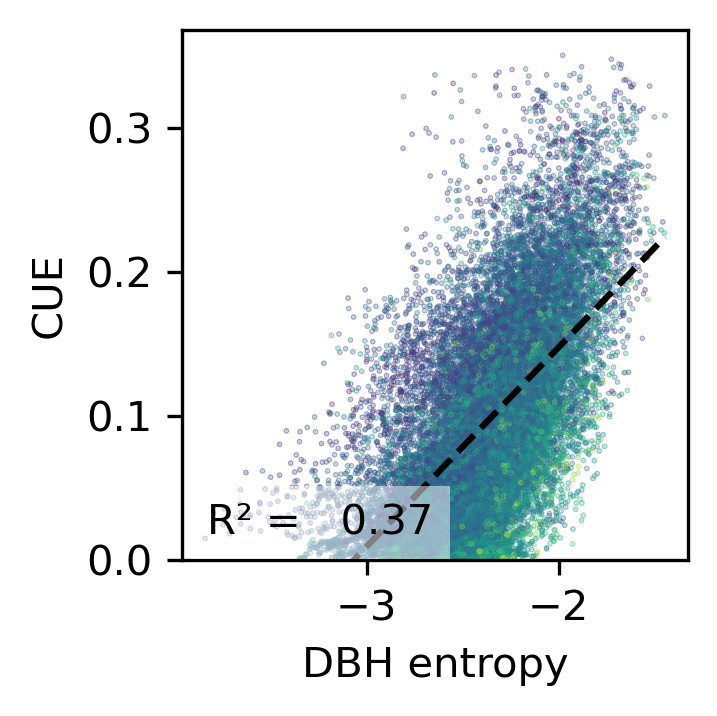}}\caption{Relationship between the basal area proportion of immature trees and
(a) the CUE, (b) the GPP, and (c) the respiration. The CUE is proportional
to the basal area of immature trees. Though the CUE can be directly
computed from the GPP and respiration, a similar relationship is not
visible for these, indicating that they are not the drivers behind
the proportionality. (d) Relationship between the DBH entropy and
the CUE. Though this relationship is weaker than that between the
proportion of immature trees and and CUE, the DBH entropy may serve
as a proxy for the CUE.\protect \\
All subfigures were created on the $\protect\un{0.04}{ha}$ scale.
The colour corresponds to the basal area (dark: low, light: high).
\label{fig:Ba-growing_Entropy_CUE}}
\end{figure}

The CUE was proportional to the proportion of immature trees in the
forest (Fig. \ref{fig:Ba-growing_Entropy_CUE}). The regression analysis
yielded an intercept of $3.16\e{-3}$ on the small and $2.182\e{-4}$
on the large scale, with $R^{2}$ values of $0.88$ and $0.82$, respectively.
The proportionality constants (slopes of the fitted curves) were $0.29$
and $0.28$. The relationship between CUE and DBH entropy was also
significant, but weaker ($R^{2}=0.3$ on the small and $R^{2}=0.28$
on the large scale). The GPP and respiration did not show a strong
correlation with the proportion of immature trees ($R^{2}=0$ and
$R^{2}=006$, respectively on the small scale).

\section{Discussion}

We suggested a simple framework of ``mature'' and ``immature''
trees to study forest productivity in old-growth forests. Using this
framework, we found that the drivers of NPP and NEE were distinct
from those determining GPP. While the latter was strongly correlated
with the total basal area, NPP and NEE were related to the basal area
of immature trees only. This indicates that the increased respiratory
losses of mature trees play a major role in forests' carbon balance:
despite having a significant GPP, mature trees do not contribute considerably
to wood production but rather reduce the productivity of other trees
via competition. Hence, tree maturity may be a major driver of the
difference between NPP and GPP, making GPP-related covariates insufficient
to explain local variations in NPP and NEE.

This conclusion is supported by the observed proportionality between
the CUE and the basal area share of immature trees: carbon usage was
more efficient the more the forest was dominated by immature trees.
 The proportionality can be explained by the strong connection between
the individual-level basal area and GPP in conjunction with the negligible
NPP of mature trees. On the stand level, however, neither the GPP
nor the respiration were correlated with the proportion of immature
trees (Fig. \ref{fig:Ba-growing_Entropy_CUE}), showing that the proportionality
was not driven by the decreased GPP or increased respiration of forests
with a high share of mature trees. 

To study how these results find expression in relationships between
forest structure and productivity, we suggested the DBH as a proxy
for the prevalence of immature trees and thereby NPP and NEE. In fact,
the DBH entropy was positively correlated with the basal area of immature
trees and negatively correlated with the basal area of mature trees,
but its relationships to NPP and NEE were even stronger. This indicates
that the predictive capacity of the DBH entropy stems not only from
its correlation with the prevalence of immature trees but also from
other mechanisms. This supports previous studies identifying structural
diversity as a major driver of forest productivity \citep{danescu_structural_2016,bohn_importance_2017,silvapedroDisentanglingEffectsCompositional2017,bohn_species_2018,park_influence_2019,larue_structural_2023}.
Note that our DBH entropy index differs from the classic entropy-based
measures for structural diversity \citep{staudhammer_introduction_2001}
by the basal-area-based weighting \citep{park_influence_2019}, which
improved its predictive capability (SI~\ref{subsec:DBH-entropy-parameterization}). 

Remarkably, the height standard deviation, another measure for structural
diversity, did not have a significant positive correlation to any
of the productivity measures. The height standard deviation depends
on the width of the height spectrum, i.e., the difference between
the height of the smallest and the largest tree. Hence, forests with
a high diversity of tree heights may not exhibit a high standard deviation
and vice versa. This contrasts with the entropy, which measures how
many different tree sizes there are without regarding their absolute
values. The strong negative relationship between the height standard
deviation and GPP can be explained by the weighting we applied. Weighting
the tree heights by basal area decreases the standard deviation in
forest stands with many large trees, which in turn have a large GPP.

The Shannon diversity of PFTs was not strongly related to any of
the forest productivity measures. This was due to the differences
between stem count and biomass of the PFTs. Four PFTs contributed
significantly to the forest's stem count and thus the Shannon diversity.
In contrast, the biomass was dominated by two PFTs only, which consequently
contributed most to the production. Hence, the Shannon diversity of
PFTs was a poor predictor for productivity. However, if the Shannon
diversity was computed based on tree species rather than PFTs, it
may yield useful information on the diversity of the DBH limits,
because they are species dependent. Setting this diversity of limits
into relation with the actual diversity (or entropy) of DBH values
could hence improve NPP estimates. 

Changing the spatial scale from $\un{0.04}{ha}$ to $\un 1{ha}$ did
not alter most of the relationships we considered. By construction,
the coefficient of determination is insensitive to the addition of
independently identically distributed random variables. As the interactions
between forest patches were weak and the basal area, GPP, NPP, and
NEE are additive measures, their respective correlations were not
affected by the scale. The same applied to the height standard deviation,
which is additive if the weighted mean height is approximately constant
in all small-scale patches. The Shannon diversity of PFTs did not
show strong patterns on any scale. The DBH entropy, however, was most
informative on a small scale (e.g. $\un{0.04}{ha}$). On large scales
(e.g. $\un 1{ha}$), the entropy increases and varies less between
forest sections, since more trees are considered. This result is significant,
as many previous studies considered entropy-based diversity indices
on larger scales (often $\geq\un{0.5}{ha}$; \citealt{danescu_structural_2016,silvapedroDisentanglingEffectsCompositional2017,park_influence_2019}).
In line with our results, a loss of information on larger scales was
noticed by \citet{chisholm_scale-dependent_2013} with respect to
the Shannon index. Nonetheless, if the scale is smaller than that
of plant interactions, the DBH entropy cannot reflect information
on competition and dominance, and the similarities between mature
trees cannot be incorporated. 

\subsection{Model parameterization and limitations}

Our individual-based modelling approach allowed us to differentiate
immature trees from those that have reached their maximal sizes and
to analyze carbon fluxes on small spatial scales. Measuring GPP and
NEE on small scales is challenging, since eddy covariance measurements,
for example, typically apply to the whole stand level only, are costly
and bound to one location, as the measure towers are not mobile. We
introduced a number of innovations in model design and parameter estimation.
Our likelihood-based fitting method allowed us to estimate parameters
based on forest characteristics on the small scale (here: $\un{0.04}{ha}$)
despite their stochastic variations. The distribution of local stem
counts and biomass yields information on local interactions and consequently
the range and diversity of local states a forest can attain. This
information is typically lost on larger scales. Circumventing the
need to reduce stochasticity via aggregation over several hectares
of forest (see e.g. \citealp{rodig_spatial_2017}) allowed us to
estimate parameters affecting the small-scale forest dynamics, and
we could optimize $18$ parameters on regeneration, light response,
optimal growth, and respiration. Applying a parameterization framework
focusing on the tree-level carbon use efficiency guaranteed a balanced
parameterization of the individual-level NPP and GPP. 

Our fitting approach also circumvented challenges typically arising
in the Bayesian framework. Bayesian methods, such as approximate Bayesian
computation (ABC; \citealp{beaumont_approximate_2002,csillery_approximate_2010}),
require the evaluation of many parameter combinations. This is computationally
costly in models for old-growth forests, as the entire succession
has to be simulated. Furthermore, the stochastic search performed
in ABC and classical Markov Chain Monte Carlo may fail to find good
parameter combinations when the parameter space is large. Hence, our
methodological advances can also benefit future forest models.

The good match between the biomass and stem count distributions in
the simulated forest and the inventory indicates that the model replicates
the forest structure well. Validation via independent estimates of
biomass, GPP, NPP, and LAI showed furthermore that the model reproduces
major forest dynamics. Nonetheless, the model underestimated biomass,
GPP, and NPP. The low biomass estimates resulted partially from our
focus on trees' main stems. In the absence of allometry data for individual
branches, we could have included the stems as separate trees. This,
however, would have led to overestimates of the LAI, which in turn
would have made it difficult to fit the model to field data. The
partly underestimated biomass along with our assumption that mature
trees stop growing may also have caused the underestimated NPP and
GPP. Nonetheless, these quantitative differences do not invalidate
the strong qualitative results we obtained.

The strong correlation we observed between basal area and GPP may
stem from our assumption that leaf area and basal area are proportional
within a PFT. Though this assumption is in line with theoretical and
empirical findings \citep{west_general_1999,xu_modelling_2021}, local
conditions and competition can blur this relationship in practice,
weakening it in field observations. As an alternative, the GPP could
be estimated from stand-level LAI values \citep[see e.g.][]{xie_assessment_2019}.

The relationship between basal area and GPP could also be weakened
by competition for water and other resources, which might also yield
other interactions between mature and immature trees. Added competition
may strengthen the negative effect of mature trees on forest productivity,
so that the basal area of mature trees may need to be considered in
addition to the basal area of immature trees to accurately estimate
NPP and NEE. In special cases, mature trees could also have positive
effects on smaller trees, for example by providing shelter \citep{lett_global_2018}
and improving soil conditions \citep{yunusa_plants_2003}. In forests
whose dynamics are driven by sink limitations (i.e., limitations affecting
carbon allocation to growth) rather than source limitations (limitations
affecting carbon supply), such effects could induce a positive effect
of mature trees on NPP.

Our analysis built on the assumption that trees have maximal sizes.
We modelled this via a transition from the growing to the mature stage,
which is a common approach in forest modelling \citep{shugart_gap_2018}.
In reality this transition can be gradual, and trees may require minimal
DBH increments to maintain the function of their vascular system \citep{prislanPhenologicalVariationXylem2013}.
Nonetheless, for the purpose of our analysis, the growth of mature
trees may be neglected as long as it is significantly reduced. Though
the concept of growth limitations acting on the individual scale is
subject to an ongoing debate \citep{stephenson_rate_2014,fosterPredictingTreeBiomass2016,sheilDoesBiomassGrowth2017,forresterDoesIndividualtreeBiomass2021,andersonteixeira_joint_2022},
there is strong evidence that the NPP and / or CUE decrease with the
age of forest stands \citep{gower_aboveground_1996,tangSteeperDeclinesForest2014,collaltiForestProductionEfficiency2020},
indicating that tree age or size have a significant effect on individual
biomass increment \citep{west_increasing_2020}. Modelling tree maturity
via a gradual growth slow-down would add significant complexity to
the model without changing the mechanism behind the results. Hence,
we expect that the observed relationships continue to hold with maturity
defined via growth slowdown.

We considered a forest under uniform environmental conditions to study
the within-stand productivity variations and their connection with
forest structure. On a regional scale, climate, soil, species composition,
and other factors will affect forest productivity \citep{munne-bosch_limits_2018,west_increasing_2020,geaizquierdo_forest_2022};
the NEE may be increased by disturbances and climate-induced increases
in mortality. Hence, for regional scales, our results would need to
be combined with appropriate stand-level covariates to obtain productivity
estimates. Nevertheless, our findings may be applicable to extended
areas with comparable climatic conditions.

\subsection{Outlook}

Using the concept of tree maturity to classify trees could become
a useful framework to understand forest productivity on local scales.
The phenomenological distinction between significantly growing and
mature trees may be conducted irrespective of the mechanism behind
the size limitations, be it increased respiratory losses \citep{oleary_core_2019},
sink limitations \citep{potkayTurgorlimitedPredictionsTree2022},
limited nutrient or water availability \citep{munne-bosch_limits_2018},
or even genetic predisposition \citep{liuLinkingIndividualLevel2016}.
As we used a generic forest model and our results were robust across
scales, our observations may hint towards a universal relationship
between tree maturity and forest productivity. This connection could
be used to develop new theory that could eventually lead to accurate
predictions of NPP and NEE based on general forest characteristics.
Such predictions have proven difficult in the past \citep{chisholm_scale-dependent_2013,rodig_importance_2018}
but could be highly relevant for a broad spectrum of applied and theoretical
questions in forest ecosystem science. Here, the DBH entropy could
prove particularly useful, as it can be easily obtained from inventory
data and may serve both as a measure for forests' structural diversity
on the local scale and as a proxy for net forest productivity in old-growth
forests. 

Confirming and generalizing the observed relationships between tree
maturity, DBH entropy, NPP, and NEE is a promising endeavour for both
theoretical and field studies. Further modelling studies could assess
the expected strength of the relationships in forests in different
successional stages, under varying environmental conditions, and in
the presence of additional stressors such as competition for nutrients
and water. Field studies could attempt to validate these findings.
Typical DBH maxima are documented for many species from temperate
forests and could serve as a first proxy for maturity \citep{aibaCrownArchitectureLifeHistory1997,kohyamaTreeSpeciesDifferentiation2003,russellMaximumLargestCrown2011,delrioTreeAllometryVariation2019}.
Combining the gained insights with large-scale predictors for forest
productivity could then lead to a unified theory of forest productivity.

\section{Conclusion}

We applied a modelling approach to investigate how the prevalence
of mature (full-grown) trees and forest structure explain within-stand
variations of forest productivity. We found that NPP and NEE are mainly
driven by the basal area of immature trees, whereas the GPP depends
on the total basal area. This suggests that loss-induced limitations
rather than variations in GPP determine NPP and NEE.

The forest stand CUE was proportional to the basal area share of immature
trees. We suggested and tested the basal-area-weighted DBH entropy
as an easy-to-compute proxy for both the prevalence of mature trees
and NPP and NEE. Other measures for structural diversity, namely the
height standard deviation and the Shannon entropy of functional types,
had much smaller predictive power. Our results were robust across
spatial scales, and due to their solid mechanistic foundation and
our generic model, our findings yield promising hypotheses for field
studies and new theoretical work. 

Understanding the drivers of forest productivity is key for an accurate
assessment of forests' role in the global carbon cycle. Yet, despite
significant research effort, it is not fully understood how the productivity
of a forest can be deduced from its stand structure. This is partially
due to the challenge of accounting for increased carbon losses of
mature trees in structure-productivity relationships. We suggest to
tackle this problem by identifying the share and structure of immature
trees within mature forests and show that this approach could significantly
improve estimates of forests\textquoteright{} net productivity. As
it is challenging to assess tree maturity for each individual in the
field, we suggest an easy-to-compute stand-level proxy for the prevalence
of mature trees, yielding the theoretical basis for future field studies
improving our understanding of structure-productivity relationships.

\section*{Acknowledgements}

The authors would like to thank the members of the vegetation modelling
group at the UFZ for helpful discussions and feedback. This research
was conducted as part of the project ``The role of species traits
and forest structure on spatial carbon dynamics of temperate forests''
(ForCTrait), established within the cooperation ``China-NSFC-DFG
2019'' between the Deutsche Forschungsgemeinschaft (DFG, German Research
Foundation) and the Natural Science Foundation of China (NSFC). This
work was funded by the Deutsche Forschungsgemeinschaft (DFG) -- 43150473.

\section*{Author contributions}

SMF and AH jointly conceived the study. XW contributed the field data.
SMF parameterized the model with substantial input by AH and conducted
the data analysis. SMF and AH jointly conceived the manuscript; SMF
wrote the manuscript; AH revised the manuscript. All authors approved
the manuscript.

\section*{Competing interests}
The authors declare no competing interests.

\bibliographystyle{copernicus}

\newpage{}

\appendix
\renewcommand{\theequation}{S\arabic{equation}}
\renewcommand{\thefigure}{S\arabic{figure}}
\renewcommand{\thetable}{S\arabic{table}}
\setcounter{page}{1}
\setcounter{figure}{0}
\setcounter{equation}{0}

\part*{Supplementary Information}

\section{Data availability by species\label{sec:Data-availability}}

The table below displays basic information and data availability for
each species in the inventory of $2014$. The rows are sorted by the
species' respective basal areas (including minor stems). The column
``PFT'' indicates the plant functional types the species were assigned
to; the PFT numbers correspond to those provided in SI~\ref{subsec:Classification-of-species}.
The column ``Allometry data available'' shows whether DBH-dependent
data on allometric properties, such as tree height or crown length,
were available. The column ``Biomass equation available'' indicates
whether we found a suitable DBH biomass relationship in \citet{chojnacky_updated_2014}.
Species not present in the inventory of $2014$ were omitted.

The allometry dataset contained DBH values, heights, crown diameters,
and crown lengths for individual trees. On average, the dataset contained
$73$ individuals per species covered in the dataset. 

{\footnotesize \singlespacing \setstretch{1.3}%
\begin{longtable}[c]{>{\centering}m{4cm}>{\centering}m{2cm}>{\centering}m{2cm}>{\centering}m{3cm}>{\centering}m{2cm}>{\centering}m{2cm}}
\hline
\toprule 
Species & Basal area in inventory$\left[\frac{\mt m^{2}}{\mt{ha}}\right]$ & PFT & Shade tolerance & Allometry data available & Biomass equation available\tabularnewline
\midrule
\hline
\endhead
\textit{Tilia amurensis}  & 12.556  & 6  & Shade tolerant  & $\checkmark$  & $\checkmark$ \tabularnewline
\midrule 
\textit{Pinus koraiensis}  & 9.870  & 4  & Midtolerant  & $\checkmark$  & $\checkmark$ \tabularnewline
\midrule 
\textit{Quercus mongolica}  & 6.748  & 3  & Light demanding  & $\checkmark$  & $\checkmark$ \tabularnewline
\midrule 
\textit{Fraxinus mandshurica}  & 6.098  & 4  & Midtolerant  & $\checkmark$  & $\checkmark$ \tabularnewline
\midrule 
\textit{Acer mono}  & 2.552  & 6  & Shade tolerant  & $\checkmark$  & $\checkmark$ \tabularnewline
\midrule 
\textit{Ulmus japonica}  & 1.867  & 4  & Midtolerant  & $\checkmark$  & $\checkmark$ \tabularnewline
\midrule 
\textit{Acer pseudo-sieboldianum}  & 1.254  & 5  & Shade tolerant  & $\checkmark$  & $\checkmark$ \tabularnewline
\midrule 
\textit{Populus ussuriensis}  & 1.212  & 2  & Light demanding  & $\checkmark$  & $\checkmark$ \tabularnewline
\midrule 
\textit{Tilia mandshurica}  & 0.345  & 6  & Shade tolerant  & $\checkmark$  & $\checkmark$ \tabularnewline
\midrule 
\textit{Maackia amurensis}  & 0.285  & 4  & Midtolerant  & $\checkmark$  & -- \tabularnewline
\midrule 
\textit{Populus koreana}  & 0.203  & 2  & Light demanding  & $\checkmark$  & $\checkmark$ \tabularnewline
\midrule 
\textit{Acer barbinerve}  & 0.199  & 5  & Shade tolerant  & --  & -- \tabularnewline
\midrule 
\textit{Betula platyphylla}  & 0.179  & 2  & Light demanding  & $\checkmark$  & $\checkmark$ \tabularnewline
\midrule 
\textit{Corylus mandshurica}  & 0.151  & 5  & Shade tolerant  & --  & -- \tabularnewline
\midrule 
\textit{Acer triflorum}  & 0.120  & 6  & Shade tolerant  & $\checkmark$  & -- \tabularnewline
\midrule 
\textit{Acer tegmentosum}  & 0.120  & 5  & Shade tolerant  & $\checkmark$  & -- \tabularnewline
\midrule 
\textit{Syringa reticulata}  & 0.110  & 1  & Light demanding  & $\checkmark$  & -- \tabularnewline
\midrule 
\textit{Malus baccata}  & 0.103  & 6  & Shade tolerant  & $\checkmark$  & -- \tabularnewline
\midrule 
\textit{Phellodendron amurense}  & 0.100  & 2  & Light demanding  & $\checkmark$  & -- \tabularnewline
\midrule 
\textit{Acer mandshuricum}  & 0.083  & 6  & Shade tolerant  & $\checkmark$  & -- \tabularnewline
\midrule 
\textit{Prunus padus}  & 0.073  & 6  & Shade tolerant  & $\checkmark$  & -- \tabularnewline
\midrule 
\textit{Ulmus laciniata}  & 0.068  & 4  & Midtolerant  & $\checkmark$  & -- \tabularnewline
\midrule 
\textit{Betula costata}  & 0.053  & 2  & Light demanding  & $\checkmark$  & -- \tabularnewline
\midrule 
\textit{Populus davidiana}  & 0.031  & 2  & Light demanding  & $\checkmark$  & -- \tabularnewline
\midrule 
\textit{Pyrus ussuriensis}  & 0.026  & 2  & Light demanding  & $\checkmark$  & -- \tabularnewline
\midrule 
\textit{Abies nephrolepis }  & 0.026  & 4  & --  & $\checkmark$  & -- \tabularnewline
\midrule 
\textit{Rhamnus ussuriensis}  & 0.017  & 5  & Shade tolerant  & --  & -- \tabularnewline
\midrule 
\textit{Cerasus maximowiczii}  & 0.010  & 1  & Light demanding  & --  & -- \tabularnewline
\midrule 
\textit{Acer ginnala}  & 0.009  & 5  & Shade tolerant  & --  & -- \tabularnewline
\midrule 
\textit{Sorbus alnifolia}  & 0.006  & 5  & Shade tolerant  & $\checkmark$  & -- \tabularnewline
\midrule 
\textit{Philadelphus schrenkii}  & 0.004  & 5  & Shade tolerant  & --  & -- \tabularnewline
\midrule 
\textit{Rhamnus davurica}  & 0.004  & 5  & Shade tolerant  & --  & -- \tabularnewline
\midrule 
\textit{Crataegus maximowiczii }  & 0.002  & 1  & --  & --  & -- \tabularnewline
\midrule 
\textit{Euonymus pauciflorus}  & 0.001  & 5  & Shade tolerant  & --  & -- \tabularnewline
\midrule 
\textit{Euonymus alatus}  & 0.001  & 5  & Shade tolerant  & --  & -- \tabularnewline
\midrule 
\textit{Acanthopanax senticosus}  & 0.000  & 5  & Shade tolerant  & --  & -- \tabularnewline
\midrule 
\textit{Sambucus williamsii}  & 0.000  & 1  & Light demanding  & --  & -- \tabularnewline
\midrule 
\textit{Lonicera chrysantha}  & 0.000  & 5  & --  & --  & -- \tabularnewline
\midrule 
\textit{Viburnum sargenti}  & 0.000  & 5  & Shade tolerant  & --  & -- \tabularnewline
\midrule 
\textit{Actinidia kolomikta}  & 0.000  & 5  & --  & --  & -- \tabularnewline
\midrule 
\textit{Viburnum bureiaeticum}  & 0.000  & 5  & Shade tolerant  & --  & -- \tabularnewline
\midrule 
\textit{Rhamnus diamantiaca}  & 0.000  & 5  & Shade tolerant  & --  & -- \tabularnewline
\midrule 
\textit{Euonymus macropterus}  & 0.000  & 5  & Shade tolerant  & --  & -- \tabularnewline
\midrule 
\textit{Vitis amurensis}  & 0.000  & 5  & --  & --  & -- \tabularnewline
\midrule 
\textit{Aralia elata}  & 0.000  & 5  & Shade tolerant  & --  & -- \tabularnewline
\midrule 
\textit{Deutzia amurensis}  & 0.000  & 5  & --  & --  & -- \tabularnewline
\midrule 
\textit{Sorbaria sorbifolia}  & 0.000  & 4  & --  & --  & -- \tabularnewline
\bottomrule
\end{longtable}

}

\section{Parameterization of the forest model \noun{Formind} \label{sec:Parameterization}}

The forest model \noun{Formind} is described in detail by \citet{bohn_climate_2014}
and \citet{fischer_lessons_2016}. Below we focus on those aspects
of the model that deviate from this description, and we provide details
about the parameter choice and model fitting procedure. We based our
analysis on forest inventory data from an old-growth temperate forest
in the Changbaishan National Nature Reserve in northeastern China.
The surveyed area consists of $\un{25}{ha}$ of conifer/broad-leaf
mixed forest with $47$ species, a total biomass of $\unitfrac[302]{t\,ODM}{ha}$
\citep{piponiot_distribution_2022}. The inventory data contain the
position, diameter at breast height (DBH) and species of each tree
with $\mt{DBH}\geq1\mt{cm}$ for the census years $2004$, $2009$,
and $2014$. Each tree is uniquely identified with an ID number. For
trees that had multiple stems at breast height, we focused on the
main stem (maximal DBH) in our analysis and we disregarded minor stems. 

\subsection{Time step and simulation area}

We ran the model using a yearly time step. We simulated a square-shaped
forest area of one hectare, subdivided into $25$ patches of $20\mt m\times20\mt m$,
in which light competition occurs. Plants in different patches interact
via tree falling only. For this interaction, we assume torus boundary
conditions to minimize boundary effects. 

\subsection{Classification of species to plant functional types (PFTs)\label{subsec:Classification-of-species}}

We assigned the $47$ tree species into $6$ plant functional types
(PFTs) according to their shade tolerance and their maximal height.
In addition, we considered the species Mongolian Oak (\emph{Quercus
mongolica}) individually, as it had a unique DBH distribution in the
forest, making it difficult to assign it to other PFTs without major
information loss. We considered the following $6$ PFTs:
\begin{enumerate}
\item Small shade intolerant species (pioneers with maximal diameter at
breast height (DBH) below $30\mt{cm}$). 
\item Large shade intolerant species 1 (pioneers with maximal DBH exceeding
$30\mt{cm}$). 
\item Large shade intolerant species 2 (Mongolian oak).
\item Large mid-tolerant species (intermediate species with maximal DBH
exceeding $30\mt{cm}$).
\item Small shade tolerant species (climax species with maximal DBH below
$30\mt{cm}$).
\item Large shade tolerant species (climax species with maximal DBH exceeding
$30\mt{cm}$).
\end{enumerate}
We did not consider a PFT of small mid-tolerant species, because there
were no mid-tolerant species with maximal DBH below $30\mt{cm}$.

\subsubsection*{Classification of species with unknown shade tolerance class}

We assigned species for which shade tolerance classification data
were not readily available to the PFTs via a likelihood-based cluster
analysis. For this analysis, we determined the median DBH change after
5 years for each species' individuals observed in the inventory. We
used this value along with numerical shade tolerance data  as covariates. 

\begin{figure}
\begin{centering}
\includegraphics[width=0.8\textwidth]{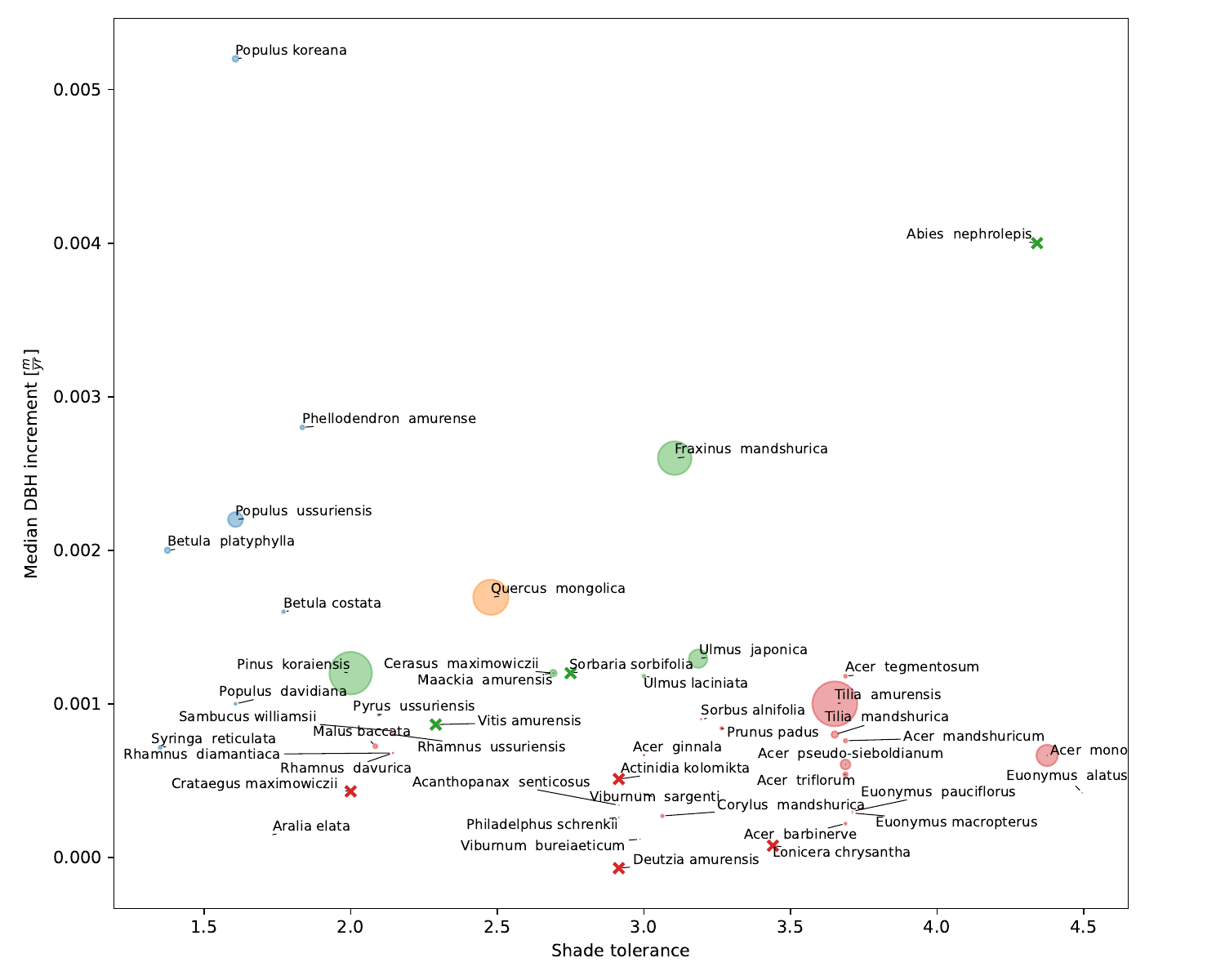}
\par\end{centering}
\caption{Assignment of species to PFTs with unknown shade tolerance based on
a cluster analysis. Each colour corresponds to a shade tolerance type:
shade intolerant (blue), mid-tolerant (green), shade tolerant (red).
\emph{Q. mongolica} is drawn in a separate colour (orange), as it
is a PFT on its own. Circles indicate species with a known shade tolerance
type. The size of the circles correspond to the respective species'
basal area in the inventory. Crosses depict species for which the
shade tolerance type was assigned via the cluster analysis.\label{fig:PFT-Cluster}}

\end{figure}

We assumed that the covariates follow a multivariate normal distribution
$\normal{\mu_{i}}{\Sigma_{i}}$ for each shade tolerance class $i$.
We estimated the means $\mu_{i}$ and covariance matrices $\Sigma_{i}$
using the method of moments. For each shade tolerance group $i$,
we determined the mean values $\mu_{ij}$ of the covariates $j$ and
covariances $\Sigma_{ij_{1}j_{2}}$ between covariate $j_{1}$ and
$j_{2}$. We assigned each species $s$ with unknown shade tolerance
type to the class for which the likelihood based on the derived distributions
was maximized. That is, with $x_{s}$ being the covariate vector of
species $s$ and $f_{\mathcal{N}}$ the density function of the two-dimensional
multivariate normal distribution, we set
\begin{equation}
\mt{class}\ap s=\argmaxo{\text{classes }i}f_{\mathcal{N}}\ap{x_{j};\mu_{i},\Sigma_{i}}.
\end{equation}
Fig. \ref{fig:PFT-Cluster} depicts the classification of the species
into shade tolerance classes.

\subsection{Allometric relationships\label{subsec:Allometric-relationships}}

\global\long\def\alloset{\mathcal{A}}%
\global\long\def\invset{\mathcal{I}}%
\global\long\def\specset{\mathcal{S}}%

We determined allometric relationships for the six PFTs based on allometry
data for the individual species. Specifically, we estimated the relationships
between DBH and the geometric properties tree height, crown length,
and crown radius, respectively. Let $p$ be the index of a geometric
property, $i$ the considered PFT, $\th_{pi}$ a parameter vector
and $x_{pi}$ the value of allometric property $p$ for PFT $i$,
$d$ be the DBH and $g_{p}:\R_{+}\rightarrow\R_{+}$ an injective
function. Then we set
\begin{align}
x_{pi} & =g_{p}\ap{d;\,\th_{pi}}.
\end{align}

To estimate the parameter vectors $\th_{pi}$, we used a dataset containing
tuples of tree DBH and tree height, crown length, and crown radius
for several species. As we desired to find the allometric relationships
that best represent the considered forest in Changbaishan, we weighted
the data according to the frequency of trees with similar species
and DBH in the inventory (see subsection \ref{subsec:Computing-the-weights}).
Then we fitted the parameters $\th_{pi}$ based on the weighted likelihood,
assuming that the data were subject to a normally distributed error
with constant variance $\sigma_{pi}^{2}$:
\begin{align}
X_{pi} & \sim\normal{g_{p}\ap{d;\,\th_{pi}}}{\sigma_{pi}^{2}},
\end{align}
where $X_{pi}$ denotes the observed geometry values. This reduces
to a weighted least squares method. That is, the objective function
can be expressed as 
\begin{equation}
\bar{\ell}\ap{\th_{pi}}=-\smo{k\in\alloset_{i}}\left(x_{pik}-g_{p}\ap{d_{k};\,\th_{pi}}\right)^{2}w_{k},\label{eq:allometry-objective}
\end{equation}
where $\alloset_{i}$ is the set of entries in the allometry dataset
corresponding to trees of PFT $i$, $x_{pk}$ is the observed value
for the geometric property $p$ in entry $k$, $d_{k}$ is the corresponding
DBH value, and $w_{k}$ is the weight. To maximize the likelihood,
we used a Newton-Raphson-type trust region approach \citep{conn_trust-region_2000}
as provided in the Python library Scipy \citep{jones_scipy:_2001}. 

\subsubsection{Height}

For the relationship between DBH and tree height, we used the model
\begin{equation}
g_{\mt{height}}\ap{d;\,\th_{\mt{height},i}}=\frac{\th_{\mt{height},i,0}\th_{\mt{height},i,1}d}{d\cdot\th_{\mt{height},i,0}+\th_{\mt{height},i,1}},\label{eq:DBH-to-height}
\end{equation}
where $\th_{\mt{height},i,0}$ is the initial slope of $g_{\mt{height}}$
and $\th_{\mt{height},i,1}$ is the height asymptote. The fitted parameter
values are displayed in Table \ref{tab:Parameters-allometry}; the
fitted curves are shown in Fig. \ref{fig:height-dbh}..

\begin{table}
\begin{centering}
\footnotesize%
\begin{tabular}{cc>{\centering}p{1.8cm}>{\centering}p{1.8cm}>{\centering}p{1.8cm}>{\centering}p{1.8cm}>{\centering}p{1.8cm}>{\centering}p{1.8cm}}
\toprule 
 & Unit & Small shade intolerant & Large shade intolerant 1 & Large shade intolerant 2 & Large mid-tolerant & Small shade tolerant & Large shade tolerant\tabularnewline
\midrule
\midrule 
$\th_{\mt{height},i,0}$ & $\frac{\mt m}{\mt m}$ & 127.81 & 198.06 & 174.00 & 143.19 & 167.82 & 129.98\tabularnewline
$\th_{\mt{height},i,1}$ & $\mt m$ & 29.25 & 30.59 & 31.22 & 36.75 & 19.87 & 38.92\tabularnewline
\midrule 
$\th_{\mt{\text{crown-l}},i}$ & $\frac{\mt m}{\mt m}$ & 0.35 & 0.33 & 0.36 & 0.36 & 0.37 & 0.35\tabularnewline
\midrule 
$\th_{\mt{\text{crown-d}},i,0}$ & $\frac{\mt m}{\mt m^{\th_{\mt{\text{crown-d}},i,1}}}$ & 12.01 & 13.35 & 14.34 & 11.27 & 15.72 & 12.72\tabularnewline
$\th_{\mt{\text{crown-d}},i,1}$ & $1$ & 0.49 & 0.63 & 0.59 & 0.52 & 0.48 & 0.48\tabularnewline
\bottomrule
\end{tabular}
\par\end{centering}
\caption{Parameter values for allometric relationships.\label{tab:Parameters-allometry}}

\end{table}

\subsubsection{Crown length }

We used a linear relationship to model the relationship between tree
height and crown length:
\begin{equation}
\tilde{g}_{\mt{\text{crown-l}}}\ap{h;\,\th_{\mt{\text{crown-l}},i}}=\th_{\mt{\text{crown-l}},i}\cdot h,\label{eq:height-to-crown-length}
\end{equation}
where $h$ is the tree height. With equation (\ref{eq:DBH-to-height}),
equation (\ref{eq:height-to-crown-length}) can also be expressed
as a function of the DBH: 
\begin{equation}
g_{\mt{\text{crown-l}}}\ap{d;\,\th_{\mt{height},i},\th_{\mt{\text{crown-l}},i}}=\th_{\mt{\text{crown-l}},i}\cdot g_{\mt{height}}\ap{d;\,\th_{\mt{height},i}}.\label{eq:DBH-to-crown-length}
\end{equation}
The fitted parameter values are displayed in Table \ref{tab:Parameters-allometry};
the fitted curves are depicted in Fig. \ref{fig:crownL-height}.

\subsubsection{(Maximal) crown diameter}

\begin{figure}
\subfloat[\hspace*{\fill}\label{fig:height-dbh}]{\includegraphics[width=0.49\textwidth]{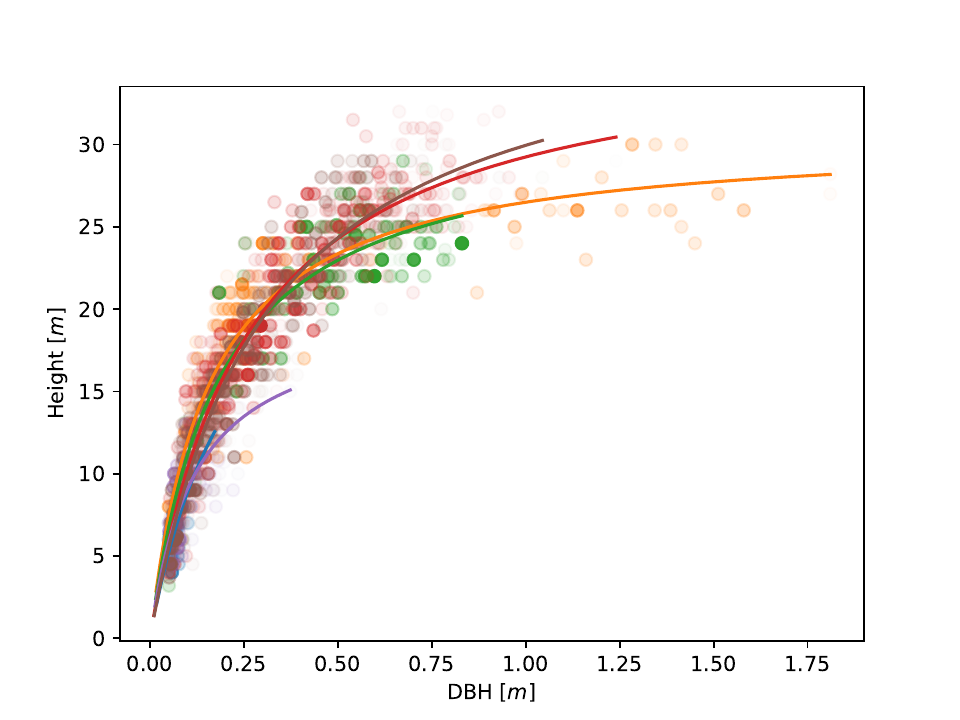}

}\hspace*{\fill}\subfloat[\hspace*{\fill}\label{fig:crownL-height}]{\includegraphics[width=0.49\textwidth]{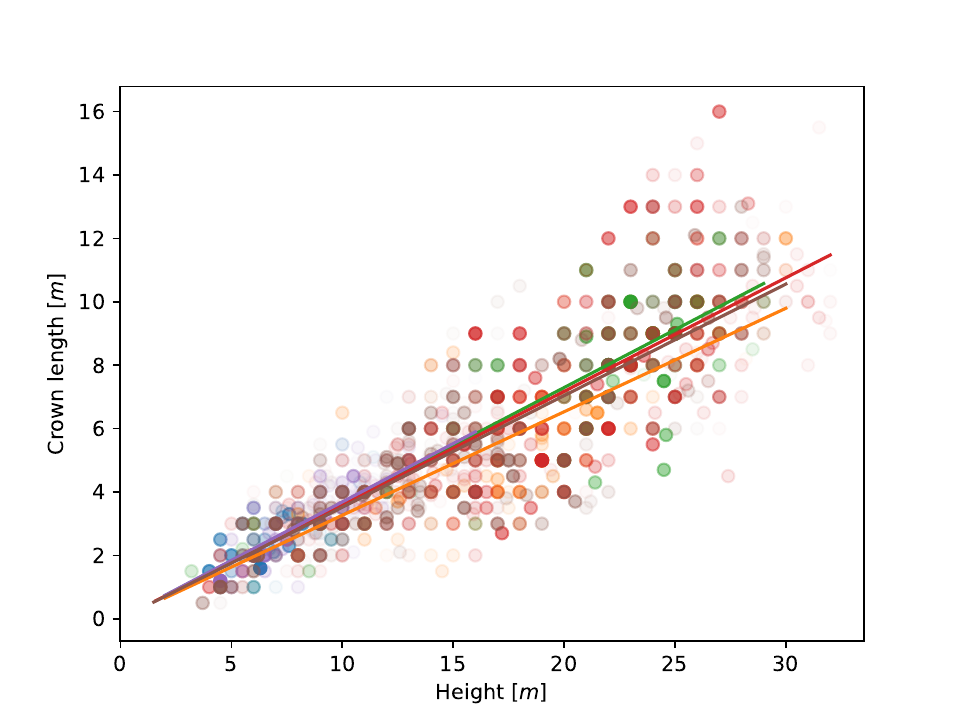}

}

\subfloat[\hspace*{\fill}\label{fig:crownD-dbh}]{\includegraphics[width=0.49\textwidth]{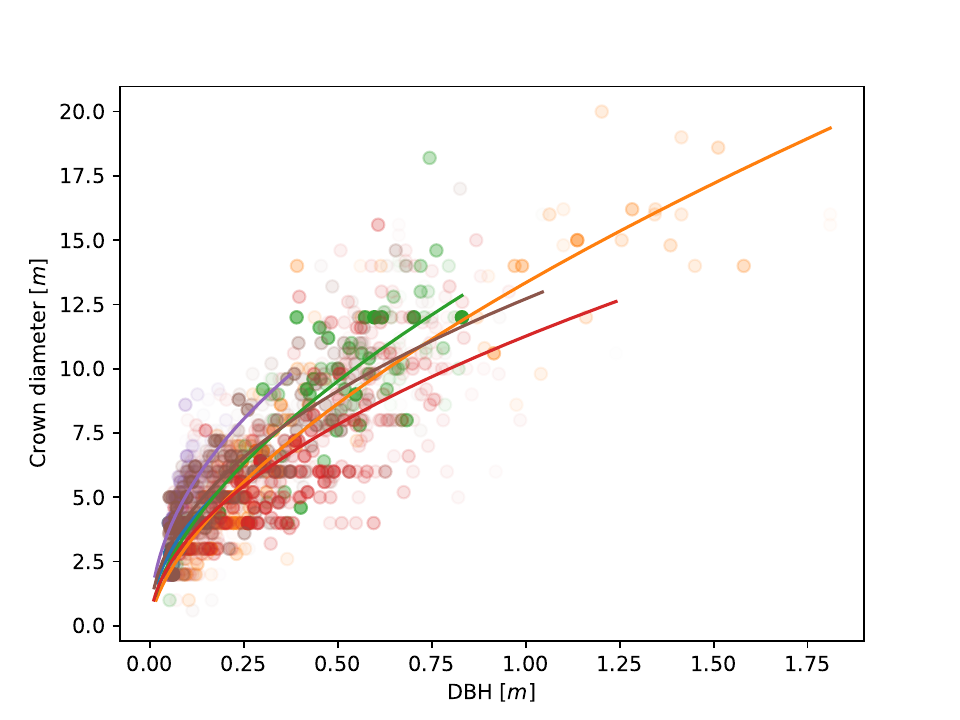}

}\hspace*{\fill}\subfloat[\hspace*{\fill}\label{fig:lai-dbh}]{\includegraphics[width=0.49\textwidth]{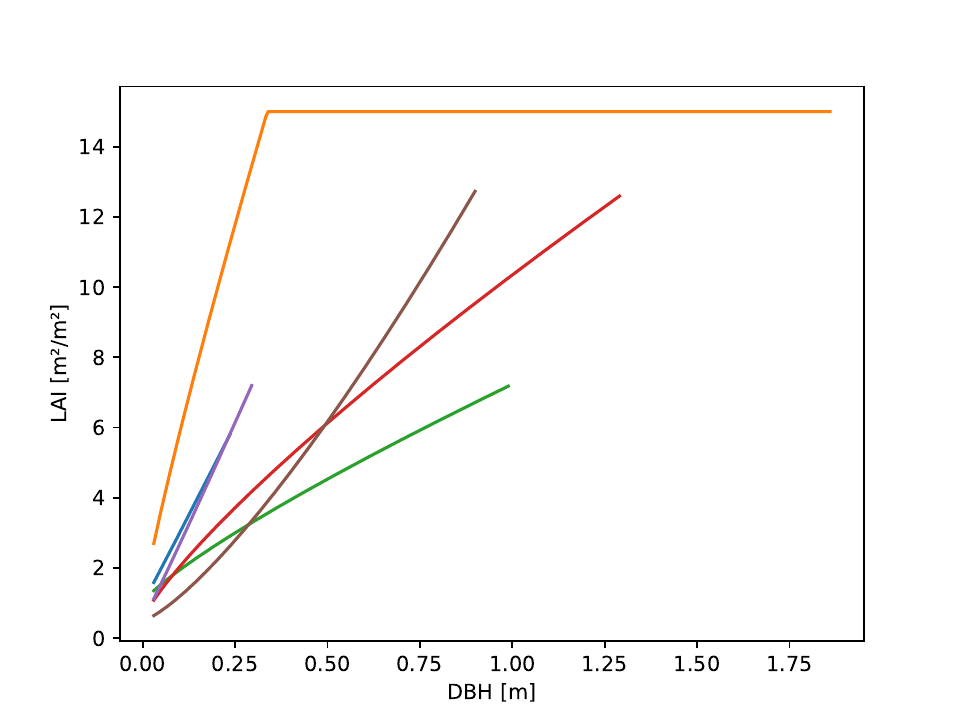}

}

\caption{Size-dependent plant traits. The circles depict data points from the
allometry dataset; their opacity shows their weight. Each colour corresponds
to a different PFT: small shade intolerant (blue), large shade intolerant
1 (orange) large shade intolerant 2 (green), large mid-tolerant (red),
small shade tolerant (purple), large shade tolerant (brown).\label{fig:Size-dependent-plant-traits}}
\end{figure}

For the relationship between DBH and crown diameter, we used a power-law
model: 
\begin{equation}
g_{\mt{\text{crown-d}}}\ap{d;\,\th_{\mt{\text{crown-d}},i}}=\th_{\mt{\text{crown-d}},0}\cdot d^{\th_{\mt{\text{crown-d}},1}},
\end{equation}
where $\th_{\mt{\text{crown-d}},0}$ is the scaling factor and $\th_{\mt{\text{crown-d}},1}$
is the exponent. The fitted parameter values are displayed in Table
\ref{tab:Parameters-allometry} and the fitted curves in Fig. \ref{fig:crownD-dbh}.

Typically, the crown diameter of trees varies with height, and the
available allometry data represent \emph{maximal} crown diameters.
In this parameterization of \noun{Formind}, however, crowns are assumed
to have cylindrical shapes, with diameters constant along the vertical
axis. Hence, using the observed maximal crown diameters as diameters
of the cylindrical shapes used in the model would lead to an overestimation
of crown volumes and, as a result, the LAI. To correct for this potential
bias, we assumed that the trees from which the data were taken had
crowns shaped like ellipsoids, rotationally symmetric around the vertical
axis. A cylinder with the same volume and height as this ellipsoid
must have a diameter scaled by factor $\sqrt{\frac{2}{3}}$ as compared
to the length of the horizontal semi-axis of the ellipsoid. Hence,
we parameterized the model with the scaled DBH-crown-diameter relationship
\begin{equation}
g_{\mt{\text{crown-d}}}\ap{d;\,\th_{\mt{\text{crown-d}},i}}=\sqrt{\frac{2}{3}}\th_{\mt{\text{crown-d}},0}\cdot d^{\th_{\mt{\text{crown-d}},1}}.\label{eq:crown-d}
\end{equation}

\subsubsection{Stem volume and form factor\label{subsec:Form-factor}}

To compute the stem volume $V_{\mt{stem},i}$, we used the formula
\begin{equation}
V_{\mt{stem},i}\ap d=\frac{\pi}{4}d^{2}g_{\mt{height}}\ap{d;\,\th_{\mt{height},i}}\nu_{i}\ap d,
\end{equation}
where $d$ is the DBH, $g_{\mt{height}}\ap{d;\,\th_{\mt{height},i}}$
is the height (see equation (\ref{eq:DBH-to-height})), and $\nu_{i}\ap d$
is a DBH- and PFT-dependent form factor. A form factor $\nu_{i}\ap d=1$
corresponds to a cylindrical stem shape, $\nu_{i}\ap d=\frac{1}{3}$
to a cone, $\nu_{i}\ap d\in\left(1,\frac{1}{3}\right)$ to a convex
cone-like shape, and $\nu_{i}\ap d\in\left(1,\frac{1}{3}\right)$
to a concave cone-like shape of the stem. In line with earlier parameterizations
of \noun{Formind} \citep{dislich_simulating_2009}, we chose 
\begin{equation}
\nu_{i}\ap d=\theta_{\mt{form},i,0}d^{\theta_{\mt{form},i,1}}
\end{equation}
with $\theta_{\mt{form},i,0}=0.336\,\mt m^{-\theta_{\mt{form},i,1}}$
and $\theta_{\mt{form},i,1}=-0.18$ for all PFTs $i$.

\subsection{Plant traits \label{subsec:Plant-traits}}

Besides geometric relationships, the \noun{Formind} model requires
information about the maximal size of trees, their wood density, and
their leaf area index (LAI).

\subsubsection{Maximal DBH\label{subsec:Maximal-DBH}}

We assumed that each tree $t$ has its own site-dependent maximal
DBH $d_{t}^{\mt{max}}$. As this value may depend on the tree's species,
which is neglected when species are summarized to PFTs, we constructed
the distribution of maximal heights based on each species' maximal
DBH and the species' frequency in the inventory. Below we provide
a detailed description of our approach.

Let $s$ be a species and $\text{\ensuremath{\alloset_{s}}}$ and
$\text{\ensuremath{\invset_{s}}}$ the subsets of the allometry and
inventory dataset, respectively, that correspond to species $s$.
We determine the maximal DBH of species $s$ based on the maximal
DBH observed in the allometry dataset and the $99.5\text{th}$ percentile
of the inventory:
\begin{equation}
\bar{d}_{s}^{\mt{max}}=\max\left\{ F_{\left\{ d_{k},k\in\alloset_{s}\right\} }^{-1}\ap{0.995},\,\maxo{t\in\invset_{s}}d_{t}\right\} ,
\end{equation}
where $F_{\cdot}^{-1}\ap{\cdot}$ is the observed percentile function.

There were some cases in which the maximal DBH from the inventory
dataset was more than $10\%$ lower than the corresponding maximum
from the allometry dataset (here: $10\%\,\hat{=}\,\un{15}{cm}$ difference).
This may indicate that for these species, local conditions are unfavourable,
which in turn should be reflected in the parameterization. In cases
where we had enough (more than $1000$) trees in the inventory to
suggest that the maximal DBHs in the inventory coincide with the maximal
DBH reachable the study site, we therefore used the value $F_{\left\{ d_{k},k\in\alloset_{s}\right\} }^{-1}\ap{0.995}/0.9$.
These cases are shown in Table \ref{tab:maxDBH-exceptions}.
\begin{table}
\begin{centering}
\footnotesize%
\begin{tabular}{c>{\centering}p{2.8cm}>{\centering}p{2.8cm}>{\centering}p{2.8cm}>{\centering}p{2.8cm}}
\toprule 
Species & Number of trees with DBH $\geq\un 5{cm}$ in the inventory & Maximal diameter estimated from the inventory data $\left[\mt m\right]$ & Maximal diameter estimated from the allometry data $\left[\mt m\right]$ & Value used in the model $\left[\mt m\right]$\tabularnewline
\midrule
\midrule 
\textit{Acer mono} & $2469$ & $0.38$ & $0.61$ & $0.43$\tabularnewline
\textit{Acer pseudo-sieboldianum} & $1722$ & $0.20$ & $0.37$ & $0.23$\tabularnewline
\textit{Pinus koraiensis} & $2236$ & $0.79$ & $0.98$ & $0.88$\tabularnewline
\textit{Tilia amurensis} & $2115$ & $0.76$ & $1.04$ & $0.85$\tabularnewline
\bottomrule
\end{tabular}
\par\end{centering}
\caption{Maximal DBH values for species where the estimates from the inventory
and the allometry data deviate strongly. \label{tab:maxDBH-exceptions}}
\end{table}

We determined the frequency of each species in the inventory based
on its total basal area. Based on this, we constructed a discrete
probability distribution for the maximal tree height of a tree $t$.
Let $\specset_{i}$ be the species belonging to PFT $i$. We obtained
the following probability mass function for the maximal height of
a tree of PFT $i$: 
\begin{equation}
p_{i}^{\mt{max}}\ap d=\frac{\smo{s\in\specset_{i}}\I{\left\{ \bar{d}_{s}^{\mt{max}}\right\} }dw_{s}}{\smo{s\in\specset_{i}}w_{s}}
\end{equation}
with 
\begin{equation}
w_{s}=\smo{t\in\ensuremath{\invset_{s}}}d_{t}^{2}\label{eq:BA-weights}
\end{equation}
 and the indicator function $\I{\left\{ X\right\} }x$, which is $1$
id $x\in X$ and $0$ otherwise.

\begin{figure}
\begin{centering}
\includegraphics[width=1\textwidth,height=0.95\textheight,keepaspectratio]{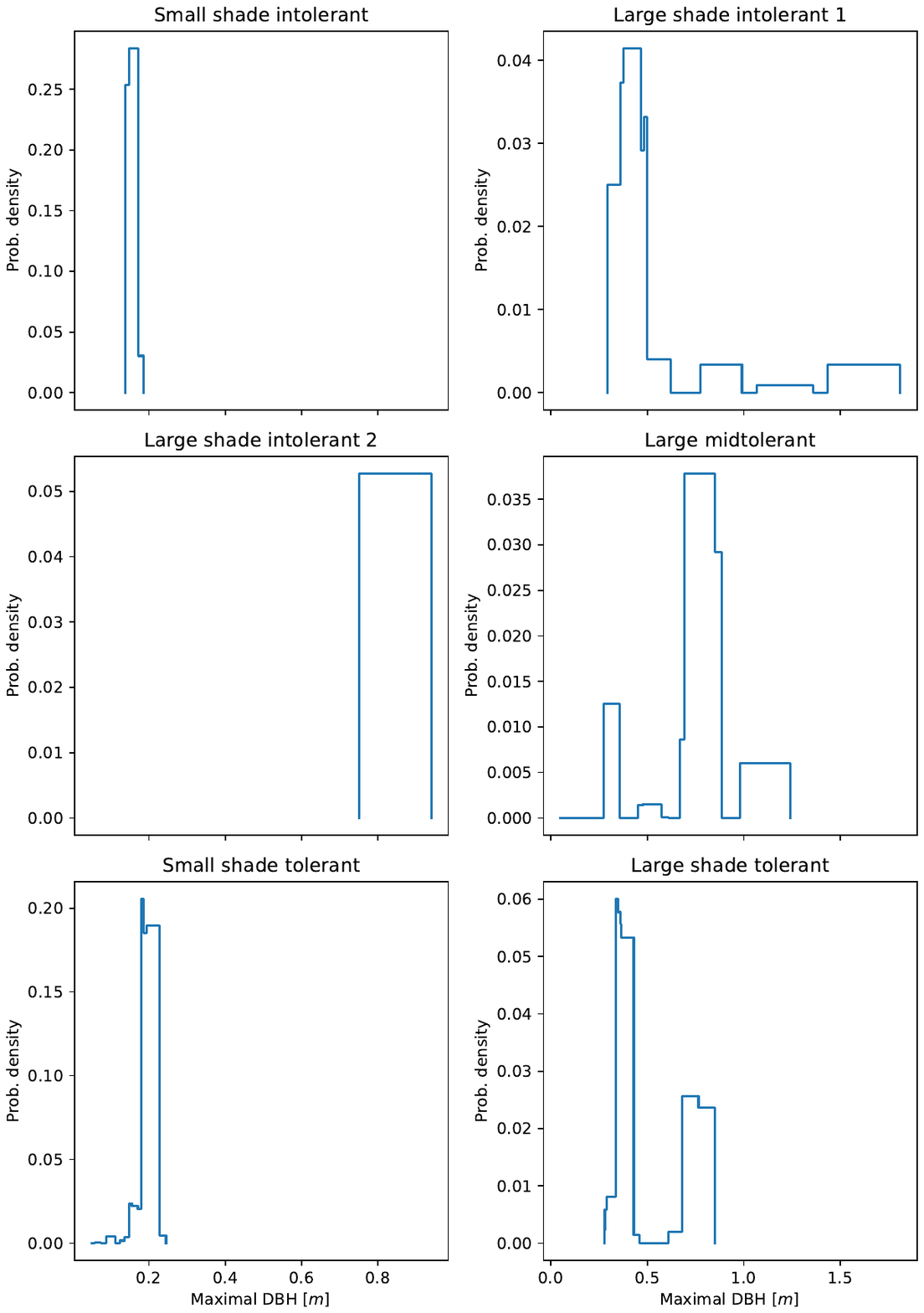}
\par\end{centering}
\caption{Distribution of the maximal DBH for the six PFTs.\label{fig:max-dbh-dist}}

\end{figure}

Since even trees of the same species may have different site-dependent
growth limits and to reduce a potential model artifact arising from
drawing the maximal DBHs from discrete distributions, we constructed
continuous distributions for the maximal diameters by blurring the
distribution below the maximal DBH values $\bar{d}_{s}^{\mt{max}}$.
That way, we obtained a continuous distribution with probability density
function
\begin{equation}
f_{i}^{\mt{max}}\ap d=c\smo{s\in\specset_{i}}\I{\left[\left(1-\beta\right)\bar{d}_{s}^{\mt{max}},\bar{d}_{s}^{\mt{max}}\right]}d\frac{w_{s}}{2\beta\bar{d}_{s}^{\mt{max}}},\label{eq:max-d-dist}
\end{equation}
where $\beta$ is a measure for the relative within-species variation
of the maximal diameter and $c$ is a normalization constant. We assumed
that the maximal diameter for each species can take values $\beta=20\%$
below the observed maximum. The resulting probability density functions
are displayed in Fig. \ref{fig:max-dbh-dist}.

For technical reasons, we used a discretized version of distribution
(\ref{eq:max-d-dist}). To that end, we considered $200$ potential
maximal DBH values homogeneously distributed in the interval $\left[\mino{s\in\specset_{i}}\left(1-\beta\right)\bar{d}_{s}^{\mt{max}},\maxo{s\in\specset_{i}}\left(1-\beta\right)\bar{d}_{s}^{\mt{max}}\right]$. 

\subsubsection{Wood density\label{subsec:Wood-density}}

We computed the wood density of each PFT by taking a weighted average
of species-specific wood densities. As weights, we used each species'
basal area in the inventory. That is, with weights $w_{s}$ defined
as in equation (\ref{eq:BA-weights}), we computed the wood density
$\rho_{i}$ of PFT $i$ via 
\begin{equation}
\rho_{i}=\frac{\smo{s\in\specset_{i}}\rho_{s}w_{s}}{\smo{s\in\specset_{i}}w_{s}},
\end{equation}
where $\rho_{s}$ is the wood density of species $s$.

\subsubsection{LAI}

Both theoretical and empirical studies suggest that a tree's leaf
area is roughly proportional to its basal area \citep{west_general_1999,xu_modelling_2021}.
However, as it is difficult and costly to determine the leaf area
of individual trees, empirical individual-level data on leaf area,
leaf number, or LAI are sparse. \citet{xu_modelling_2021} estimated
leaf numbers as functions of the DBH based on measurements on different
branch levels \citep{liu_nested_2009} for three species common in
our study area. In the absence of more direct measurements, we used
a simple approximation based on results by \citet{xu_modelling_2021}
to parameterize the PFTs to which the species they considered belong
and used generic estimates for the other PFTs.

As general ansatz for the relationship between DBH $d$ and LAI $L$,
we used the following function:
\begin{equation}
L\ap d=\th_{\mt{LAI},0,i}+\th_{\mt{LAI},1,i}\frac{d^{\th_{\mt{LAI},2,i}}}{A_{i}\ap d},\label{eq:LAI-equation}
\end{equation}
where $\th_{\mt{LAI},0,i}$, $\th_{\mt{LAI},1,i}$, $\th_{\mt{LAI},2,i}$
are parameters for PFT $i$ and 
\begin{equation}
A_{i}\ap d=\frac{\pi}{4}g_{\mt{\text{crown-d}}}\ap{d;\,\th_{\mt{\text{crown-d}},i}}^{2}
\end{equation}
 is the corresponding crown projection area (see also equation (\ref{eq:crown-d})).
The division by the crown projection area $A_{i}\ap d$ transforms
leaf area values to LAI values. The intercept parameter is necessary,
because trees require a minimal LAI to growth as much as observed
in the field. As the crown projection area is roughly proportional
to the DBH, the LAI would converge to $0$ for small trees if $\th_{\mt{LAI},0,i}=0$. 

We used the exponents $\th_{\mt{LAI},2,i}$ reported by \citet{xu_modelling_2021}
for \emph{Betula platyphylla}, \emph{Pinus koraiensis} and \emph{Tilia
amurensis} for the large shade intolerant 1, large mid-tolerant, and
large shade tolerant PFT, respectively. For the other PFTs, we assumed
a generic value of $2$ \citep{west_general_1999}. We computed the
remaining parameters $\th_{\mt{LAI},1,i}$ based on mean LAI values
$L_{\mt{ref},i}$ reported by \citet{xu_modelling_2021}. For PFTs
with unknown mean LAI, we used a generic value of $3$. As it was
unclear, to which DBH values the reported mean values corresponded,
we set reference DBH values $d_{\mt{ref},i}$ dependent on the maximal
tree sizes: $0.1\mt m$ for small PFTs and $0.25\mt m$ for large
PFTs. Setting $L\ap{d_{\mt{ref},i}}=L_{\mt{ref},i}$, we obtained
$\th_{\mt{LAI},1,i}$ with a simple manipulation of equation (\ref{eq:LAI-equation}).
The resulting parameter values are displayed in Table \ref{tab:Parameters-LAI}.
The resulting curves are visible in Fig. \ref{fig:lai-dbh}.

\begin{table}
\begin{centering}
\footnotesize%
\begin{tabular}{cc>{\centering}p{1.8cm}>{\centering}p{1.8cm}>{\centering}p{1.8cm}>{\centering}p{1.8cm}>{\centering}p{1.8cm}>{\centering}p{1.8cm}}
\toprule 
 & Unit & Small shade intolerant & Large shade intolerant 1 & Large shade intolerant 2 & Large mid-tolerant & Small shade tolerant & Large shade tolerant\tabularnewline
\midrule
\midrule 
$d_{\mt{ref},i}$ & $\mt m$ & $0.1$ & $0.25$ & $0.25$ & $0.25$ & $0.1$ & $0.25$\tabularnewline
$L_{\mt{ref},i}$ & $\frac{\mt m^{2}}{\mt m^{2}}$ & $3$ & $11.79${*} & $3$ & $3.717${*} & $3$ & $2.622${*}\tabularnewline
\midrule 
$\th_{\mt{LAI},0,i}$ & $\frac{\mt m^{2}}{\mt m^{2}}$ & $1$ & $1$ & $1$ & $0.5$ & $0.5$ & $0.5$\tabularnewline
$\th_{\mt{LAI},1,i}$ & $\frac{\mt m^{2}}{\mt m^{\th_{\mt{LAI},2,i}}}$ & $1581$ & $3373$ & $671.1$ & $654.9$ & $3086$ & $1191$\tabularnewline
$\th_{\mt{LAI},2,i}$ & $1$ & $2$ & $2.132${*} & $2$ & $1.847${*} & $2$ & $2.27${*}\tabularnewline
\bottomrule
\end{tabular}
\par\end{centering}
\caption{Parameter values for the relationship between DBH and LAI. Values
marked with an asterisk ({*}) were taken from \citet{xu_modelling_2021}.
\label{tab:Parameters-LAI}}
\end{table}

\subsubsection{Light extinction and transmission}

To compute the light climate in the forest, parameters for the light
extinction and light transmission of leafs are needed. We assumed
that these coefficients are independent of the PFTs. For the light
extinction coefficients we assumed a value of $0.5$ and for the light
transmission coefficients a value of $0.1$. 

\subsubsection{Mean stem biomass proportion\label{subsec:Mean-stem-biomass-prop}}

In \noun{Formind}, the biomass of a tree is computed by scaling the
stem biomass by an expansion factor, which reflects that some biomass
is allocated in branches and leaves. This expansion factor may depend
on the tree size and PFT. In our parameterization, we computed the
factor based on a submodel described in section \ref{subsec:Growth-allocation}
below. However, to parameterize this submodel, we needed information
about the mean stem biomass proportions. 

We determined the mean proportions of above-ground biomass in the
tree crown (i.e., branches and leafs) by comparing independent biomass
estimates \citep{piponiot_distribution_2022} for the Changbaishan
forest plot with the biomass estimates obtained via the allometric
relationships estimated in the sections above. As the estimates by
\citet{piponiot_distribution_2022} correspond to the entire forest
only, we reimplemented their approach, which is based on the allometric
biomass equations presented in Table 5 in \citet{chojnacky_updated_2014}.
We mapped the species found in the inventory data to the taxa found
in the table and used the corresponding biomass equations to estimate
the species' respective total biomasses in the study area. For species
for which we could not find a matching biomass equation, we used the
equation corresponding to Aceraceae with specific gravity below $0.5$.
We then adjusted the mean stem biomass proportions until our biomass
estimates matched the ones obtained via the equations by \citet{chojnacky_updated_2014}.
The resulting stem biomass proportions are displayed in Table \ref{tab:stem-B-prop}.

\begin{table}
\begin{centering}
\footnotesize%
\begin{tabular}{c>{\centering}p{1.8cm}>{\centering}p{1.8cm}>{\centering}p{1.8cm}>{\centering}p{1.8cm}>{\centering}p{1.8cm}>{\centering}p{1.8cm}}
\toprule 
 & Small shade intolerant & Large shade intolerant 1 & Large shade intolerant 2 & Large mid-tolerant & Small shade tolerant & Large shade tolerant\tabularnewline
\midrule
\midrule 
Mean stem biomass prop. & $0.8$ & $0.5$ & $0.6$ & $0.7$ & $0.7$ & $0.75$\tabularnewline
\bottomrule
\end{tabular}
\par\end{centering}
\caption{Estimated mean stem biomass proportion for the different PFTs. \label{tab:stem-B-prop}}
\end{table}

\subsection{Seed production and mortality}

We assume that there is a constant external seed influx to the forest.
This assumption holds approximately if the considered forest is part
of a larger forest area and seed availability does not depend on local
species abundances. The seeds are distributed evenly among the patches.
 Seeds that do not establish to small trees accumulate in a ``seed
bank'' and may establish in later years. However, seeds in the seed
bank are subject to a mortality of $50\%$ per year. 

We determined the number $n_{\mt{seeds},i}$ of incoming seeds of
PFT $i$ by fitting the model to forest inventory data (see section
\ref{subsec:Fitting-procedure} for details). The resulting values
are displayed in Table \ref{tab:Parameters-seeds}.

\begin{table}
\begin{centering}
\footnotesize%
\begin{tabular}{cc>{\centering}p{1.8cm}>{\centering}p{1.8cm}>{\centering}p{1.8cm}>{\centering}p{1.8cm}>{\centering}p{1.8cm}>{\centering}p{1.8cm}}
\toprule 
 & Unit & Small shade intolerant & Large shade intolerant 1 & Large shade intolerant 2 & Large mid-tolerant & Small shade tolerant & Large shade tolerant\tabularnewline
\midrule
\midrule 
$n_{\mt{seeds},i}$ & $\frac{1}{\mt{ha}\cdot\mt{yr}}$ & $1.297$ & $9.997$ & $1.28$ & $4.346$ & $2.603$ & $3.409$\tabularnewline
\midrule 
$\th_{\mt{est},0}$ & $1$ & $0.0714$ & $0.202$ & $0.0807$ & $0.0091$ & $0.0405$ & $3.36\e{-4}$\tabularnewline
\bottomrule
\end{tabular}
\par\end{centering}
\caption{Parameters for seed influx and establishment. \label{tab:Parameters-seeds}}
\end{table}

\subsection{Ingrowth\label{subsec:Ingrowth}}

\begin{figure}
\begin{centering}
\includegraphics[width=0.6\textwidth]{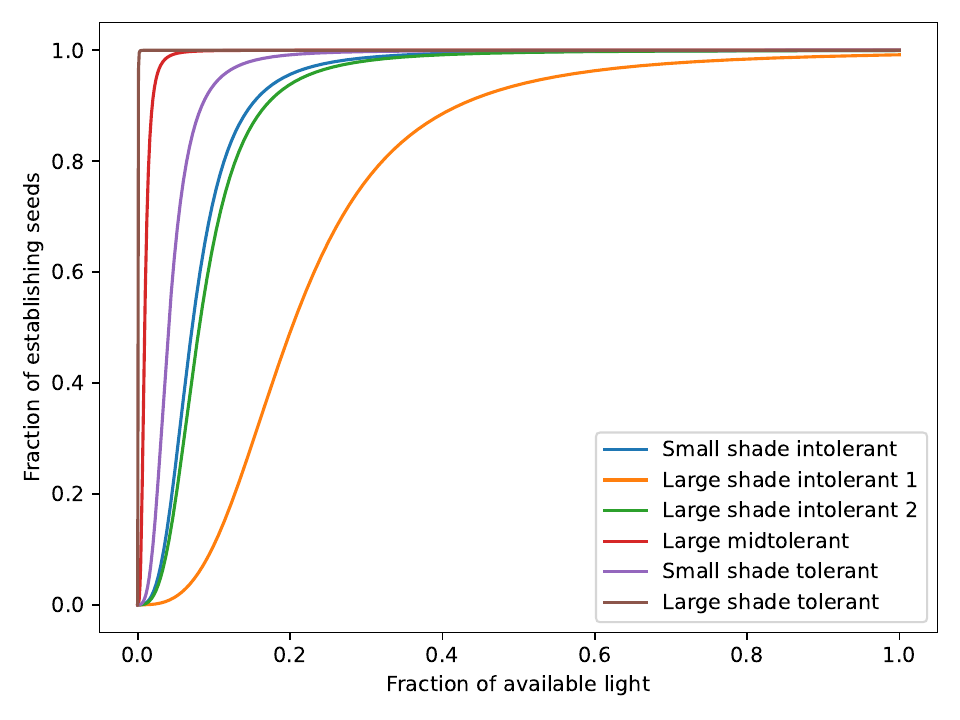}
\par\end{centering}
\caption{Ingrowth functions for the six PFTs. The fraction of seeds in the
seedbank that establish depends on the fraction of irradiation reaching
the bottom of the forest as compared to the incoming irradiation.\label{fig:ingrowth}}
\end{figure}

Seeds establish to small trees dependent on the light available at
the forest's ground and the length of the productive season. The fraction
$\phi_{\mt{seed},i}$ of seeds of PFT $i$ that establish is computed
using a Hill function:
\begin{equation}
\phi_{\mt{seed},i}\ap{\phi_{\mt{light},j}}=\frac{\phi_{\mt{light}}^{\th_{\mt{est},1}}}{\phi_{\mt{light}}^{\th_{\mt{est},1}}+\th_{\mt{est},0,i}^{\th_{\mt{est},1}}},
\end{equation}
where $\phi_{\mt{light},j}\in\left[0,1\right]$ is the fraction of
the incoming irradiance that reaches the ground in patch $j$, the
parameter $\th_{\mt{est},0,i}\in\left[0,1\right]$ is the irradiance
at which half of the seeds of PFT $i$ germinate, and $\th_{\mt{est},1}$
is a parameter controlling how steep the transition from unfavourable
to favourable germination conditions is. We estimated the parameters
$\th_{\mt{est},0,i}$ and $\th_{\mt{est},1}$ by fitting the model
to forest inventory data (section \ref{subsec:Fitting-procedure}).
The resulting values for $\th_{\mt{est},0,i}$ are displayed in Table
\ref{tab:Parameters-seeds}; the threshold sharpness was not fitted
PFT-specifically and assumed a value of $\th_{\mt{est},1}=3$. The
resulting curves are shown in Fig. \ref{fig:ingrowth}.

The number $n_{\mt{seedling},i,j}$ of newly establishing trees of
PFT $i$ in patch $j$ is computed by rounding the product of the
number $n_{\mt{seedbank},i,j}$ of seeds in the corresponding seed
bank and the number of establishing seeds $\phi_{\mt{seed},i}\ap{\phi_{\mt{light},j}}$:
\begin{equation}
n_{\mt{seedling},i,j}:=\left\lfloor n_{\mt{seedbank},i,j}\phi_{\mt{seed},i}\ap{\phi_{\mt{light},j}}+0.5\right\rfloor 
\end{equation}
All newly established trees have an initial DBH of $0.05\mt m$ irrespective
of the PFT.

\subsection{Growth\label{subsec:Growth}}

\begin{figure}
\begin{centering}
\includegraphics[width=1\textwidth]{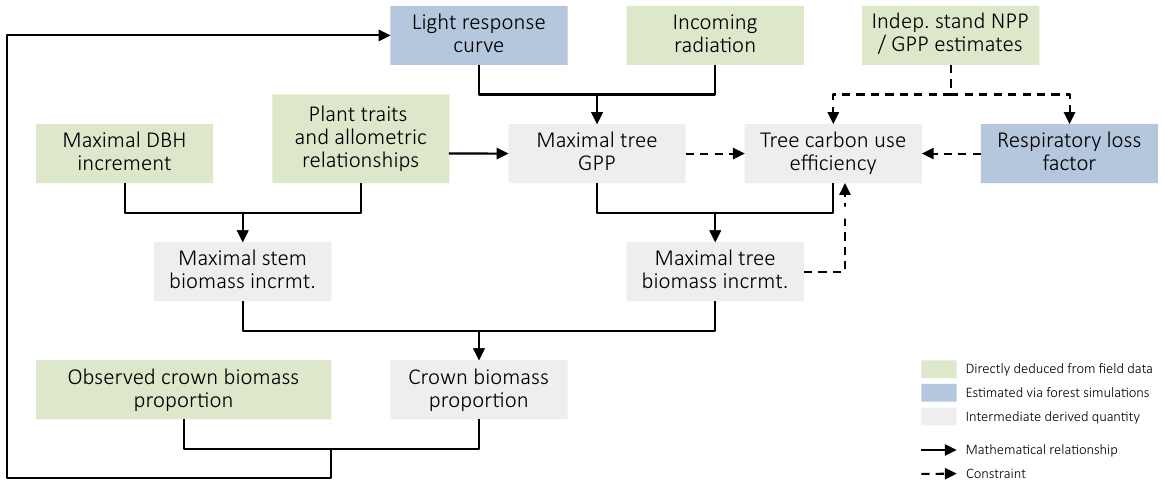}
\par\end{centering}
\caption{Overview of the model components and intermediate results used to
fit the submodels for the growth of individual trees. Solid arrows
depict direct mathematical relationships, whereas dashed arrows denote
constraints. Submodels and quantities that could be estimated independently
from the full model are drawn in green. Submodels with parameters
that could only be estimated from the full model are shown in blue.
Quantities that were derived from other components are depicted in
grey. \label{fig:Flow-chart-growth-calibration}}
\end{figure}

In \noun{Formind}, the growth of a tree is modelled using multiple
interacting submodels, which we calibrated partly jointly and partly
independently from one another (see Fig. \ref{fig:Flow-chart-growth-calibration}
for an overview). Key idea of our approach was to focus on trees growing
under the best possible conditions found on site. Focusing on optimal
conditions reduces the complexity while at the same time setting a
frame for the possible model behaviour. Below we briefly summarize
our approach before providing details in the succeeding sections.

Based on the forest inventory data, we estimated the PFT- and DBH-dependent
DBH increment under optimal conditions (section \ref{subsec:optimal-DBH-increment})
and used this along with the estimated allometric relationships (section
\ref{subsec:Allometric-relationships}) and plant traits (section
\ref{subsec:Plant-traits}) to approximate the stem biomass increment
under optimal conditions. At the same time, we used our model to estimate
the GPP (section \ref{subsec:Light-response}) and carbon use efficiency
(section \ref{subsec:CUE}) of trees under optimal growth conditions.
In a second step, we computed the aboveground wood production, which
we could use along with the observed stem biomass increments to deduce
the biomass allocated to the crown (section \ref{subsec:Growth-allocation}).
We compared these values with field estimates, in turn, to refine
the parameters that we used to compute the GPP. Parameters that could
not be estimated with this procedure were estimated by fitting the
full forest model to the forest inventory data (section \ref{subsec:Fitting-procedure}).

\subsubsection{DBH increment under optimal conditions\label{subsec:optimal-DBH-increment}}

We estimated the DBH increment under optimal conditions based on the
DBH increments observed in consecutive forest inventory data.  We
modelled the DBH increment as observed in the inventory data via a
simple stochastic model and used this as a baseline to derive the
optimal DBH increment.

We assumed that the DBH increment $\Delta d_{i_{k}}\ap{d_{k}}$ of
a tree $k$ with PFT $i_{k}$ and DBH $d_{k}$ follows a Gamma distribution.
Specifically, 
\begin{equation}
\Delta d_{i}\ap d\sim\gammadist{\frac{\mu_{\mt{\Delta DBH},i}\ap d}{\th_{\mt{\Delta DBH}}}}{\th_{\mt{\Delta DBH}}},\label{eq:deltaDBH-dist}
\end{equation}
where $\mu_{\mt{\Delta DBH},i}\ap d$ is the DBH-dependent mean DBH
increment, and $\th_{\mt{\Delta DBH}}$ is a scale parameter controlling
the distributions mean to variance ratio, which we assumed to be independent
of the DBH. We assumed that 
\begin{equation}
\mu_{\mt{\Delta DBH},i}\ap d=\th_{\mt{\Delta DBH},i,0}+\th_{\mt{\Delta DBH},i,1}d+\th_{\mt{\Delta DBH},i,2}d^{2}+\th_{\mt{\Delta DBH},i,3}d^{3}
\end{equation}
 is a cubic polynomial satisfying the following constraints:
\begin{align}
\mu_{\mt{\Delta DBH},i}\ap{d_{i}^{\mt{max}}} & =0,\label{eq:mu-dbh-dMax}\\
\mu_{\mt{\Delta DBH},i}\ap 0 & \geq0,\label{eq:mu-dbh-0}\\
\mu'_{\mt{\Delta DBH},i}\ap 0 & \geq0,\label{eq:mu-dbh-der0}\\
\mu'_{\mt{\Delta DBH},i}\ap{d_{i}^{\mt{max}}} & \leq0,\label{eq:mu-dbh-derDmax}
\end{align}
where $d_{i}^{\mt{max}}$ is the maximal DBH a tree of PFT $i$ can
assume. Constraint (\ref{eq:mu-dbh-dMax}) reflects that trees with
DBH $d_{i}^{\mt{max}}$ cannot grow even under optimal conditions.
Together with constraints (\ref{eq:mu-dbh-0})-(\ref{eq:mu-dbh-derDmax}),
it follows that $\mu'_{\mt{\Delta DBH},i}$ is always non-negative
and at most unimodal in the interval $\left[0,d_{i}^{\mt{max}}\right]$.
Note that constraint (\ref{eq:mu-dbh-dMax}) implies that one of the
parameters $\th_{\mt{\Delta DBH},i,0},\dots,\th_{\mt{\Delta DBH},i,3}$
can be expressed in terms of the other ones, reducing the degree of
freedom when fitting the model. 

\begin{figure}
\begin{centering}
\includegraphics[width=1\textwidth,height=0.95\textheight]{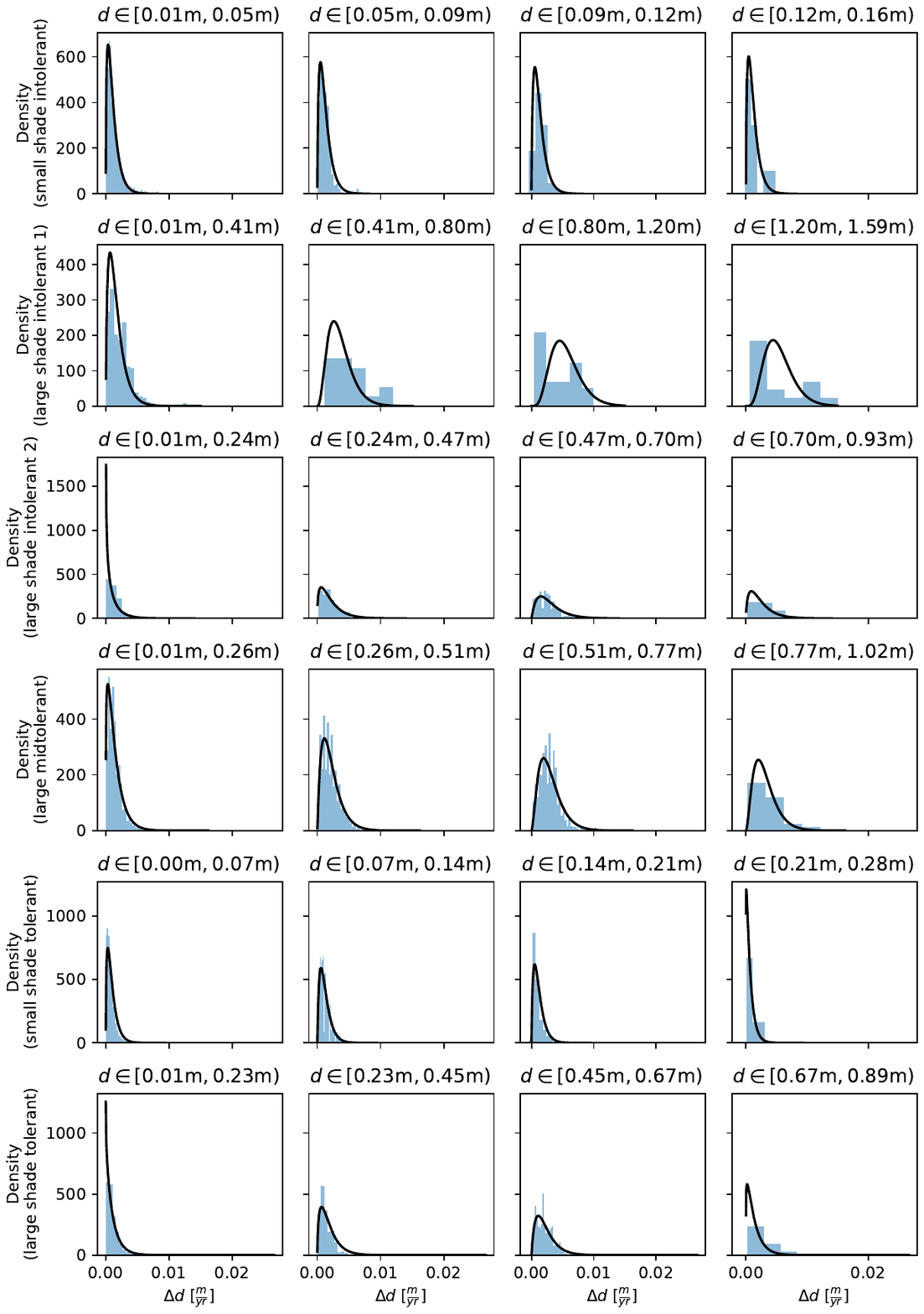}
\par\end{centering}
\caption{Histogram of observed DBH increments (blue histograms) and fitted
DBH increment used in the model (black lines) for the six PFTs and
plant sizes. Each panel corresponds to a PFT (indicated in the row
description) and a size class (range of considered DBH indicated in
the panel heading). The observed DBH increments are averages over
five year periods. The plotted probability densities correspond to
the DBHs in the centre of the respective considered DBH interval.
\label{fig:dbh-inc-dist}}
\end{figure}

We estimated the parameters by maximizing the likelihood given data
from consecutive forest inventories, conducted in intervals of five
years. For each tree $k$ that appeared in two consecutive inventories,
we determined the observed DBH difference 
\begin{equation}
\Delta d_{k}^{\mt{obs}}=d_{k,t_{2}}-d_{k,t_{1}},
\end{equation}
where $d_{k,t}$ is the observed DBH of tree $k$ in year $t$ and
$\Delta t=t_{2}-t_{1}=\un 5{yr}$. As empirical data may always be
prone to error, we disregarded all data that were more than $5$ standard
deviations apart from the mean DBH increment, taken over all individuals
of the considered PFT. Afterwards, we also excluded all negative values
$\Delta d_{k}^{\mt{obs}}<0$.  We estimated the parameters for the
optimal DBH growth by fitting the distribution (\ref{eq:deltaDBH-dist})
to the values 
\begin{equation}
\Delta\bar{d}_{k}^{\mt{obs}}=\frac{\Delta d_{k}^{\mt{obs}}}{\Delta t}.
\end{equation}
The resulting parameter estimates are displayed in Table \ref{tab:Parameters-deltaDBH}.
In Fig. \ref{fig:dbh-inc-dist}, we show histograms for the observed
DBH increments and the density functions of the corresponding fitted
Gamma distributions.

\begin{table}
\begin{centering}
\footnotesize%
\begin{tabular}{cc>{\centering}p{1.8cm}>{\centering}p{1.8cm}>{\centering}p{1.8cm}>{\centering}p{1.8cm}>{\centering}p{1.8cm}>{\centering}p{1.8cm}}
\toprule 
 & Unit & Small shade intolerant & Large shade intolerant 1 & Large shade intolerant 2 & Large mid-tolerant & Small shade tolerant & Large shade tolerant\tabularnewline
\midrule
\midrule 
$\th_{\mt{\Delta DBH}}$ & $1$ & $7.607\e{-4}$ & $9.939\e{-4}$ & $1.361\e{-3}$ & $1.110\e{-3}$ & $6.554\e{-4}$ & $1.171\e{-3}$\tabularnewline
\midrule 
$\th_{\mt{\Delta DBH},i,0}$ & $\mt m$ & $1.013\e{-3}$ & $1.299\e{-3}$ & $5.552\e{-4}$ & $1.276\e{-3}$ & $7.098\e{-4}$ & $9.300\e{-4}$\tabularnewline
\midrule 
$\th_{\mt{\Delta DBH},i,1}$ & $1$ & $3.354\e{-3}$ & $0$ & $0$ & $0$ & $7.866e\e{-3}$ & $0$\tabularnewline
\midrule 
$\th_{\mt{\Delta DBH},i,2}$ & $\frac{1}{\mt m}$ & $1.665\e{-2}$ & $9.564\e{-3}$ & $1.771\e{-2}$ & $9.321\e{-3}$ & $-2.317\e{-2}$ & $1.324\e{-2}$\tabularnewline
\midrule 
$\th_{\mt{\Delta DBH},i,3}$ & $\frac{1}{\mt m^{2}}$ & $-2.084\e{-1}$ & $-5.344\e{-3}$ & $-1.848\e{-2}$ & $-7.820\e{-3}$ & $-3.957\e{-2}$ & $-1.601\e{-2}$\tabularnewline
\bottomrule
\end{tabular}
\par\end{centering}
\caption{Parameters for the DBH increment distributions. \label{tab:Parameters-deltaDBH}}
\end{table}

\begin{figure}
\begin{centering}
\includegraphics[width=1\textwidth,height=0.95\textheight]{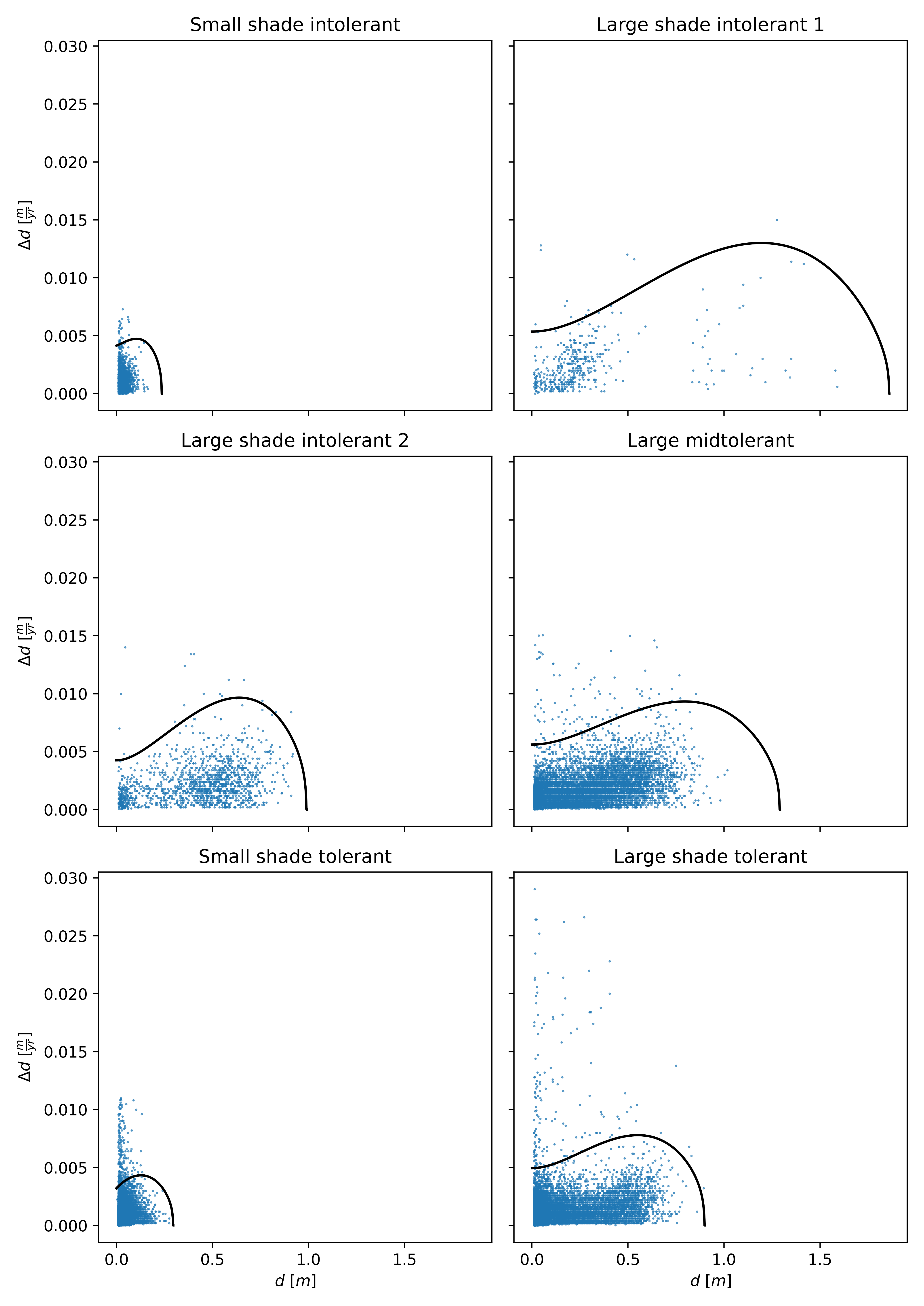}
\par\end{centering}
\caption{Observed yearly DBH increments (blue dots) and maximal DBH increment
used in the model (black lines) for the six PFTs.\label{fig:max-dbh-inc}}
\end{figure}

The Gamma distribution can take arbitrarily large values. Our goal,
however, was to determine some ``maximal'' DBH increment. We assumed
that the maximal DBH increment is given by some (high) quantile $q_{\mt{\Delta DBH}}$
of the fitted DBH increment distribution: 
\begin{equation}
\Delta d_{\mt{max},i}\ap d=F_{\Delta d_{i}\ap d}^{-1}\ap{q_{\mt{\Delta DBH}}},
\end{equation}
where $F_{\Delta d_{i}\ap d}^{-1}$ is the inverse cumulative probability
density function of $\Delta d_{i}\ap d$. That is, the DBH under optimal
conditions is the value chosen so that a fraction of $q_{\mt{\Delta DBH}}$
of the DBH increments of similar trees are expected to be lower. Whereas
we estimated the distribution of the DBH increments from forest inventory
data, we fitted the parameter $q_{\mt{\Delta DBH}}$ along with other
parameters based on a dynamic forest simulation (see section \ref{subsec:Fitting-procedure}).
We obtained a value of $q_{\mt{\Delta DBH}}=0.991$. The resulting
curves for the DBH-dependent optimal DBH increment are displayed in
Fig. \ref{fig:max-dbh-inc}. 

\subsubsection{Reference conditions}

We assumed that the estimated optimal DBH increments (section \ref{subsec:optimal-DBH-increment})
were obtained under the best possible conditions found at the Changbaishan
site. To link these observed DBH increments to the modelled GPP, we
needed to model these reference conditions explicitly. We assumed
that the optimal growth conditions correspond to the best light conditions
observed in the forest inventory. For large trees, this is equivalent
to being unshaded by other trees. However, there may be no unshaded
small trees in the inventory for some PFT, requiring us to adjust
the reference light conditions accordingly. This issue was not considered
in previous paramterizations of \noun{Formind}. This may have led
to underestimated growth of small trees. 

\begin{figure}
\begin{centering}
\includegraphics[width=0.6\textwidth]{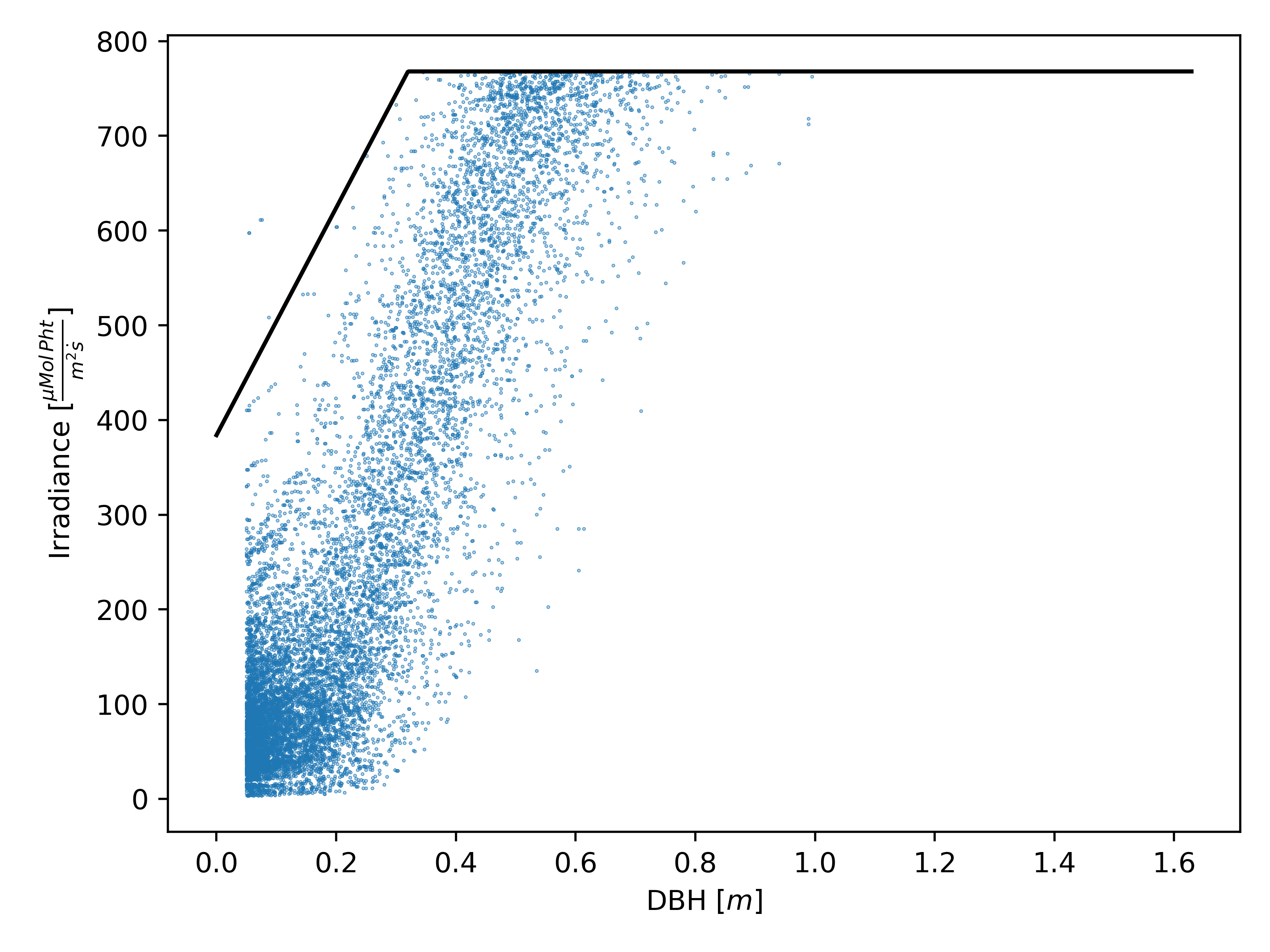}
\par\end{centering}
\caption{Reference light conditions dependent on the DBH. Each blue dot corresponds
to a tree in the inventory and shows its DBH and the irradiance that
it received according to the shading model in \noun{Formind}. The
black line depicts the irradiance that is used as \textquotedblleft optimal\textquotedblright{}
reference in the parameterization. As there are no unshaded small
trees, the estimated maximal observed DBH increment (section \ref{subsec:optimal-DBH-increment})
does not correspond to unshaded trees. Therefore, the black curve
starts at an irradiance $50\%$ below the irradiance received by unshaded
trees.\label{fig:Reference-light-conditions}}
\end{figure}

Here, we made an ad-hoc correction to account for the range of light
conditions found for trees in the inventory. We initialized \noun{Formind}
with the forest inventory data, computed the incoming light for all
trees (Fig. \ref{fig:Reference-light-conditions}), and determined
a simple piecewise linear function that yields for each DBH the maximal
fraction of incoming radiation observed for trees with this DBH
\[
\phi_{\mt{light}}\ap d=\min\ap{\th_{\mt{light},0}+\th_{\mt{light},1}d,\,1}
\]
where $\th_{\mt{light},0}$ is the most favourable fraction of irradiance
received by small plants and $\th_{\mt{light},1}$ is the initial
slope of the reference light fraction. We fitted this curve via visual
inspection, observing (1) the approximate maximal irradiance received
by small trees and (2) the DBH at which some trees received the full
irradiance. We obtained the values $\th_{\mt{light},0}=0.5$ and $\th_{\mt{light},1}=1.5625\mt m^{-1}$.
The resulting relation is displayed in Fig. \ref{fig:Reference-light-conditions}.

\subsubsection{Light response curve\label{subsec:Light-response}}

In \noun{Formind}, a tree's GPP is determined based on the light response
curve mapping the incoming radiation $I_{\mt{leaf}}$ of a leaf of
PFT $i$ to its photosynthetic rate $\gpp_{\mt{leaf}}$
\begin{equation}
\gpp_{\mt{leaf},i}\ap{I_{\mt{leaf}}}=\frac{\th_{\mt{production},i,0}I_{\mt{leaf}}}{\th_{\mt{production},i,1}+I_{\mt{leaf}}},
\end{equation}
where $\th_{\mt{production},i,0}$ is the maximal possible photosynthetic
rate and $\th_{\mt{production},i,1}$ the irradiation at which half
of the maximally possible photosynthetic rate is achieved. We fitted
the parameters $\th_{\mt{production},i,1}$ based on model simulations
and the forest inventory data (section \ref{subsec:Fitting-procedure}).
For each given value of $\th_{\mt{production},i,1}$, we computed
the corresponding parameter $\th_{\mt{production},i,0}$ by determining
how large the production needs to be to let the trees of PFT $i$
attain their observed crown biomass proportions based on our assumptions
on the carbon use efficiency (see section \ref{subsec:CUE}) and stem
biomass allocation. Details are provided in section \ref{subsec:Growth-allocation}.
\begin{table}
\begin{centering}
\footnotesize%
\begin{tabular}{cc>{\centering}p{1.8cm}>{\centering}p{1.8cm}>{\centering}p{1.8cm}>{\centering}p{1.8cm}>{\centering}p{1.8cm}>{\centering}p{1.8cm}}
\toprule 
 & Unit & Small shade intolerant & Large shade intolerant 1 & Large shade intolerant 2 & Large mid-tolerant & Small shade tolerant & Large shade tolerant\tabularnewline
\midrule
\midrule 
$\th_{\mt{production},i,0}$ & $\frac{\mu\mt{Mol}\,\mt{CO}_{2}}{\mt m^{2}\cdot\mt s}$ & $13.677$ & $4.864$ & $5.274$ & $3.459$ & $3.215$ & $11.553$\tabularnewline
\midrule 
$\th_{\mt{production},i,1}$ & $\frac{\mu\mt{Mol}\,\mt{photon}}{\mt m^{2}\cdot\mt s}$ & $500$ & $118.56$ & $100$ & $70.82$ & $274.15$ & $492.73$\tabularnewline
\bottomrule
\end{tabular}
\par\end{centering}
\caption{Parameters for the light response curve. \label{tab:Parameters-light}}
\end{table}

\subsubsection{Carbon use efficiency and respiration\label{subsec:CUE}}

We define the carbon use efficiency (CUE; in formulas $\cue_{k}$)
of a tree $k$ as the fraction of its primary production (GPP; in
formulas $\gpp_{k}$) that is used for net (aboveground) primary production
(NPP, in formulas $\npp_{k}$):
\begin{equation}
\cue_{k}=\frac{\npp_{k}}{\gpp_{k}}.
\end{equation}
The NPP, in turn, can be written as the difference of GPP and respiration:
\begin{equation}
\npp_{k}=\gpp_{k}-R_{k}.
\end{equation}
We considered two types of respiratory losses: the maintenance respiration
$R_{\mt{maint},i}\ap d$, dependent on the tree size but independent
of the GPP, and other losses and limitations $R_{\mt{loss},k}$, proportional
to the NPP but otherwise independent of the tree size:
\[
R_{k}=R_{\mt{maint},i}\ap{d_{k}}+R_{\mt{loss},k}=R_{\mt{maint},k}+\frac{\gamma_{i_{k}}}{1-\gamma_{i_{k}}}\npp_{k}=R_{\mt{maint},k}+\gamma_{i_{k}}\left(\gpp_{k}-R_{\mt{maint},k}\right),
\]
where $\gamma_{i_{k}}$ is a PFT-dependent loss factor, modelling
how much of the production not assigned to maintenance can be used
for production. It follows
\begin{align}
\cue_{k} & =\frac{\gpp_{k}-R_{\mt{maint},k}-\gamma_{i_{k}}\left(\gpp_{k}-R_{\mt{maint},k}\right)}{\gpp_{k}}\nonumber \\
 & =\left(1-\gamma_{i_{k}}\right)\left(1-\frac{R_{\mt{maint},k}}{\gpp_{k}}\right)\label{eq:CUE-loss-constraint}
\end{align}
Note that the maintenance respiration represents the tree's minimal
respiratory needs and thus cannot be reduced even if the tree is under
stress. Hence, if the maintenance respiration is large compared to
the other losses, already a moderate reduction of the GPP (e.g. due
to shading) can entail that a tree cannot satisfy its respiratory
needs and stops growing or dies. 

As no data on the optimal CUE on single-tree level were available
to us, we created a phenomenological model for the \emph{optimal}
CUE (below: OCUE) based on a number of observations:
\begin{enumerate}
\item The OCUE decreases as trees grow in size.
\item The OCUE must be sufficiently large that trees can reach the estimated
optimal biomass increment.\label{enu:cue-opt-binc}
\item The CUE must suffice that most trees observed in the inventory can
satisfy their minimal respiratory needs.\label{enu:cue-inventory}
\item The order of magnitude of the OCUE must be chosen so that the values
of GPP and NPP match field measurements on the stand level approximately.
\label{enu:cue-observed}
\item The OCUE is subject to additional limitations and carbon losses independent
of the maintenance respiration. Hence the OCUE cannot exceed $1-\gamma_{i}$.\label{enu:cue-loss-const}
\end{enumerate}
As baseline for the OCUE model, we used the following formula:
\begin{equation}
\cue_{\mt{base},i}\ap d=\th_{\mt{OCUE},0,i}-\th_{\mt{OCUE},1,i}d^{\th_{\mt{OCUE},2,i}},\label{eq:CUE-basis}
\end{equation}
where $i$ is the PFT, $d$ is the DBH, and $\th_{\mt{OCUE},0,i}$,
$\th_{\mt{OCUE},1,i}$, and $\th_{\mt{OCUE},2,i}$ are parameters.
However, to guarantee that constraint \ref{enu:cue-opt-binc} is satisfied,
we also computed the minimal required CUE so that the trees can grow
as much as observed under optimal conditions. Let 
\begin{equation}
\npp_{\mt{stem},i}^{\mt{opt}}\ap d=\rho_{i}\left(V_{\mt{stem},i}\ap{d+\Delta d_{\mt{max},i}\ap d}-V_{\mt{stem},i}\ap d\right)\label{eq:stem-B-inc}
\end{equation}
be the stem biomass increment under optimal conditions, where $d$
is the current DBH, $\rho_{i}$ is the wood density, $V_{\mt{stem},i}$
the stem volume, and $\Delta d_{\mt{max},i}$ the DBH increment under
optimal conditions. We assumed that, under optimal conditions, at
least a factor $\kappa_{\mt{min}}=0.1$ of the NPP is allocated to
crown growth. Hence, the NPP under optimal conditions must be at least
$\frac{1}{1-\kappa}\npp\npp_{\mt{stem},i}^{\mt{opt}}\ap d$. Consequently,
we adjusted the OCUE correspondingly:
\begin{equation}
\cue_{\mt{opt},i}\ap d=\max\ap{\cue_{\mt{base},i}\ap d,\,\frac{1}{1-\kappa_{\mt{min}}}\npp_{\mt{stem},i}^{\mt{opt}}\ap d}.\label{eq:CUE-opt}
\end{equation}

\begin{figure}
\begin{centering}
\includegraphics[width=1\textwidth,height=0.95\textheight]{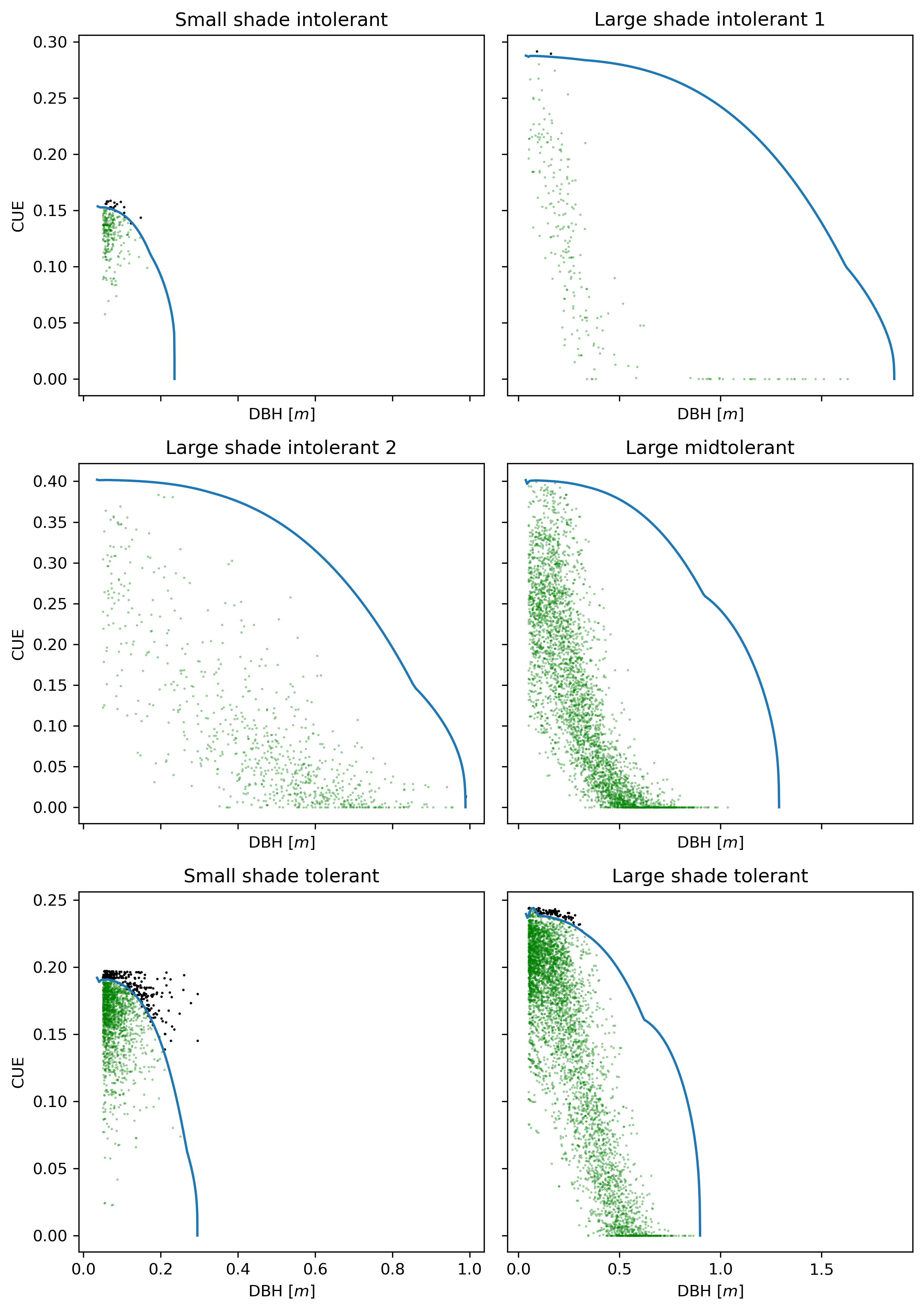}
\par\end{centering}
\caption{\scriptsize The optimal carbon use efficiency OCUE for the different
PFTs. The OCUEs used in the model are depicted as solid blue lines.
The points show estimated lower bounds for the required CUEs obtained
for trees in the inventory via the shading module of \noun{Formind}.
Each point corresponds to a tree; the colour shows whether the tree
could satisfy its respiratory needs according to the model (green:
yes; black: no). The OCUE curves were chosen so that as many of the
points are below the blue curves. The sharp transitions between the
curve sections are due to constraint \ref{enu:cue-opt-binc} imposing
a different shape of the curve for large DBH values (see also equation
(\ref{eq:CUE-opt})). \label{fig:cue}}
\end{figure}

We assumed that the OCUE is monotonously decreasing as trees grow.
With constraint \ref{enu:cue-loss-const}, we obtain that $\th_{\mt{OCUE},0,i}\leq\gamma_{i}$.
At the same time, constraint \ref{enu:cue-inventory} requires that
$R_{\mt{maint},i}\ap d$ is small for small trees, as small shaded
trees observed in the inventory could not survive otherwise. Hence,
we set 
\begin{equation}
\th_{\mt{OCUE},0,i}=\gamma_{i}-0.01.
\end{equation}
Similarly, applying the shading module of \noun{Formind} to the inventory
data, we observed that the OCUE must decrease slowly for small trees
(Fig. \ref{fig:cue}), which in turn requires a sufficiently large
exponent $\th_{\mt{OCUE},2,i}$. We therefore set $\th_{\mt{OCUE},2,i}=3$
for all PFTs $i$. Lastly, we determined $\th_{\mt{OCUE},1,i}$ so
that the largest possible trees of PFT $i$ have an OCUE of $0$ at
their maximal DBH. That is, if 
\begin{equation}
d_{\mt{max},i}=\sup\left\{ d;\,f_{i}^{\mt{max}}\ap d>0\right\} 
\end{equation}
is the maximal DBH a tree of PFT $i$ can attain (cf. equation (\ref{eq:max-d-dist})),
then 
\begin{equation}
\th_{\mt{OCUE},1,i}=\th_{\mt{OCUE},0,i}d_{\mt{max},i}^{-\th_{\mt{OCUE},2,i}}.
\end{equation}

We estimated the loss factors $\gamma_{i}$ by fitting the full model
to the inventory data (section \ref{subsec:Fitting-procedure}). However,
to satisfy constraint \ref{enu:cue-observed}, we constrained the
loss factors $\gamma_{i}$ to the interval $\left[0.6,1\right]$ to
match the relatively low CUE values observed in the Changbaishan mountain
area in independent studies \citep{piponiot_distribution_2022}.
The resulting parameter estimates are displayed in Table \ref{tab:Parameters-resp-loss}
\begin{table}
\begin{centering}
\footnotesize%
\begin{tabular}{cc>{\centering}p{1.8cm}>{\centering}p{1.8cm}>{\centering}p{1.8cm}>{\centering}p{1.8cm}>{\centering}p{1.8cm}>{\centering}p{1.8cm}}
\toprule 
 & Unit & Small shade intolerant & Large shade intolerant 1 & Large shade intolerant 2 & Large mid-tolerant & Small shade tolerant & Large shade tolerant\tabularnewline
\midrule
\midrule 
$\gamma_{i}$ & $1$ & $0.15$ & $0.285$ & $0.4$ & $0.4$ & $0.189$ & $0.236$\tabularnewline
\bottomrule
\end{tabular}
\par\end{centering}
\caption{Scaling factors relating the NPP to respiratory losses other than
the maintenance respiration. \label{tab:Parameters-resp-loss}}
\end{table}

As we assume that the maintenance respiration is independent of a
tree's productivity, equation (\ref{eq:CUE-loss-constraint}) must
in particular hold for trees under optimal growth conditions. Hence,
after inserting the fitted OCUE $\cue_{\mt{opt},i}\ap d$ and GPP
under optimal conditions, equation (\ref{eq:CUE-loss-constraint})
can be manipulated to derive the maintenance respiration for a tree
of given PFT and DBH.

\subsubsection{Growth allocation\label{subsec:Growth-allocation}}

Based on the OCUE and the GPP under optimal conditions, denoted $\cue_{i}$
and $\gpp_{i}^{\mt{opt}}$, respectively, we could compute the corresponding
NPP $\npp_{i}^{\mt{opt}}\ap d$ for trees of a given DBH and PFT.
Based on the estimated DBH increment under optimal conditions, determined
the respective stem biomass increment $\npp_{\mt{stem},i}^{\mt{opt}}\ap d$
(see equation (\ref{eq:stem-B-inc})). If $B_{i}\ap d$ is the biomass
of a tree of PFT $i$ with DBH $d$ and $\zeta_{i}$ is the corresponding
stem biomass proportion, then 
\begin{align}
\npp_{i}^{\mt{opt}}\ap d & =\cue_{i}\ap d\gpp_{i}^{\mt{opt}}\ap d\nonumber \\
 & =B_{i}\ap{d+\Delta d_{\mt{max},i}\ap d}-B_{i}\ap d\nonumber \\
 & =\frac{B_{\mt{stem},i}\ap{d+\Delta d_{\mt{max},i}\ap d}}{\zeta_{i}\ap{d+\Delta d_{\mt{max},i}\ap d}}-\frac{B_{\mt{stem},i}\ap d}{\zeta_{i}\ap d}\\
\dad\nonumber \\
\zeta_{i}\ap{d+\Delta d_{\mt{max},i}\ap d} & =\frac{\zeta_{i}\ap dB_{\mt{stem},i}\ap{d+\Delta d_{\mt{max},i}\ap d}}{B_{\mt{stem},i}\ap d+\zeta_{i}\ap d\cue_{i}\ap d\gpp_{i}^{\mt{opt}}\ap d}.\label{eq:stem-B-prop-diff-eq}
\end{align}
We used this difference equation to compute the stem biomass proportion
for all DBHs and PFTs. We provide details below.

Equation (\ref{eq:stem-B-prop-diff-eq}) requires knowledge of the
previous stem biomass proportion $\zeta_{i}\ap d$. Hence, we needed
initial values for the interval $\left[d_{0},d_{0}+\Delta d_{\mt{max},i}\ap d\right]$
with $d_{0}$ being the stem diameter of new saplings. These initial
values may be chosen arbitrarily. Using a shifted exponential ansatz
for the initial condition yielded well-behaved smooth results for
$\zeta_{i}$: 
\begin{equation}
\zeta_{i}\ap d=a_{0,i}+a_{1,i}\exp\ap{a_{2,i}\cdot d}\hspace{2cm}\text{if }d<d_{0}+\Delta d_{\mt{max},i}\ap{d_{0}}.
\end{equation}
We chose the coefficients $a_{0i}$, $a_{1i}$, $a_{2i}$ so that
the curve $\zeta_{i}\ap d$ is continuous, approximately differentiable,
and starts at a given initial value $\zeta_{0i}=\zeta_{i}\ap{d_{0}}$. 

To see how the coefficients were determined, first note that in practice,
the curve $\zeta_{i}$ is computed numerically and hence evaluated
at a discrete set of sampling points only. We chose the sampling points
so that they have a constant distance to one another. Intermediate
values were obtained via linear interpolation between these points.
Now, let $d_{1i}=d_{0}+\Delta d_{\mt{max},i}\ap{d_{0}}$, let $\bar{d}_{1i}>d_{1i}$
be the smallest sampling point larger than $d_{1i}$, and choose $\bar{d}_{0i}$
so that $\bar{d}_{1i}=\bar{d}_{0i}+\Delta d_{\mt{max},i}\ap{\bar{d}_{0i}}$.
Furthermore, define (evaluating equation (\ref{eq:stem-B-prop-diff-eq})
at $d_{0i}$ and $\bar{d}_{0i}$)
\begin{alignat}{1}
\zeta_{1i}= & \frac{\zeta_{0i}B_{\mt{stem},i}\ap{d_{1}}}{B_{\mt{stem},i}\ap{d_{0}}+\zeta_{0i}\cue_{i}\ap d\gpp_{i}^{\mt{opt}}\ap{d_{0}}},\\
\bar{\zeta}_{1i}= & \frac{\zeta_{0i}B_{\mt{stem},i}\ap{\bar{d}_{1}}}{B_{\mt{stem},i}\ap{\bar{d}_{0i}}+\zeta_{0i}\cue_{i}\ap d\gpp_{i}^{\mt{opt}}\ap{\bar{d}_{0i}}}.
\end{alignat}
 Now we imposed the following conditions
\begin{alignat}{2}
\zeta_{i}\ap{d_{0}}= & a_{0i}+a_{1i}\exp\ap{a_{2i}\cdot d_{0}} & = & \zeta_{0i},\\
\zeta_{i}\ap{d_{1i}}= & a_{0i}+a_{1i}\exp\ap{a_{2i}\cdot d_{1i}} & = & \zeta_{1i},\\
\zeta_{i}\ap{\bar{d}_{1i}}= & a_{0i}+a_{1i}\exp\ap{a_{2i}\cdot\bar{d}_{1i}} & = & \bar{\zeta}_{1i}\label{eq:condition-end-section}
\end{alignat}
and obtained 
\begin{alignat}{1}
a_{1i} & =\frac{\zeta_{1i}-\zeta_{0i}}{\exp\ap{a_{2i}\cdot d_{1i}}-\exp\ap{a_{2i}\cdot d_{0}}}\label{eq:a1}\\
a_{0i} & =\zeta_{0i}-a_{1i}\exp\ap{a_{2i}\cdot d_{0}}.\label{eq:a0}
\end{alignat}
We computed the remaining unknown coefficient $a_{2i}$ via a binary
search on equation (\ref{eq:condition-end-section}) using the values
for $a_{0i}$ and $a_{1i}$ from equations (\ref{eq:a1})-(\ref{eq:a0}).

We approximated the mean of the curves $\zeta_{i}\ap d$ by taking
the man of the functions values at $50$ equidistant points in the
intervals $\left[0.1\mt m,\,d_{\mt{max},i}\right]$, respectively.
We then conducted a binary search in the maximal possible photosynthetic
rate $\th_{\mt{production},i,0}$ (see section \ref{subsec:Light-response})
until the approximate mean values matched the mean stem biomass proportions
estimated from the field data (section \ref{subsec:Mean-stem-biomass-prop}).

\begin{figure}
\begin{centering}
\includegraphics[width=0.8\textwidth]{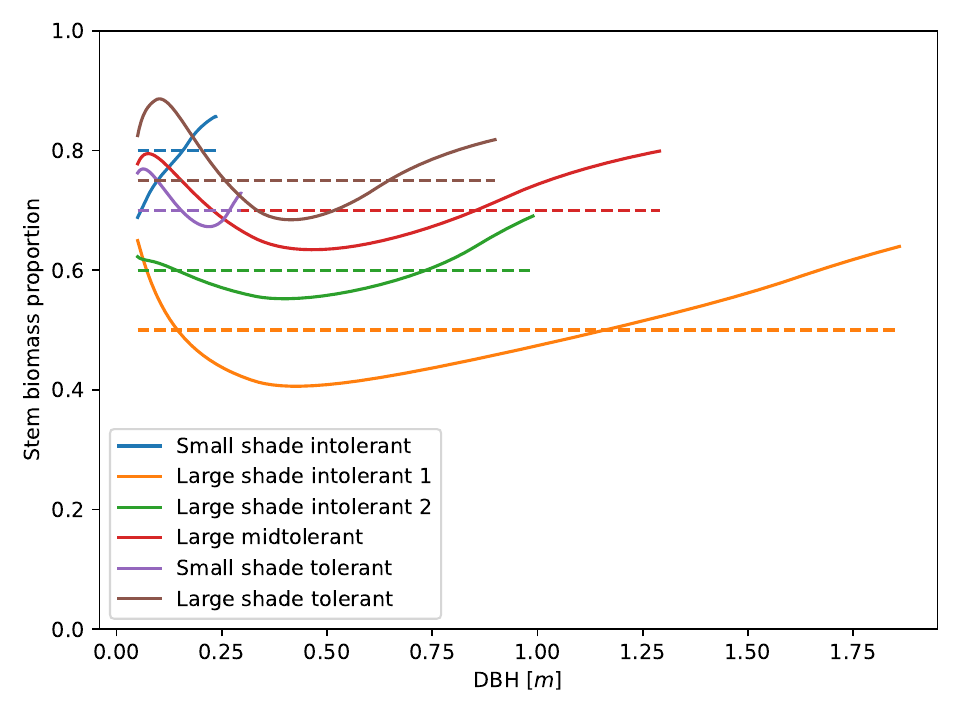}
\par\end{centering}
\caption{Stem biomass proportions of the six PFTs. The solid lines depict the
stem biomass proportions used in the model (obtained via the approach
described in section \ref{subsec:Growth-allocation}). The dashed
lines show the independently estimated mean values (see section \ref{subsec:Mean-stem-biomass-prop}).\label{fig:stem-b-prop}}
\end{figure}

\subsubsection{Defoliation\label{subsec:Defoliation}}

If trees are shaded, it can happen that their maintenance respiration
exceeds their GPP. In these cases, we assumed that parts of the crown
die until the remaining tree can be maintained. Here, we assumed that
for a tree of given DBH, the maintenance respiration is proportional
to its biomass. That is, a tree $k$ with insufficient production
$\gpp_{k}$, maintenance respiration $R_{\mt{maint},k}$, and biomass
$B_{k}$ will reduce its biomass to 
\begin{equation}
\tilde{B}_{k}=B_{k}\frac{\gpp_{k}}{R_{\mt{maint},k}},
\end{equation}
and its maintenance respiration will be set to $\gpp_{k}$. As we
assume the biomass is lost in the crown only, the stem biomass proportion
is adjusted accordingly to a value $\tilde{\zeta}_{k}$. 

We assumed that the loss in crown biomass also affects the tree's
number of leafs and thereby the LAI. We reduced the LAI proportional
to the crown completeness 
\begin{align}
\eta_{k} & =\frac{\tilde{B}_{\mt{crown},k}}{B_{\mt{crown},i_{k}}\ap{d_{k}}}\nonumber \\
 & =\frac{\tilde{B}_{k}-B_{\mt{stem},i_{k}}\ap{d_{k}}}{B_{i_{k}}\ap{d_{k}}-B_{\mt{stem},i_{k}}\ap{d_{k}}},
\end{align}
where $\tilde{B}_{\mt{crown},k}$ is the reduced crown biomass and
$B_{i_{k}}\ap{d_{k}}$ the biomass of a tree with complete crown and
DBH $d_{k}$. As a result, trees with incomplete crowns have reduced
GPP and shade other trees less. Trees without any crown biomass ($\eta_{k}=0$)
cannot recover and die. 

We assumed that if the light conditions for a tree with incomplete
crown improve, the new biomass is first allocated to ``refill''
the crown until $\eta_{k}=1$. Any remaining new biomass is allocated
to the usual tree growth with corresponding DBH increment. 

\subsection{Competition\label{subsec:Competition}}

We assumed that trees solely compete for light. In particular, we
did not apply any space competition. Instead, the forest density is
self-regulated via crown defoliation and the resulting tree death. 

\subsection{Mortality\label{subsec:Mortality}}

We assumed that trees die randomly with probabilities dependent on
their PFT and DBH. As model for the mortality, we used a linear combination
of exponentials:
\begin{equation}
p_{\mt{mort},i}\ap d=\th_{\mt{\mt{mort},0,i}}+\th_{\mt{\mt{mort},1,i}}\exp\ap{\th_{\mt{\mt{mort},2,i}}d}+\th_{\mt{\mt{mort},3,i}}\exp\ap{\th_{\mt{\mt{mort},4,i}}d},\label{eq:p-mort}
\end{equation}
where $p_{\mt{mort},i}\ap d$ is the probability that a tree of PFT
$i$ and DBH $d$ dies within a year. This model may take a variety
of shapes including mortality increasing or decreasing with plant
size or a ``bathtub'' shape, where the mortality is lowest for plants
with intermediate sizes.

\begin{figure}
\begin{centering}
\includegraphics[width=1\textwidth,height=0.95\textheight]{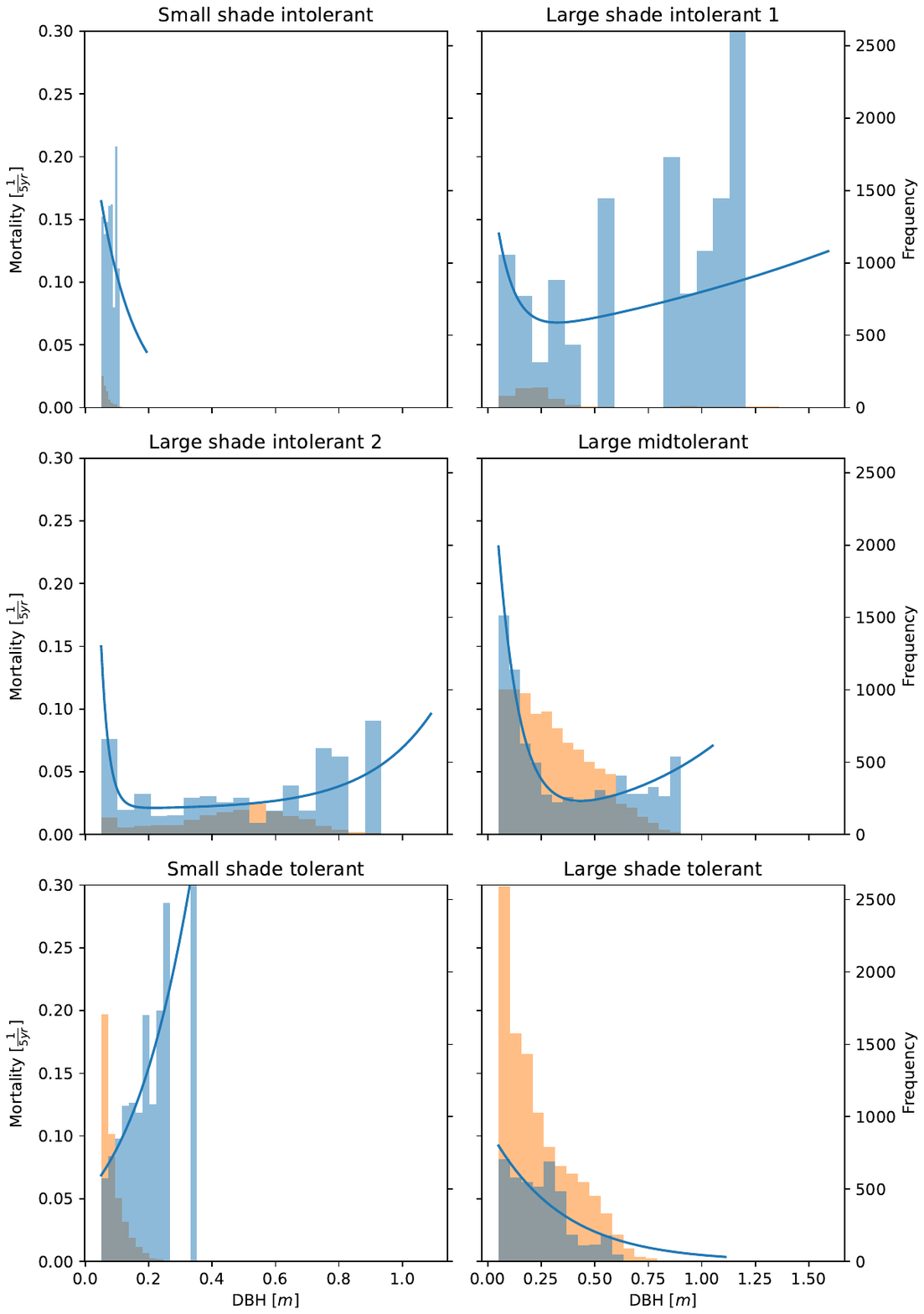}
\par\end{centering}
\caption{Mortality by PFT (field data and model). The blue curves (primary
axis) depict the modelled DBH-dependent probabilities that a tree
dies within a year. The blue bars correspond to field estimates of
the death probabilities (number of dead trees divided by the total
number of trees in the inventory). The orange bars (secondary axis)
are a histogram for the tree sizes, indicating where the mortality
estimates have the strongest empirical support.\label{fig:mort}}
\end{figure}

We estimated the parameters in equation (\ref{eq:p-mort}) using data
from consecutive forest inventories. We determined which trees died
in the intermediate time by comparing which trees that were present
in the first inventory were also present in the second inventory.
For simplicity, we assumed that the tree DBH does not change significantly
during the $5$ year period between two censuses and that random mortality
is the only death mechanism at play. If $d_{k,t_{1}}$ is the DBH
observed in the inventory in year $t_{1}$, the probability that the
tree survived until the year $t_{2}$ of the second inventory is approximately
\begin{equation}
p_{\mt{mort},i_{k}}^{\mt{obs}}=\left(1-p_{\mt{mort},i_{k}}\ap{d_{k,t_{1}}}\right)^{t_{2}-t_{1}}.
\end{equation}
We used this to construct the likelihood for the observed death and
survival events. We then estimated the parameters in equation (\ref{eq:p-mort})
for the different PFTs. The resulting parameters are displayed in
Table \ref{tab:Parameters-mort} and the resulting curves in Fig.
\ref{fig:mort}.
\begin{table}
\begin{centering}
\footnotesize%
\begin{tabular}{cc>{\centering}p{1.8cm}>{\centering}p{1.8cm}>{\centering}p{1.8cm}>{\centering}p{1.8cm}>{\centering}p{1.8cm}>{\centering}p{1.8cm}}
\toprule 
 & Unit & Small shade intolerant & Large shade intolerant 1 & Large shade intolerant 2 & Large mid-tolerant & Small shade tolerant & Large shade tolerant\tabularnewline
\midrule
\midrule 
$\th_{\mt{\mt{mort},0,i}}$ & $1$ & $0$ & $0$ & $0.004$ & $0$ & $0$ & $0$\tabularnewline
\midrule 
$\th_{\mt{\mt{mort},1,i}}$ & $1$ & $0.0568$ & $0.0345$ & $0.2415$ & $0.0829$ & $0.0106$ & $0.0224$\tabularnewline
\midrule 
$\th_{\mt{\mt{mort},2,i}}$ & $1$ & $-9.5049$ & $-12.6546$ & $-43.2337$ & $-10.6963$ & $5.6635$ & $-3.0862$\tabularnewline
\midrule 
$\th_{\mt{\mt{mort},3,i}}$ & $1$ & $0$ & $0.0112$ & $6.823\e{-5}$ & $0.002$ & $0$ & $0$\tabularnewline
\midrule 
$\th_{\mt{\mt{mort},4,i}}$ & $1$ & $0$ & $0.5339$ & $5.0046$ & $1.8684$ & $0$ & $0$\tabularnewline
\bottomrule
\end{tabular}
\par\end{centering}
\caption{Parameters for the mortality probabilities. \label{tab:Parameters-mort}}
\end{table}

Besides the random mortality, we trees may die due to strong light
competition (see section \ref{subsec:Defoliation}) or by falling
large trees. We assumed that trees larger than $\un{0.1}m$ may fall
with a probability of $0.4$ and kill smaller trees. Details of this
mechanism are described in \citet{fischer_lessons_2016}.

\subsection{Climate\label{subsec:Climate}}

We used a static climate in our simulations. Advanced features such
as the soil water module, temperature effects, and daily changes to
the climate were not included. Instead, we used averaged values, which
we provide below.
\begin{description}
\item [{Evapotranspiration.}] For the mean actual evapotranspiration, we
used a value of $600\frac{\mt{mm}}{\text{\ensuremath{\mt{yr}}}}$.
This is in line with independent estimates for the Changbaishan region
\citep{sun_spatial_2004} and earlier parameterizations of the model
for temperate forests \citep{bohn_climate_2014}.
\item [{Growing~season.}] We defined the growing season as the months
with positive mean temperature. This were the months March until October
\citep{wang_study_2020}. 
\item [{Irradiance.}] We computed the mean yearly light intensity (``PAR'')
above the canopy during daytime in the growing season based on the
WFDEI forcing dataset \citep{weedon_wfdei_2014}. We obtained a value
of $768\frac{\mu\mt{Mol}\,\text{photons}}{\mt m^{2}}$.
\item [{Day~length.}] We computed the average length of a day in the growing
season and obtained a value of $13.39\mt h$. 
\end{description}

\subsection{Fitting procedure\label{subsec:Fitting-procedure}}

\begin{figure}
\begin{centering}
\includegraphics[width=1\textwidth]{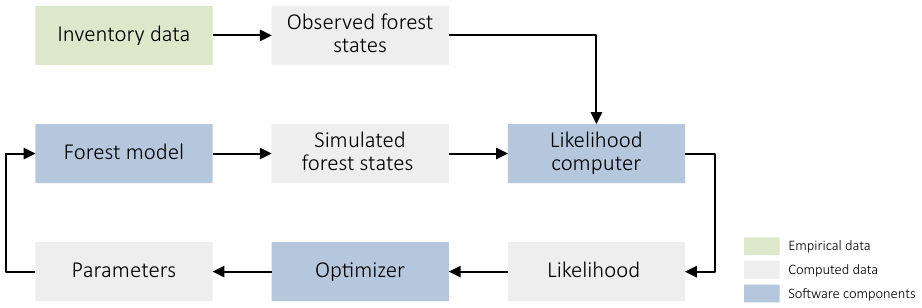}
\par\end{centering}
\caption{Overview of the model fitting procedure.\label{fig:Flow-chart-model-fit}}
\end{figure}

Some parameters were not available from the literature and could not
be determined directly from the available data. We estimated these
parameters based on dynamical forest simulations and the inventory
data (see Fig. \ref{fig:Flow-chart-model-fit}). After a burn-in period,
we generated a sample of forest states via simulations. Then, we used
the generated sample to estimate the likelihood for the parameters
given the inventory data via kernel density estimation (KDE). We then
optimized the parameters by maximizing the likelihood. Below we provide
details for each of the steps involved.

\subsubsection{Forest state characterization}

We characterized the forest's state by determining the stem count
and the biomass in the considered patch for each PFT. That is, the
state space was $12$-dimensional. The combined information of stem
count and biomass yields basic insight into the size distribution
of trees, as a large stem count with small biomass indicates a young
forest with many small trees, and a small stem count with high biomass
indicates an old forest with few large trees. Hence, these summary
statistics provide relatively rich information about the overall forest
state.

We considered forest states on the $20\mt m\times20\mt m$ level.
For the forest plot in Changbaishan, which has a size of $25\mt{ha}$,
we therefore obtained a sample of $625$ observed forest states.

\subsubsection{Sample generation}

To generate a sample of forest states from the model, we first simulated
$1\mt{ha}$ of forest until it reached its limiting behaviour ($2000\text{\ensuremath{\mt{yr}}}$).
Then, we generated a sample of forest states on the $20\mt m\times20\mt m$
scale by sampling the forest $500$ times every $5\mt{yr}$. Via parallel
simulations, we repeated this procedure $67$ times. That way we obtained
a sample of forest states with $\nsample=837,500$ entries. 

In \noun{Formind}, the interactions between $20\mt m\times20\mt m$
patches are small (only via tree falling, which is a rare event).
Furthermore, taking samples over a relatively long period of $2500\text{\ensuremath{\mt{yr}}}$
reduces the temporal correlations between the generated samples. Therefore,
and because we also conducted $67$ mutually independent simulations,
the generated sample is approximately identically independently distributed.

\subsubsection{Likelihood estimation}

The distribution of the forest states according to the forest model
is not known in closed form and can only be studied via simulations.
Therefore, we estimated the probability density, and based on this
the likelihood for the parameters given the data, from the model-generated
sample of forest states. To this end, we used kernel density estimation
\citep[KDE;][]{wand_kernel_1995}. In KDE, the probability density
$f$ of an element $y$ of the state space is estimated as the mean
of kernel functions centred at the elements $x_{i}$ of the generated
sample: 
\begin{eqnarray}
f\ap{y;\,x_{1},\dots,x_{\nsample}}\approx\hat{f}\ap{y;\,x_{1},\dots,x_{\nsample}} & = & \frac{1}{\nsample}\smft{i=1}{\nsample}\prft{j=1}{\ndim}K\ap{y_{j},x_{ij};h_{j}},\label{eq:KDE}
\end{eqnarray}
where $\hat{f}$ is the estimated probability density, $\nsample$
is the number of generated sample points, $\ndim$ is the dimension
of the state space, $K$ is the kernel function, and $h_{j}$ is a
bandwidth parameter defining the (marginal) scale on which two points
are considered approximately similar. Due to its computational simplicity
on the log-scale, we used a Gaussian Kernel. However, since stem counts
and biomasses are constrained to the non-negative range, we applied
reflecting boundary conditions: 
\begin{equation}
K_{j}\ap{y_{j},x_{ij};h_{j}}=\begin{cases}
\frac{1}{w_{j}}\left(\exp\ap{-\frac{\left(x_{ij}-y_{j}\right)^{2}}{2h_{j}^{2}}}+\exp\ap{-\frac{\left(x_{ij}+y_{j}\right)^{2}}{2h_{j}^{2}}}\right) & \text{if }y_{j}\geq0\\
0 & \text{else}
\end{cases}\label{eq:kernel-def1}
\end{equation}
with $w_{j}=\sqrt{2\pi}h_{j}$. For the stem counts, we furthermore
needed to normalized the kernel to correctly account for the discrete
nature of the data:
\begin{equation}
K_{j}\ap{y_{j},x_{ij};h_{j}}=\begin{cases}
\frac{1}{w_{j}}\left(\exp\ap{-\frac{\left(x_{ij}-y_{j}\right)^{2}}{2h_{j}^{2}}}+\exp\ap{-\frac{\left(x_{ij}+y_{j}+1\right)^{2}}{2h_{j}^{2}}}\right) & \text{if }y_{j}\geq0\\
0 & \text{else}
\end{cases}\label{eq:kernel-def2}
\end{equation}
with
\begin{eqnarray*}
w_{j} & = & 2\smft{k=0}{\infty}\exp\ap{-\frac{k^{2}}{2h_{j}^{2}}}-1.
\end{eqnarray*}
The bandwidths we used are displayed in Table \ref{tab:Bandwidths}.

KDE provides unbiased estimate of probability densities. For likelihood
optimization, however, we need the log-likelihood, because working
with the original likelihood would require us to handle extremely
small numbers, which is numerically infeasible. As taking the logarithm
of a random variable changes its distribution and, in particular,
expected value, we applied a bias correction. First, note that for
a Gaussian kernel, the KDE's expected value $\mu_{\mt{KDE}}$ and
variance $\sigma_{\mt{KDE}}^{2}$ can be approximated as follows \citep{wand_kernel_1995}:
\begin{alignat}{1}
\mu_{\mt{KDE}} & \approx f\ap y\\
\sigma_{\mt{KDE}}^{2} & \approx\frac{\mu_{\mt{KDE}}}{2\sqrt{\pi}nh}.
\end{alignat}
We desired to find a bias correction function $g$ so that 
\begin{align}
\ev{g\ap{\hat{f}\ap y}} & =\ln\mu_{\mt{KDE}}.
\end{align}
Applying a Taylor expansion about $\mu_{\mt{KDE}}$, we find
\begin{align}
\ev{g\ap{\hat{f}\ap y}} & \approx g\ap{\mu_{\mt{KDE}}}+\frac{1}{2}g''\ap{\mu_{\mt{KDE}}}\descr{\ev{\left(\hat{f}\ap y-\mu_{\mt{KDE}}\right)^{2}}}{\sigma_{\mt{KDE}}^{2}}\nonumber \\
 & =g\ap{\mu_{\mt{KDE}}}+\frac{\sigma_{\mt{KDE}}^{2}}{2}g''\ap{\mu}\nonumber \\
 & \overset{!}{=}\ln\mu_{\mt{KDE}}.\label{eq:bias-ODE}
\end{align}
We solved differential equation (\ref{eq:bias-ODE}) to obtain the
bias correction function, into which we inserted the original results
(\ref{eq:KDE}) from the KDE. To avoid numerical issues, we performed
all these steps on the log scale.

To fit the model, we considered a $12$-dimensional state space. As
a result, the products of the kernel functions in equation (\ref{eq:KDE})
can become very small and very sensitive to stochastic differences
between simulation runs. We therefore estimated the probability density
for each PFT independently and multiplied the results to obtain the
joint density. This is equivalent to assuming that the states of different
PFTs are mutually independent. Though this assumption is inaccurate
in general, using the resulting composite likelihood still yields
consistent parameter estimates \citep{varin_composite_2008}.

\begin{table}
\begin{centering}
\footnotesize%
\begin{tabular}{>{\raggedright}m{1.2cm}cc>{\centering}p{1.8cm}>{\centering}p{1.8cm}>{\centering}p{1.8cm}>{\centering}p{1.8cm}>{\centering}p{1.8cm}>{\centering}p{1.8cm}}
\toprule 
\centering{}State variable & Unit &  & Small shade intolerant & Large shade intolerant 1 & Large shade intolerant 2 & Large mid-tolerant & Small shade tolerant & Large shade tolerant\tabularnewline
\midrule
\midrule 
\multirow{3}{1.2cm}{\centering{}Stem count} & \multirow{3}{*}{$\frac{1}{400\mt m^{2}}$} & Mean & $0.48$ & $0.376$ & $1.278$ & $6.381$ & $3.17$ & $8.229$\tabularnewline
 &  & Range & $\left[0,7\right]$ & $\left[0,8\right]$ & $\left[0,7\right]$ & $\left[0,21\right]$ & $\left[0,17\right]$ & $\left[0,25\right]$\tabularnewline
 &  & Bandw. & $0.4$ & $0.4$ & $0.6$ & $1$ & $0.6$ & $2$\tabularnewline
\midrule 
\multirow{3}{1.2cm}{\centering{}Biomass} & \multirow{3}{*}{$\frac{t\,\mt{ODM}}{400\mt m^{2}}$} & Mean & $0.06$ & $0.457$ & $2.633$ & $4.617$ & $0.062$ & $3.047$\tabularnewline
 &  & Range & $\left[0,0.1\right]$ & $\left[0,10.945\right]$ & $\left[0,17.346\right]$ & $\left[0,13.133\right]$ & $\left[0,0.607\right]$ & $\left[0,10.909\right]$\tabularnewline
 &  & Bandw. & $0.005$ & $0.05$ & $0.1$ & $0.5$ & $0.02$ & $0.5$\tabularnewline
\bottomrule
\end{tabular}
\par\end{centering}
\caption{Ranges and KDE bandwidths for the considered state variables. The
bandwidth is the scale in the state space on which a data point in
the simulated sample is considered \textquotedblleft similar\textquotedblright{}
to a point in the inventory dataset. \label{tab:Bandwidths}}
\end{table}

\subsubsection{Parameter optimization}

A challenge when maximizing the kernel density estimate of the likelihood
is that this estimate is stochastic. This requires the applied optimizers
to be robust against stochastic fluctuations. We applied the algorithm
PY-BOBYQA \citep{cartis_improving_2019} on a preconditioned version
of the log-likelihood function. To reduce numerical issues, we optimized
all parameters on the log-scale except for $q_{\mt{\Delta DBH}}$,
for which we applied an inverse logit transform to constrain it to
the open interval $\left(0,1\right)$. Then, we evaluated the log-likelihood
function $10$ times at the initial parameter guess (Table \ref{tab:Parameter-Bounds})
to estimate its standard deviation. Based on this, we conducted for
each parameter individually a rough binary search to find the scale
of change on which the log-likelihood function changed by at least
$2$ standard deviations but not more than $10$ standard deviations.
We scaled the parameters accordingly for an efficient search. This
scaling process is called preconditioning.

We constrained the parameters to ecologically reasonable ranges, respectively.
The bounds we applied are displayed in Table \ref{tab:Parameter-Bounds}.
To avoid getting stuck due to stochastic deviations, we terminated
the search algorithm after $200$ likelihood evaluations and restarted
the search until a total of $8$ runs was completed. To minimize the
risk of converging to a local minimum, we furthermore applied basin-hopping
\citep{wales_global_1997} as implemented in Scipy. This algorithm
performs repeated local optimizations with randomly perturbed initial
conditions. For the perturbation, we applied a step size of $4$ on
the preconditioned parameter scale. We ran the algorithm for $5$
iterations. After finishing this optimization process, we repeated
it, using the result as initial value and baseline for preconditioning
for the repetition.

\begin{table}
\begin{centering}
\hspace*{-0.2cm}\footnotesize%
\begin{tabular}{ccc>{\centering}p{1.7cm}>{\centering}p{1.7cm}>{\centering}p{1.7cm}>{\centering}p{1.7cm}>{\centering}p{1.7cm}>{\centering}p{1.7cm}}
\toprule 
Parameter & Unit &  & Small shade intolerant & Large shade intolerant 1 & Large shade intolerant 2 & Large mid-tolerant & Small shade tolerant & Large shade tolerant\tabularnewline
\midrule
\midrule 
\multirow{2}{*}{$n_{\mt{seeds},i}$} & \multirow{2}{*}{$\frac{1}{\mt{ha}\cdot\mt{yr}}$} & Range  & $\left[0.001,50\right]$ & $\left[0.001,50\right]$ & $\left[0.001,50\right]$ & $\left[0.001,50\right]$ & $\left[0.001,50\right]$ & $\left[0.001,50\right]$\tabularnewline
 &  & Guess & $2$ & $2$ & $2$ & $2$ & $2$ & $2$\tabularnewline
\midrule 
\multirow{2}{*}{$\th_{\mt{est},0}$} & \multirow{2}{*}{$1$} & Range  & $\left[0.01,0.4\right]$ & $\left[0.01,0.4\right]$ & $\left[0.01,0.4\right]$ & $\left[0.001,0.4\right]$ & $\left[0.0001,0.3\right]$ & $\left[0.0001,0.3\right]$\tabularnewline
 &  & Guess & $0.15$ & $0.15$ & $0.15$ & $0.05$ & $0.01$ & $0.01$\tabularnewline
\midrule 
\multirow{2}{*}{$\th_{\mt{production},i,1}$} & \multirow{2}{*}{$\frac{\mu\mt{Mol}\,\mt{phot.}}{\mt m^{2}\cdot\mt s}$} & Range  & $\left[100,500\right]$ & $\left[100,500\right]$ & $\left[100,500\right]$ & $\left[50,500\right]$ & $\left[20,300\right]$ & $\left[20,500\right]$\tabularnewline
 &  & Guess & $300$ & $300$ & $300$ & $150$ & $100$ & $100$\tabularnewline
\midrule 
\multirow{2}{*}{$\gamma_{i}$} & \multirow{2}{*}{$1$} & Range  & $\left[0.15,0.4\right]$ & $\left[0.15,0.4\right]$ & $\left[0.15,0.4\right]$ & $\left[0.15,0.4\right]$ & $\left[0.15,0.4\right]$ & $\left[0.15,0.4\right]$\tabularnewline
 &  & Guess & $0.3$ & $0.3$ & $0.3$ & $0.3$ & $0.3$ & $0.3$\tabularnewline
\midrule 
\multirow{2}{*}{$q_{\mt{\Delta DBH}}$} & \multirow{2}{*}{$1$} & Range  & \multicolumn{6}{c}{$\left[0.2,0.9999\right]$}\tabularnewline
 &  & Guess & \multicolumn{6}{c}{$0.99$}\tabularnewline
\midrule 
\multirow{2}{*}{$\th_{\mt{est},1}$} & \multirow{2}{*}{$1$} & Range  & \multicolumn{6}{c}{$\left[3,20\right]$}\tabularnewline
 &  & Guess & \multicolumn{6}{c}{$5$}\tabularnewline
\bottomrule
\end{tabular}
\par\end{centering}
\caption{Parameter bounds and initial guesses used for parameter optimization.
\label{tab:Parameter-Bounds}}
\end{table}

\section{DBH entropy}

\subsection{Derivation of the DBH entropy\label{subsec:DBH-entropy-APX}}

We used the basal-area-weighted DBH entropy as a proxy for the prevalence
of large trees in a forest patch. The entropy of the weighted DBH
distribution is defined as follows:

\begin{equation}
\tilde{\entropy}_{\mt{DBH}}=-\smo{d\in D}p_{d}\ln\ap{p_{d}},\label{eq:entropy-discrete-2}
\end{equation}
where $D$ is the set of distinct DBH values occurring in the forest
patch and 
\begin{equation}
p_{d}=\frac{\smo{k\in\mathcal{I}:d_{k}=d}d_{k}^{2}}{\smo{k\in\mathcal{I}}d_{k}^{2}}.\label{eq:d-prob-1}
\end{equation}
 is the probability to randomly select a tree with DBH $d$ from the
forest patch if the probabilities were proportional to the trees'
respective basal areas. Here, $\mathcal{I}$ is the set of trees in
the inventory and $d_{k}$ is the DBH of tree $k$.

Formula (\ref{eq:entropy-discrete-2}) is sensitive to arbitrarily
small changes in DBH values, as trees need to have exactly the same
DBH values to be considered similar in equation (\ref{eq:d-prob-1}).
This is inappropriate, as DBH values come from a continuous domain,
and will never be exactly equal in practice. To make the measure more
robust, we could consider DBH intervals instead of individual DBH
values, as suggested in the main text. However, this approach is sensitive
to the choice of interval bounds and can lead to strongly different
results for slight changes of DBH values \citep[cf.][]{wand_kernel_1995}.
We therefore used kernel density estimation to obtain a continuous
distribution of tree sizes from the inventory. Then, we considered
the entropy of the resulting distribution:
\begin{equation}
\entropy_{\mt{DBH}}=-\intft 0{\infty}{f_{\mt d}\ap{\delta}\ln f_{\mt d}\ap{\delta}}{\delta},\label{eq:entropy-discrete-1-1}
\end{equation}
where 
\begin{equation}
f_{\mt d}\ap{\delta}=\smo{d\in D}w_{d}K\ap{d,\delta;h}
\end{equation}
with weights
\begin{equation}
w_{d}=\frac{d^{\eta}}{\smo{d\in D}d^{\eta}}\label{eq:weight}
\end{equation}
is the smoothed DBH distribution in the forest patch, 
\[
K\ap{d,\delta;h}=\begin{cases}
\frac{3}{4h}\left(1-\left(\frac{d-\delta}{h}\right)^{2}\right) & \text{if }\left|d-\delta\right|\leq h\\
0 & \text{else}
\end{cases}
\]
 is the Epanechnikov kernel, $\eta$ is the exponent parameter and
$h$ is a bandwidth parameter, defining the scale on which two trees
are regarded similar. 

\subsection{DBH entropy parameterization\label{subsec:DBH-entropy-parameterization}}

The DBH entropy depends on the exponent parameter $\eta$ and the
bandwidth parameter $h$. In line with our requirements for a proxy
for the prevalence of mature trees, we chose $\eta=2$ to obtain weights
by basal area and $h=\un 1{cm}$ for a sufficiently fine-grained resolution
to distinguish tree sizes well. To validate this choice of parameters
and compare it to parameters used in other studies, we assessed the
relationship between GPP, NPP, and NEE and the DBH entropy computed
with different parameter values: $\eta=0$ (no weighting), $\eta=2$
(weighting by basal area), $\eta=3$ (higher-order weighting, potentially
similar to biomass) with $h=\un 1{cm}$, respectively, and $\eta=0$
and $\eta=2$ with $h=\un{10}{cm}$. We used the same methods as for
the analysis of the other diversity measures. 

\begin{sidewaysfigure}
\includegraphics[width=1\textheight]{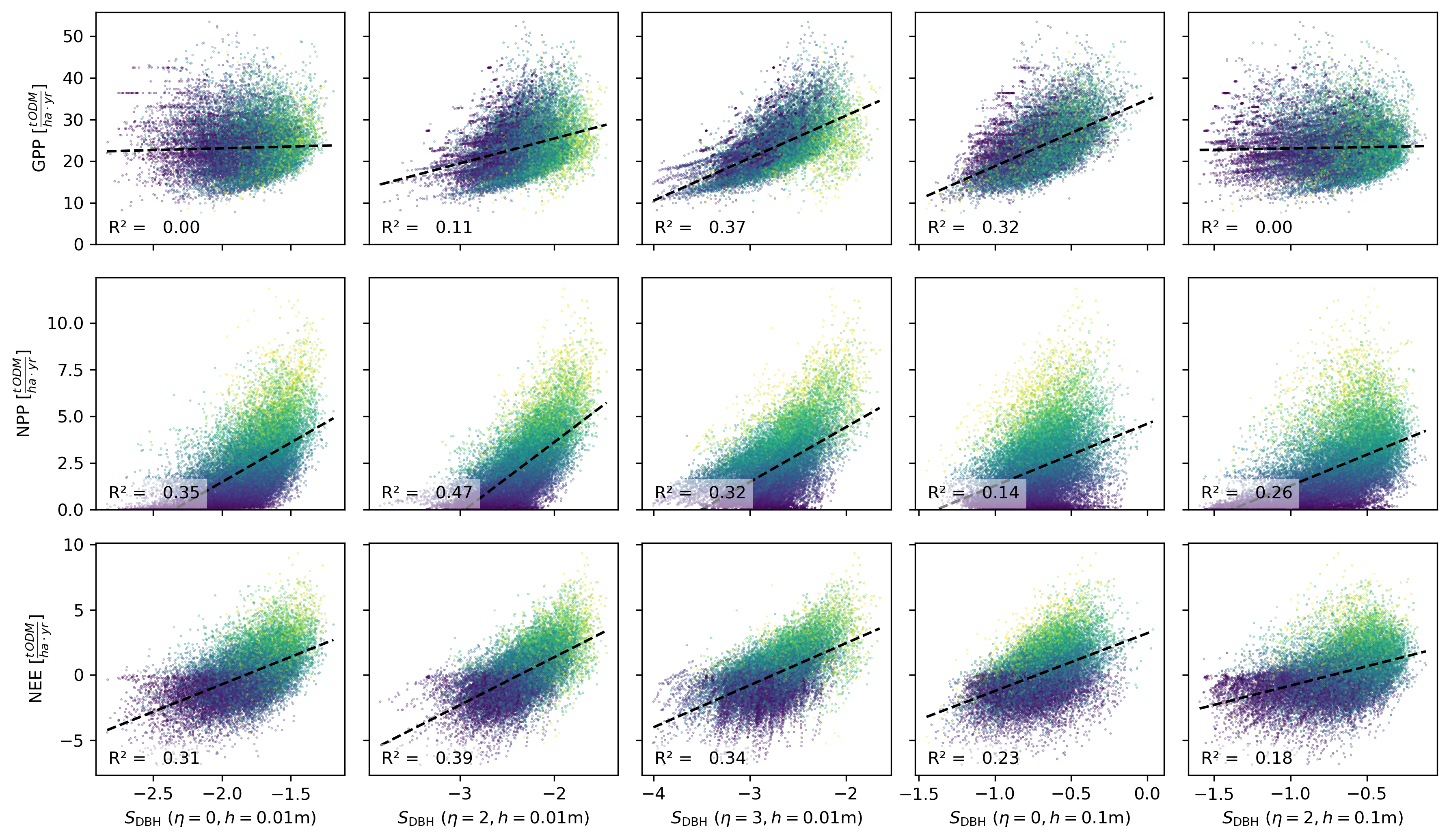}\caption{Productivity measures (GPP, NPP, and NEE) dependent on different parameters
of the DBH entropy $\protect\entropy_{\protect\mt{DBH}}$. Each dot
corresponds to a $\protect\un{0.04}{ha}$ forest patch. The colour
indicates the basal area proportion of mature trees (blue: only mature
trees; yellow: no mature trees). The relationship between entropy
and NPP or NEE is strongest if the entropy is computed with weights
based on the basal area ($\eta=2$) and a small bandwidth $h=\protect\un 1{cm}$,
at which trees are considered similar. \label{fig:DBH-Entropy-Comparison-0.04}}
\end{sidewaysfigure}

\begin{sidewaysfigure}
\includegraphics[width=1\textheight]{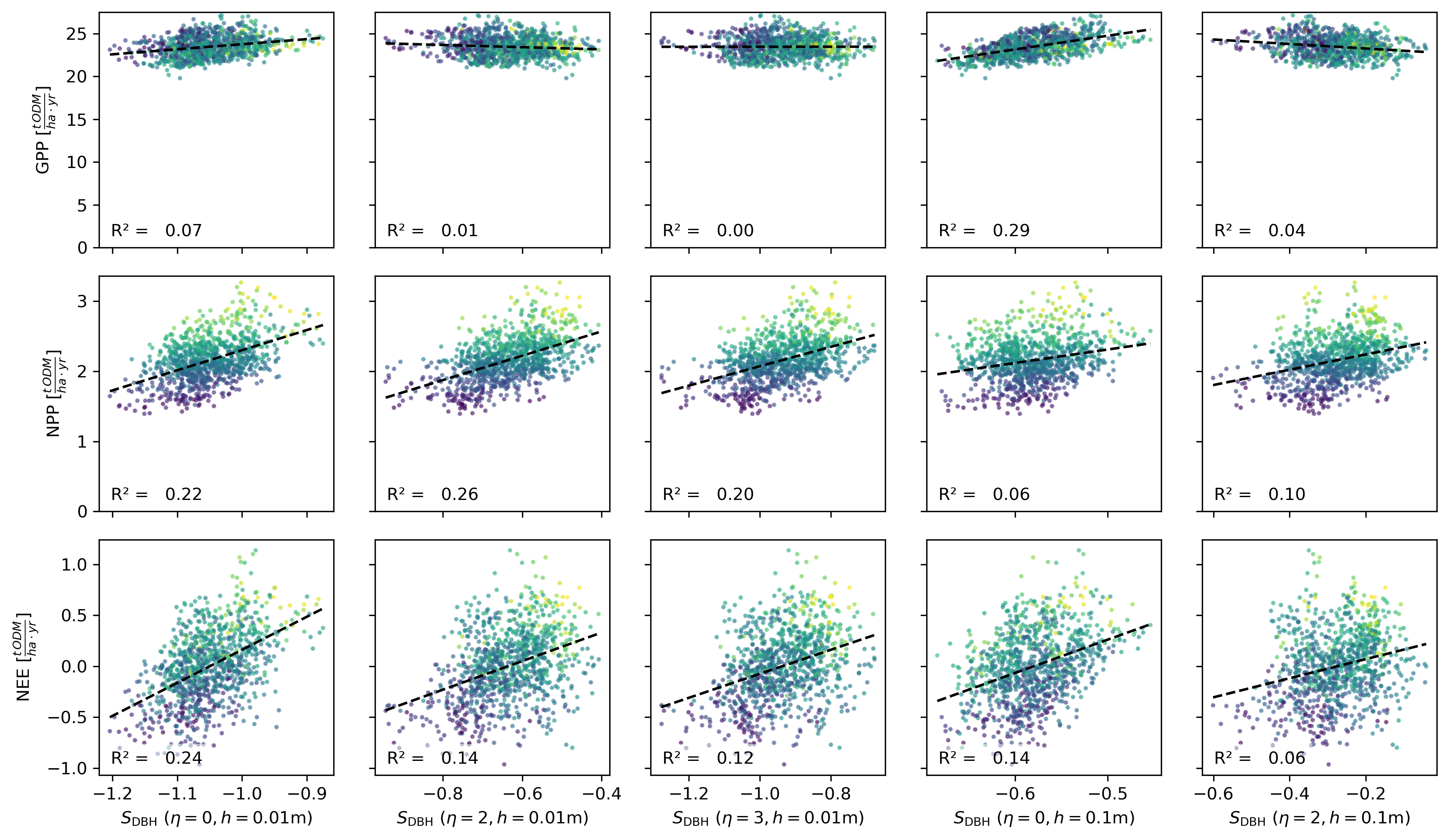}\caption{Productivity measures (GPP, NPP, and NEE) dependent on different parameters
of the DBH entropy $\protect\entropy_{\protect\mt{DBH}}$. Each dot
corresponds to a $\protect\un 1{ha}$ forest patch. The colour indicates
the basal area proportion of mature trees (blue: only mature trees;
yellow: no mature trees). The relationship between entropy and NPP
is strongest if the entropy is computed with weights based on the
basal area ($\eta=2$) and a small bandwidth $h=\protect\un 1{cm}$,
at which trees are considered similar. For the NEE, the relationship
is stronger if the unweighted DBH distribution is used ($\eta=0$).\label{fig:fig:DBH-Entropy-Comparison-1}}
\end{sidewaysfigure}

The results are displayed in Figs. \ref{fig:DBH-Entropy-Comparison-0.04}
and \ref{fig:fig:DBH-Entropy-Comparison-1} for the $\un{0.04}{ha}$
and the $\un 1{ha}$ scale, respectively. It is visible that weighting
the entropy by the basal area strengthened the relationhip with the
GPP and NEE on the fine scale; for the $\un 1{ha}$ scale the relationship
to the NEE became slightly weaker compared to the unweighted version
of the entropy. However, weighting with a higher exponent ($\eta=3$)
worsened the results. Using a larger bandwidth, i.e., counting more
trees as similar, worsened the connection between entropy and NPP
and NEE. This is notable, as many studies using the DBH entropy as
a measure for structural diversity consider the $\un 1{ha}$ scale
(or larger), use a large bandwidth (e.g. $\un{10}{cm}$; \citealp{silvapedroDisentanglingEffectsCompositional2017})
and do not weight the trees by basal area \citep[e.g.][]{danescu_structural_2016,silvapedroDisentanglingEffectsCompositional2017,park_influence_2019}.

\section{Model validation\label{sec:apx-validation}}

To verify that our optimization procedure reliably yields good fitting
results, we repeated the fitting procedure three times. We obtained
estimated log-likelihood values of $-4850.32$, $-4853.39$, and $-4866.18$,
respectively. Though already log-likelihood differences of $2$ are
significant in likelihood ratio tests and for confidence intervals,
we consider the fitting procedure successful, because the stochastic
optimization problem we needed to solve to fit the model is computationally
difficult. 

The parameter estimates we obtained in the three optimization runs
are displayed in Table \ref{tab:Parameter-estimates}. For most parameters,
the results remained in similar orders of magnitude, suggesting that
the parameters are estimable despite remaining uncertainties resulting
from the difficulty of the optimization problem. Only the parameter
$\th_{\mt{est},1}$ which controls the sharpness of the light threshold
for seedling establishment took on vastly different values. This suggests
that this parameter may not be estimable and may be set to a predefined
value without affecting the goodness of fit significantly.
\begin{table}
\begin{centering}
\hspace*{-0.4cm}\footnotesize%
\begin{tabular}{cc>{\centering}p{1.2cm}>{\centering}p{1.7cm}>{\centering}p{1.7cm}>{\centering}p{1.7cm}>{\centering}p{1.7cm}>{\centering}p{1.7cm}>{\centering}p{1.7cm}}
\toprule 
Parameter & Unit & Optimiza-tion run & Small shade intolerant & Large shade intolerant 1 & Large shade intolerant 2 & Large mid-tolerant & Small shade tolerant & Large shade tolerant\tabularnewline
\midrule
\midrule 
\multirow{3}{*}{$n_{\mt{seeds},i}$} & \multirow{3}{*}{$\frac{1}{\mt{ha}\cdot\mt{yr}}$} & 1 & $1.297$ & $9.997$ & $1.28$ & $4.346$ & $2.603$ & $3.409$\tabularnewline
 &  & 2 & $1.721$ & $7.384$ & $0.492$ & $4.019$ & $2.863$ & $3.415$\tabularnewline
 &  & 3 & $1.288$ & $12.061$ & $1.358$ & $4.367$ & $2.941$ & $3.318$\tabularnewline
\midrule 
\multirow{3}{*}{$\th_{\mt{est},0}$} & \multirow{3}{*}{$1$} & 1 & $0.0714$ & $0.202$ & $0.0807$ & $0.0091$ & $0.0405$ & $3.36\e{-4}$\tabularnewline
 &  & 2 & $0.0786$ & $0.1626$ & $0.0432$ & $0.0066$ & $0.0449$ & $1\e{-4}$\tabularnewline
 &  & 3 & $0.0737$ & $0.1962$ & $0.0864$ & $0.0116$ & $0.048$ & $0.002$\tabularnewline
\midrule 
\multirow{3}{*}{$\th_{\mt{production},i,1}$} & \multirow{3}{*}{$\frac{\mu\mt{Mol}\,\mt{phot.}}{\mt m^{2}\cdot\mt s}$} & 1 & $500$ & $118.56$ & $100$ & $70.82$ & $274.15$ & $492.73$\tabularnewline
 &  & 2 & $499.49$ & $112.94$ & $100$ & $61.11$ & $217.99$ & $500$\tabularnewline
 &  & 3 & $497.23$ & $107.87$ & $100$ & $61.05$ & $245.69$ & $401.76$\tabularnewline
\midrule 
\multirow{3}{*}{$\gamma_{i}$} & \multirow{3}{*}{$1$} & 1 & $0.15$ & $0.285$ & $0.4$ & $0.4$ & $0.189$ & $0.236$\tabularnewline
 &  & 2 & $0.15$ & $0.168$ & $0.4$ & $0.4$ & $0.15$ & $0.307$\tabularnewline
 &  & 3 & $0.15$ & $0.151$ & $0.376$ & $0.4$ & $0.193$ & $0.392$\tabularnewline
\midrule 
\multirow{3}{*}{$q_{\mt{\Delta DBH}}$} & \multirow{3}{*}{$1$} & 1 & \multicolumn{6}{c}{$0.991$}\tabularnewline
 &  & 2 & \multicolumn{6}{c}{$0.99$}\tabularnewline
 &  & 3 & \multicolumn{6}{c}{$0.985$}\tabularnewline
\midrule 
\multirow{3}{*}{$\th_{\mt{est},1}$} & \multirow{3}{*}{$1$} & 1 & \multicolumn{6}{c}{$3$}\tabularnewline
 &  & 2 & \multicolumn{6}{c}{$20$}\tabularnewline
 &  & 3 & \multicolumn{6}{c}{$3.547$}\tabularnewline
\bottomrule
\end{tabular}
\par\end{centering}
\caption{Parameter estimates resulting from the three optimization runs. For
most of the parameters, the estimates remained in the same order of
magnitude, indicating that they are estimable. Only the parameter
$\protect\th_{\protect\mt{est},1}$ took on largely different values.
This suggests that this parameter is not estimable. \label{tab:Parameter-estimates}}
\end{table}

To validate that our model fits the biomass and stem count distributions
from the forest inventory well, we compared a model-generated sample
of these values to the sample from the inventory data that was also
used in the fitting procedure. We simulated $\un 1{ha}$ of forest
for a burn-in period of $\un{2000}{yr}$ and sampled $25$ patches
($\un{0.04}{ha}$) of the simulated forest $1000$ times in time intervals
of $\un 5{yr}$. We repeated this procedure $8$ times, obtaining
a sample with $200,000$ entries, corresponding to a forest of $\un{8000}{ha}$. 

\begin{figure}
\begin{centering}
\includegraphics[width=1\textwidth,height=0.95\textheight]{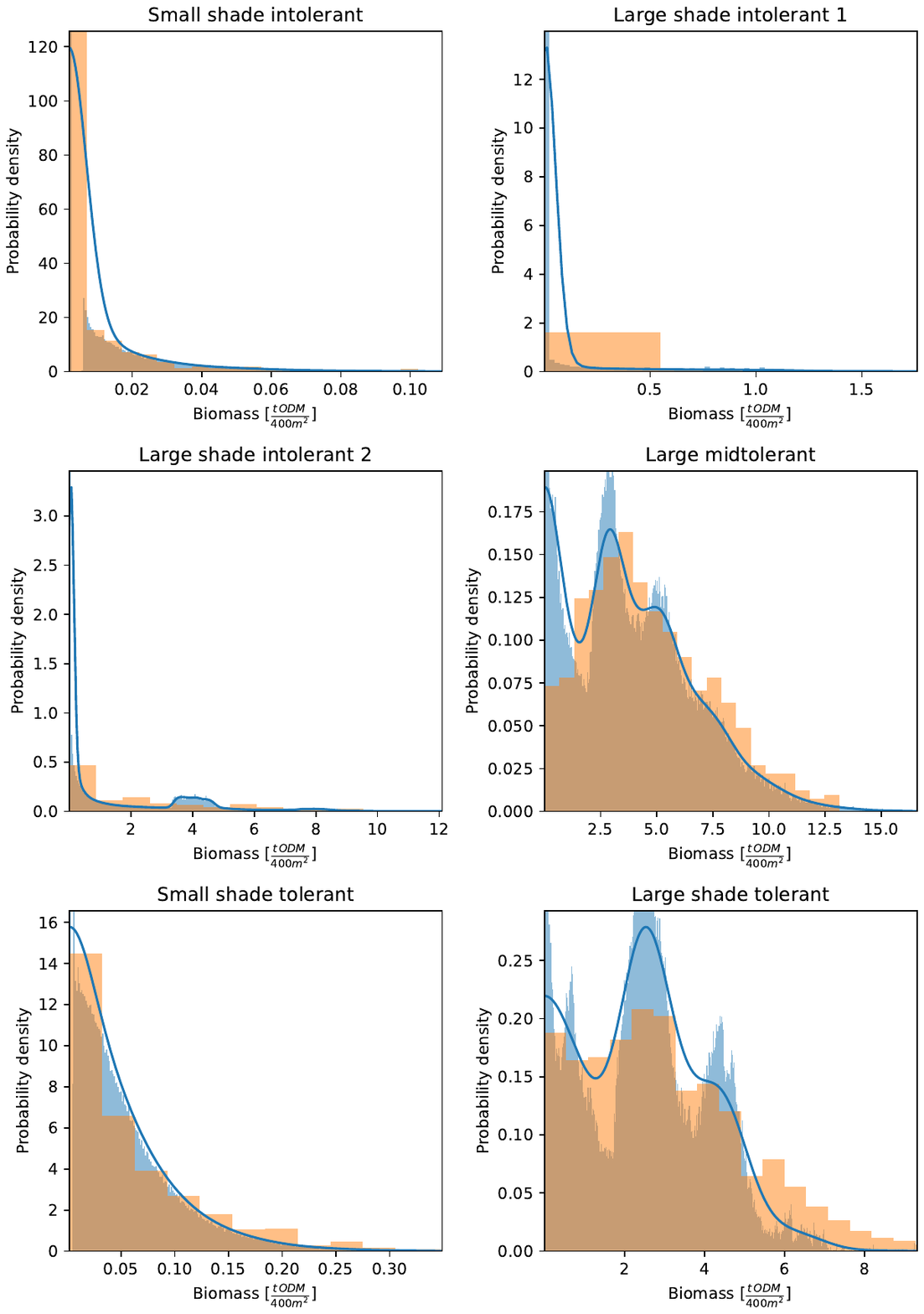}
\par\end{centering}
\caption{\scriptsize Comparison of the simulated biomass distribution with
field data. The figure displays the marginal biomass distribution
by PFT on the $\protect\un{0.04}{ha}$ scale. The orange bars form
histograms of the biomass estimates generated based on the field data
from Changbaishan. The blue bars form histograms of the biomass distributions
generated from the model. The blue curves depict the kernel-smoothed
density of the distribution used to estimate the likelihood. The distributions
obtained from the model generally match the corresponding distributions
of the field data well.\label{fig:biomass-dist}}
\end{figure}

\begin{figure}
\begin{centering}
\includegraphics[width=1\textwidth,height=0.95\textheight]{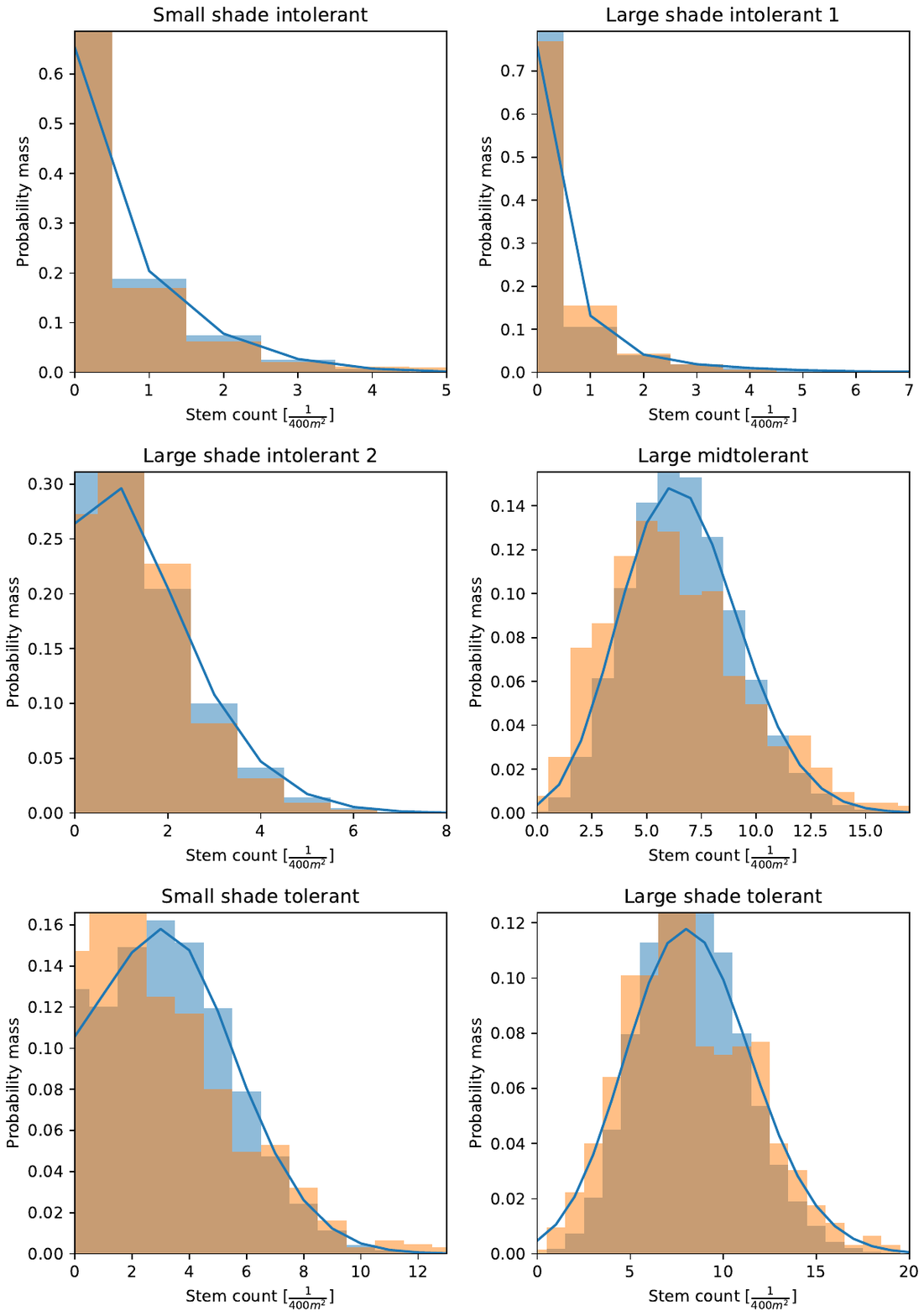}
\par\end{centering}
\caption{\scriptsize Comparison of the simulated stem count distribution with
field data. The figure displays the marginal stem count distribution
by PFT on the $\protect\un{0.04}{ha}$ scale. The orange bars form
histograms of the stem count estimates generated based on the field
data from Changbaishan. The blue bars form histograms of the stem
count distributions generated from the model. The blue curves depict
the kernel-smoothed density of the distribution used to estimate the
likelihood. The distributions obtained from the model generally match
the corresponding distributions of the field data well.\label{fig:stem-dist}}
\end{figure}

Based on the simulated data and the field data, we created one-dimensional
histograms of the biomass and stem count for each PFT. Then we plotted
these histograms to study how well they overlap. The results are displayed
in Figures \ref{fig:biomass-dist} and \ref{fig:stem-dist}, respectively.
The distributions match reasonably well, indicating a good model fit
in light of the model's complexity and the large number of model features
fitted simultaneously.

\begin{sidewaysfigure}
\includegraphics[width=1\textheight]{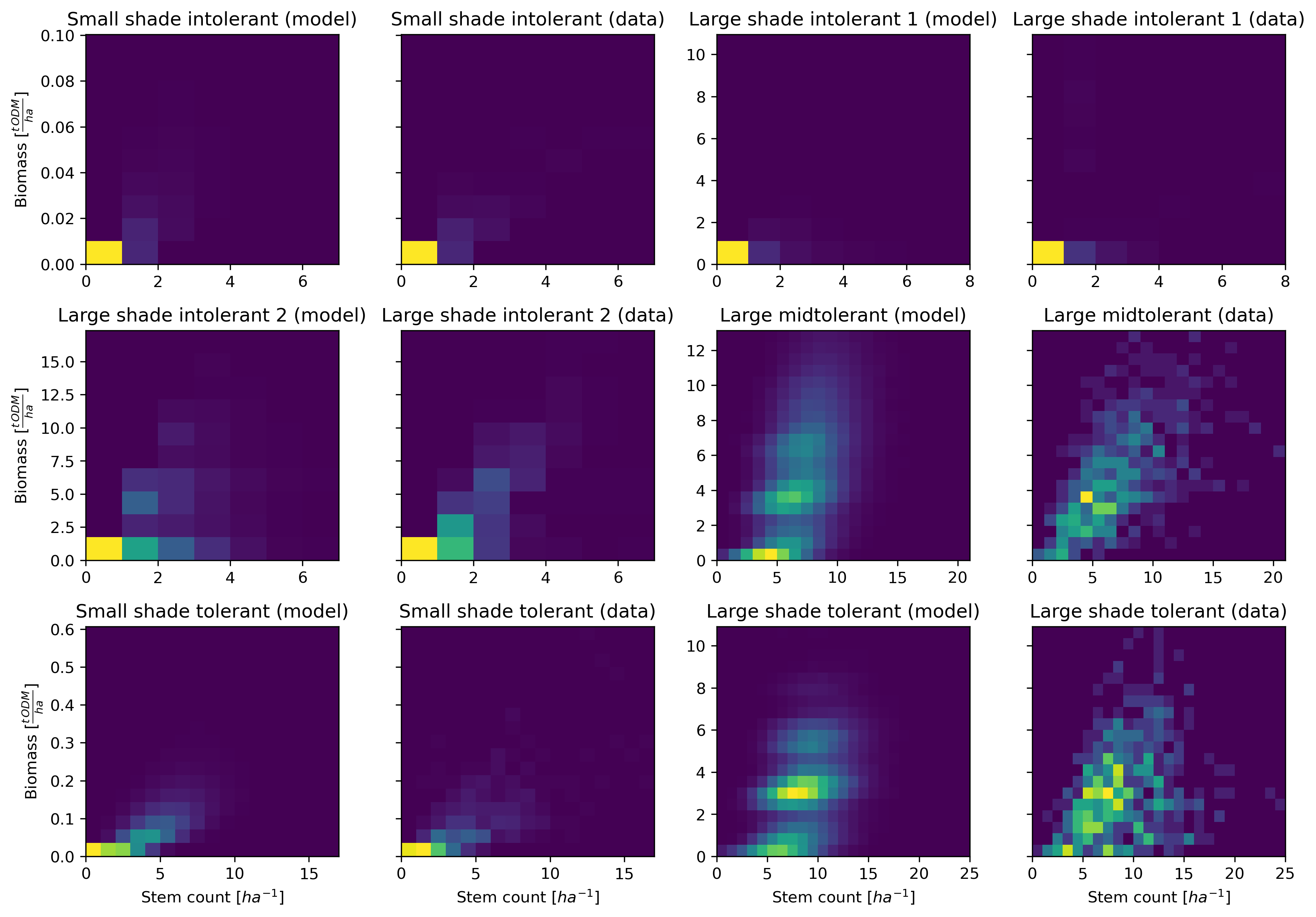}\caption{Comparison between model- and data-generated histograms of the joint
biomass and stem count distributions for the six PFTs. Columns 1 and
3 show histograms obtained from model-generated samples; column 2
and 4 show histograms obtained from the field data. The main features
of the histograms from the model and the data match, indicating a
reasonable model fit.\label{fig:joint-dist-histograms}}
\end{sidewaysfigure}

To also evaluate the model's ability to reproduce the joint distributions
of biomass and stem count for the six PFTs, we created corresponding
two-dimensional histograms, displayed in Fig. \ref{fig:joint-dist-histograms}.
The distributions from the model generally matched the patterns observed
in the field data. However, the field data often covered a broader
range of values than observed in the model simulation. This indicates
that some sources of variation are still missing in the model.

In addition to comparing the simulation results with forest inventory
data, we also computed stand-level forest characteristics (biomass,
NPP, GPP, and LAI), which we then compared to estimates from independent
studies (see main text). We considered a forest area of the same size
($\un{25}{ha}$) as the area where the inventory was conducted. We
simulated this forest for a burn-in period of $\un{1,000}{yr}$. Then,
we determined the forest characteristics of interest in each year
for a simulation period of $\un{3,000}{yr}$, yielding a quasi-independent
sample with $3,000$ entries. We then determined the sample mean and
standard deviation of each of the considered characteristics and used
the resulting values for model validation. 

\section{Further technical details\label{sec:Further-technical-details}}

\subsection{Computing the weights of the tree species in the inventory\label{subsec:Computing-the-weights}}

To derive allometric relationships for the different PFTs, we used
data available for individual species, weighted according to their
respective prevalence in the inventory. Here we describe how we computed
these weights.

For every sufficiently large tree $K$ in the forest inventory, we
added one unit of weight to the data points in the allometry dataset
that corresponded to trees of the same species with most similar DBH.
Trees with DBH below $5\mt{cm}$ were ignored, as they are not considered
in the model (see section \ref{subsec:Ingrowth}). Let $\alloset$
be an index set for the allometry dataset (ignoring entries with DBH
below $5\mt{cm}$) and $s_{k}$ be the species corresponding to $k\in\alloset$.
Let furthermore $\alloset_{k}^{+}=\left\{ \tilde{k}\in\alloset\,:\,\tilde{k}=s_{k},\,d_{\tilde{k}}>d_{k}\right\} $
the entries in the allometry dataset that correspond to the same species
and a larger DBH, and define $\alloset_{k}^{=}=\left\{ \tilde{k}\in\alloset\,:\,\tilde{k}=s_{k},\,d_{\tilde{k}}=d_{k}\right\} $
and $\alloset_{k}^{-}=\left\{ \tilde{k}\in\alloset\,:\,\tilde{k}=s_{k},\,d_{\tilde{k}}<d_{k}\right\} $
correspondingly for entries with equal or smaller DBH, respectively.
Define 
\begin{equation}
d_{k}^{+}=\begin{cases}
\mino{\tilde{k}\in\alloset_{k}^{+}}d_{\tilde{k}} & \text{if }\alloset_{k}^{+}\neq\emptyset\\
d_{k} & \text{else}
\end{cases}
\end{equation}
and
\begin{equation}
d_{k}^{+}=\begin{cases}
\maxo{\tilde{k}\in\alloset_{k}^{-}}d_{\tilde{k}} & \text{if }\alloset_{k}^{-}\neq\emptyset\\
d_{k} & \text{else}
\end{cases}
\end{equation}
as the smallest larger and the largest smaller DBH of an entry in
the allometry dataset corresponding to the same species. The contribution
$v_{Kk}$ of tree $K$ in the inventory to the weight of entry $k$
in the allometry dataset is given by 
\begin{equation}
v_{Kk}=\begin{cases}
1 & \text{if }d_{K}=d_{k},\\
1 & \text{if }d_{K}>d_{k}=d_{s_{k}}^{\mt{max}},\\
1 & \text{if }d_{K}<d_{k}=d_{s_{k}}^{\mt{min}},\\
\frac{d_{k}^{+}-d_{K}}{d_{k}^{+}-d_{k}} & \text{if }d_{K}\in\left(d_{k},d_{k}^{+}\right),\\
\frac{d_{K}-d_{k}^{-}}{d_{k}-d_{k}^{-}} & \text{if }d_{K}\in\left(d_{k}^{-},d_{k}\right),\\
0 & \text{else.}
\end{cases}
\end{equation}
That is, the contribution is $1$ if the diameters are equal or if
the the tree diameter is outside the range of diameters covered in
the allometry dataset and the allometry data entry has maximal or
minimal diameter, respectively. The weights are then computed as follows:
\begin{equation}
w_{k}=c_{\mt{class}\ap{s_{k}}}\smo{K\in\invset_{s_{k}}}\frac{v_{\kappa k}}{\left|\alloset_{k}^{=}\right|},
\end{equation}
where $c_{\mt{class}\ap{s_{k}}}$ is a normalization constant for
the PFT $\mt{class}\ap{s_{k}}$ to which species $s_{k}$ belongs,
$\invset_{s_{k}}$ is the subset of trees in the inventory that are
of species $s_{k}$, and $\left|\cdot\right|$ denotes the counting
norm. The division by the cardinality of $\alloset_{k}^{=}$ distributes
the contribution of tree $K$ evenly among all allometry entries with
similar species and diameter. As a result, each tree in the inventory
makes the same total contribution to the weights.

The normalization constants $c_{\mt{class}\ap{s_{k}}}$ do not affect
parameter estimation, but we chose 
\begin{equation}
c_{j}=\frac{\left|\alloset_{j}\right|}{\smo{k\in\alloset_{j}}\smo{K\in\invset_{s_{k}}}\frac{v_{\kappa k}}{\left|\alloset_{k}^{=}\right|}}
\end{equation}
so that the sum of the weights corresponds to the size of the dataset
used to fit the allometry curve for PFT $j$. As a result, the likelihood
computed using the weights may be of the same order of magnitude as
the unweighted likelihood, which can be helpful for model comparison
and selection.

To compute the weights efficiently, we sorted both the allometry dataset
and the inventory by tree DBH and species. Then, the weights can be
computed in linear time of the inventory dataset size (assuming that
there are only few entries in the allometry dataset that have both
the same species and DBH).

\subsection{Assignment of new seeds to patches\label{subsec:Assignment-of-seeds}}

Each year, a constant number of seeds is distributed evenly to the
different modelled forest patches. If the provided seed number is
not an integer divisible by the number of simulated patches, the seed
number is rounded stochastically for each patch so that the expected
number of seeds per hectare and PFT matches the provided seed number.
That is, if $n_{\mt{seeds},i}$ is the number of seeds per hectare
for PFT $i$ and $n_{\mt{patches}}$ the number of simulated patches,
then the number of seeds for a patch $j$ is given by 
\begin{equation}
n_{\mt{seeds},i,j}=\left\lfloor \frac{n_{\mt{seeds},i}}{n_{\mt{patches}}}\right\rfloor +B_{p_{\mt{seed}}},
\end{equation}
where 
\begin{equation}
B_{p_{\mt{\mt{seed}}}}\sim\bernoullidist{p_{\mt{\mt{seed}}}}
\end{equation}
is a Bernoulli distributed random variable with success probability
\begin{equation}
p_{\mt{\mt{seed}}}=\frac{n_{\mt{seeds},i}}{n_{\mt{patches}}}-\left\lfloor \frac{n_{\mt{seeds},i}}{n_{\mt{patches}}}\right\rfloor .
\end{equation}

\end{document}